\documentclass[showpacs,onecolumn,preprintnumbers,amsmath,amsfonts,amssymb,floatfix,aps,superscriptaddress]{revtex4}
\usepackage{amsmath}
\usepackage{amsfonts}
\usepackage{amssymb}
\usepackage{physymb}
\usepackage{graphicx}
\usepackage{epsfig}
\usepackage{siunitx}
\usepackage[hang,nooneline]{subfigure}

\newcounter{fig}
\newcommand{\lbfig}[1]{\refstepcounter{fig}           .
}

\begin{document}

\title{Rogue Waves
  in Ultracold Bosonic Seas}

\author{E. G. Charalampidis}
\affiliation{Department of Mathematics and Statistics, University of Massachusetts
Amherst, Amherst, MA 01003-4515, USA}

\author{J. Cuevas-Maraver}
\affiliation{Grupo de F\'{i}sica No Lineal, Departamento de F\'{i}sica Aplicada I,
Universidad de Sevilla. Escuela Polit\'{e}cnica Superior, C/ Virgen de \'{A}frica, 7, 41011-Sevilla, Spain \\
Instituto de Matem\'{a}ticas de la Universidad de Sevilla (IMUS). Edificio Celestino Mutis. Avda. Reina Mercedes s/n, 41012-Sevilla, Spain}

\author{D. J. Frantzeskakis}
\affiliation{Department of Physics, National and Kapodistrian University of Athens, Panepistimiopolis,
Zografos, Athens 15784, Greece}

\author{P. G. Kevrekidis}
\affiliation{Department of Mathematics and Statistics, University of Massachusetts
Amherst, Amherst, MA 01003-4515, USA}

\date{\today}

\begin{abstract}
In this work, we numerically consider the initial value problem for
nonlinear Schr{\"o}dinger (NLS) type models arising in the physics of
ultracold boson gases, 
with generic Gaussian wavepacket initial data.
The corresponding Gaussian's width and, wherever relevant also its
amplitude, serve as control parameters.
First we explore the one-dimensional, standard
NLS equation with general power law nonlinearity, in which large amplitude
excitations {\it reminiscent} of Peregrine solitons or regular solitons
appear to form, as the width of the relevant Gaussian is varied.
Furthermore, the variation of the nonlinearity exponent aims
at a first glimpse of the interplay between rogue or soliton
formation and collapse features.
The robustness of the main features to noise in the initial data
is also confirmed. To better connect our study
with the physics of atomic condensates, and explore
the role of dimensionality effects, we also consider
the nonpolynomial Schr{\"o}dinger equation (NPSE),
as well as
the full three-dimensional NLS
equation, and 
examine the degree to which relevant considerations generalize.
\end{abstract}

\maketitle

\section{Motivation and Background}

Over the past decade, the study of extreme wave events and patterns
known as rogue or freak waves, has constituted
one of the focal points of both intense theoretical analysis and
a wide range of physical applications~\cite{k2a,k2b,k2c}. In particular, such structures have emerged
in a diverse host of experiments carried out in a broad array
of physical systems including, but not limited to nonlinear optics
\cite{opt1,opt2,opt3,opt4,opt5}, mode-locked lasers \cite{laser},
superfluid helium \cite{He}, hydrodynamics \cite{hydro,hydro2,hydro3}, Faraday surface ripples
\cite{fsr}, parametrically driven capillary waves \cite{cap},
and plasmas \cite{plasma}. 
In addition, an abundance of theoretical investigations followed the
seminal work of Peregrine~\cite{H_Peregrine}, Kuznetsov~\cite{kuz}, Ma~\cite{ma},
and Akhmediev~\cite{akh}, as well as Dysthe and Trulsen~\cite{dt},
which examined rational solutions of prototypical dispersive nonlinear systems,
such as the nonlinear Schr{\"o}dinger (NLS) equation.
A large part of the considerable volume of theoretical research has
by now been summarized in a number of reviews~\cite{yan_rev,solli2,onorato}.

At the same time, the last twenty years have seen a tremendous
growth of interest in the study of solitary waves in the realm
of ultracold bosonic atom gases and Bose-Einstein condensates
(BECs)~\cite{book1,book2,dumitr1,emergent,book_new}, as well as, 
more recently, damped-driven (open system) siblings, namely exciton-polariton
condensates~\cite{rmpcarus}. This is because -- especially so at the
Hamiltonian atomic setting of ultracold and dilute enough gases -- an
accurate mean-field description 
gives rise to the NLS model, typically in the presence of a trap;
in this context, the NLS is usually referred to as Gross-Pitaevskii equation (GPE)
\cite{book1,book2}. Moreover, depending on the type of interatomic interactions
(repulsive or attractive),
which is controlled by the sign of the $s$-wave scattering length,
a self-defocusing or self-focusing nonlinearity in the equation emerges,
leading to a wide array of potentially relevant nonlinear wave structures.
Among the ones that have been experimentally verified and intensely
theoretically studied, we can classify the one-dimensional (1D) bright~\cite{expb1,expb2,expb3}, gap~\cite{gap} and dark~\cite{djf} matter-wave solitons. Moreover, their higher
dimensional analogues, namely vortices~\cite{fetter1,fetter2},
in the two-dimensional (2D) setting, as well as 
vortex lines and 
vortex rings in the three-dimensional (3D) setting~\cite{komineas_rev},
are also 
particularly interesting and relevant, and 
accessible to experiments (see the recent book \cite{book_new} and references therein).

On the other hand, there is a large volume of theoretical work devoted to studies on
rogue waves in atomic BECs. Relevant investigations, following the
fundamental in this context study
of Ref.~\cite{blkon}, include studies in single-component
\cite{bec1,bec2,bec3,bec4,bec5,bec6,bec7},
binary mixtures \cite{2c1,2c2,2c3,2c4,2c5,2c6,2c7,2c8}, as well as
three-component and spinor BECs \cite{sp1,sp2,sp3}, in quasi-1D settings.
Moreover, rogue waves in higher-dimensional BECs, namely in quasi-2D
\cite{2dbec} and 3D \cite{3dbec} nonautonomous settings, were also studied.
Nevertheless, intriguingly enough, while attractive atomic BECs are
natural candidates to support rogue waves (as they are
described by focusing NLS type models) we are not aware of an effort to produce
{\it experimentally} a rogue wave -- 
e.g., a Peregrine soliton in a condensate of $^7$Li or $^{85}$Rb atoms.
This may structurally have to do with the
difficulty of preparing the initial background state leading to such a rogue wave.
In the case of the Peregrine soliton,
before it ``appears out of nowhere'' and after it ``disappears
without a trace'' (features often alluded to rogue waves \cite{a1,a2}),
the background state is an unstable uniform one
that naturally leads to instabilities and the formation of
patterns (see, e.g., also the 
discussion in Ref.~\cite{zak} and the recent work \cite{mantza}).
Hence, a natural question to ask is whether these two themes,
the rogue wave patterns and the atomic condensate realm, may
possess a nontrivial interaction point whereby a suitable
initial condition could be utilized to produce a pattern strongly
reminiscent, e.g., of a Peregrine soliton.
This is our starting point motivating the present study.

More concretely, a brief description of
our investigations and presentation of this work is as follows.
We focus on a well-defined array of physically
realistic numerical experiments in an array of models, relevant to
the BEC physics, which are summarized in Section II.
Our initial condition has the generic form of a matter wave packet
that is supported by attractive BECs \cite{book1,book2},
namely
a Gaussian one, parametrized chiefly by its width (but in some cases, as relevant
and as will be seen below, also potentially by its amplitude). When the width
of this wavepacket is large, we expect it to self-focus and potentially
yield a large amplitude event; the latter will be compared (favorably,
as will be illustrated below) to a Peregrine soliton, although
connections of the evolution with $N$-soliton solutions will also
be sought. On the other hand, if the width is small, we expect the
pattern to adjust to a solitonic wavepacket. In this way, in some
sense, we seek a rogue-wave-to-soliton transition (in Section III)
as the width of the wavepacket is varied in the system. This phenomenology
constitutes the backbone of our observations herein. Subsequently,
we explore different types of variations to this theme. Initially,
we examine the role of noisy perturbations to the initial conditions.
Here we see that such perturbations are of progressively more limited
role as the width is decreased, but overall importantly they do not
destroy the relevant phenomenology. Another dimension of our numerical
explorations in the same Section concerns the interplay of the above
features with another class of
extreme events that are potentially
supported by NLS type models, namely self-focusing and wave collapse. To that effect, we examine
also the power-law generalization of the NLS model (where the
power of the nonlinearity is characterized by a general exponent),
so as to enable the extreme wave formation to co-exist with
unstable solitons and stable collapse events. This will be observed below to potentially lead
to strong focusing events even in the case of subcritical NLS models.
In section IV, we broach more concretely
the possibility of realization of these observations
in a physical system of atomic BECs. To do so, we go beyond the
1D mean-field, NLS-based approximation to a more realistic study of
the so-called non-polynomial Schr{\"o}dinger equation (NPSE); the latter,
was introduced some time ago to take into regard
the effect of the deviation from one-dimensionality on the longitudinal
BEC dynamics~\cite{Salasnich}.
Importantly, results obtained in the framework of the NPSE model
compare favorably with ones obtained from the full 3D GPE, 
as well as experimental observations on dark solitons in single-component
repulsive BECs~\cite{kip},
dark-bright solitons in binary repulsive BECs~\cite{pul}
and, more recently, on bright solitons in single-component attractive BECs~\cite{RH};
thus, the NPSE is appreciated to be a considerably improved approximation towards
capturing the fully 3D dynamics.
For comparison here, we present both the NPSE and the 3D GPE
results; these suggest that large amplitude events with the same
type of initial data are possible, but at the same time, some of
the more ``delicate'' phenomenology of the original NLS model
appears to be lost. Finally, in Section V, we summarize our
findings and present some considerations and possibilities for
future studies.

\section{Analytical Considerations}

\subsection{The one-dimensional case: the models and analytical setup}

We start by presenting our 
1D models and the analytical setup. To this end, the first model of interest
is the 
1D NLS equation with a focusing, power-law nonlinearity given by
\begin{equation}
i\partial_{t}u =-\frac{1}{2}\partial_{x}^2 u -|u|^{2\delta}u,
\label{nlse_1d}
\end{equation}
where $u=u(x,t)$ is the (complex) field envelope and $\delta$
determines the nonlinearity power. We will start by exploring
the integrable case of $\delta=1$, and consider the cases with
$\delta>1$ afterwards (see the discussion about numerical results
in Sec.~\ref{num_res}). Here we should point out that for
$\delta=1$ Eq.~(\ref{nlse_1d}) reduces to the GPE, 
with $u(x,t)$ standing for the macroscopic wavefunction of a 1D attractive BEC \cite{emergent}.
Nevertheless, as explained above, an investigation of cases with $\delta \ne 1$
which will be discussed below, is expected to shed light on the interplay of the
emergence of extreme events with nonlinearity. This investigation is partly motivated
by the formation of matter-wave bright solitons during the collapse of attractive
higher-dimensional BECs (see the experimental work \cite{expb3} and theoretical results in
Ref.~\cite{corni}): such a collapse may also occur in the supercritical 1D NLS model
with $\delta>1$ \cite{sulem}.

In order to turn to a more realistic class of models for atomic BECs,
as discussed above, we will also consider the 
NPSE model, describing quasi-1D BECs \cite{Salasnich}; this equation is given by:
\begin{equation}
i\partial_{t}u =-\frac{1}{2}\partial_{x}^2u+V(x)u+\frac{1-(3/2)|u|^2}{\sqrt{1-|u|^2}}u.
\label{NPSE}
\end{equation}
Here, we will also include the effect of
the external potential $V(x)$, assuming the typical
harmonic form of $V(x)=\frac{1}{2}\Omega^{2} x^{2}$ with normalized
trap strength $\Omega$. The presence of the potential is a 
common ingredient for experimental realizations in BECs,
and given our aim in this work to establish comparisons
with experimental $^7$Li BECs, we naturally include this feature.
Note that $\Omega=0$ corresponds to the case where the trap is absent,
while $\Omega=\Omega_0$ with $\Omega_0=0.0193$ is a typical value used in
experiments (see, e.g., Ref.~\cite{RH})~\cite{foot1}.
%
%

A crucial feature of our exploration is the
assumed Gaussian form of our initial wavepacket, namely: 
\begin{equation}
u(x,t=0)=\alpha\exp{\left(-\frac{x^{2}}{2\sigma^{2}}\right)},
\label{gauss_init}
\end{equation}
with amplitude $\alpha$ and width $\sigma$.
Our main focus in what follows will be to consider variations of
$\sigma$ and their corresponding impact on the dynamics. However,
in some cases (especially, as concerns the realistic BEC problem),
we will also explore variations of $\alpha$.

The main emphasis of the present study will be on the classification
of the resulting pulses obtained numerically into solitons or, potentially, rogue
waves; the former, are described by a $\sech$-profile (i.e., the customary
bright or fundamental soliton), and the latter by the
algebraically decaying
Peregrine soliton \cite{H_Peregrine}. Both of these correspond to exact solutions to
the NLSE~(\ref{nlse_1d}), and we thus briefly mention the functional
form of both waveforms utilized herein. The bright soliton at hand
is given by
\begin{equation}
u(x,t) = A\sech{[A(x-x_{0})]} \exp\left[i(A^2/2)t\right],
\label{soliton_fit}
\end{equation}
with position of the pulse $x_{0}$ and amplitude $A$. In the
following, we keep fixed the position of the pulse, thus
setting $x_{0}=0$. On the other hand, and as per the rogue wave
solutions themselves, we will be utilizing a
form of the Peregrine soliton (used, e.g., in Ref.~\cite{kamal_per}),
involving a free parameter $P_{0}$, set by the boundary conditions
(i.e., $|u|^2 \rightarrow P_0$ as $x \rightarrow \pm \infty$);
%
%
this solution is of the form:
\begin{equation}
u(x,t)=\sqrt{P_{0}}\left[1-\frac{4 (1+2 i P_0 t)}{1+4 P_0 x^{2}
+4 P_0^2 t^2}\right] e^{i P_0 t}.
\label{par_peregrine}
\end{equation}
%
%

In what follows, we will be interested in investigating whether
the waveforms found numerically match a soliton or a Peregrine rogue wave pattern.
To that effect, we will isolate the above presented structures
at $t=0$ and utilize, respectively $A$ and $P_0$ as fitting
parameters to obtain the ``best fit soliton'' or the
``best fit Peregrine'' and discuss which fit is most suitable
in each case as we vary $\sigma$.

\subsection{The three-dimensional case: the model and analytical setup}
When generalizing our 1D NPSE considerations
to 3D BECs, we will examine the radially symmetric 3D NLS/GP
equation, written in cylindrical coordinates, $\left( \rho ,x\right) $, as follows:
\begin{equation}
i \frac{\partial \psi }{\partial t}=-\frac{1}{2}\left(
\frac{\partial ^{2}\psi }{\partial \rho ^{2}}+\frac{1}{\rho }\frac{\partial
\psi }{\partial \rho }+\frac{\partial ^{2}\psi }{\partial x^{2}}\right)
+V(\rho ,x)\psi -\left\vert \psi \right\vert
^{2}\psi ,  \label{dyn3D}
\end{equation}
together with the 3D potential
\begin{equation}
V(\rho ,x)=\frac{1}{2}(\rho ^{2}+\Omega^{2}x^{2}).
\end{equation}
Motivated by the cigar-shaped BEC setup analyzed in Ref.~\cite{RH},
we fix the value of the trap strength to be of $\Omega=\Omega_0=0.0193$.

In analogy to Ref.~\cite{Salasnich}, and as far as the initialization of the
dynamics is concerned in this case, we employ Gaussian-like initial conditions of the form of
\begin{equation}\label{init_3D}
\psi(\rho,x,t=0)=\frac{\exp\left[-\rho^2/2\eta(x)^2\right]}{\sqrt{\pi}\eta(x)}\phi(x),
\end{equation}
with functions $\phi$ and $\eta$ given by
\begin{equation}
\phi(x)=\alpha\exp{\left(-\frac{x^{2}}{2\sigma^{2}}\right)},\quad \eta(x)=(1-|\phi(x)|^2)^{1/4},
\end{equation}
respectively. Note that the function $\phi$ is purely a Gaussian
with amplitude $\alpha$ and width $\sigma$, which are again
the canonical parameters of variation in what follows.

\section{Numerical Results and Discussion}
\label{num_res}

\subsection{The integrable case: $\delta=1$}
In this section we present numerical results on the dynamics
of the NLS Eq.~(\ref{nlse_1d}) for $\delta=1$, which is
initialized by the Gaussian pulse~(\ref{gauss_init}) 
with $\alpha=1$. We study the underlying initial value problem (IVP) as
a function of $\sigma$ and focus on the interval $\sigma\in[0.3,30]$;
see, also, Ref.~\cite{nls_1_movie} for a complete movie of the
dynamics. Furthermore, a fitting process is employed in order
to optimally identify as well as classify the reported waves into
solitons or rogue waves, as relevant.
To that effect, the exact and stationary
solutions given by Eqs.~(\ref{soliton_fit}) and (\ref{par_peregrine})
will be utilized, with $A$ and $P_0$ as optimization parameters
as discussed above.

\begin{figure}[!pt]
\begin{center}
\mbox{\hspace{-0.5cm}
\subfigure[][]{\hspace{-1.0cm}
\includegraphics[height=.17\textheight, angle =0]{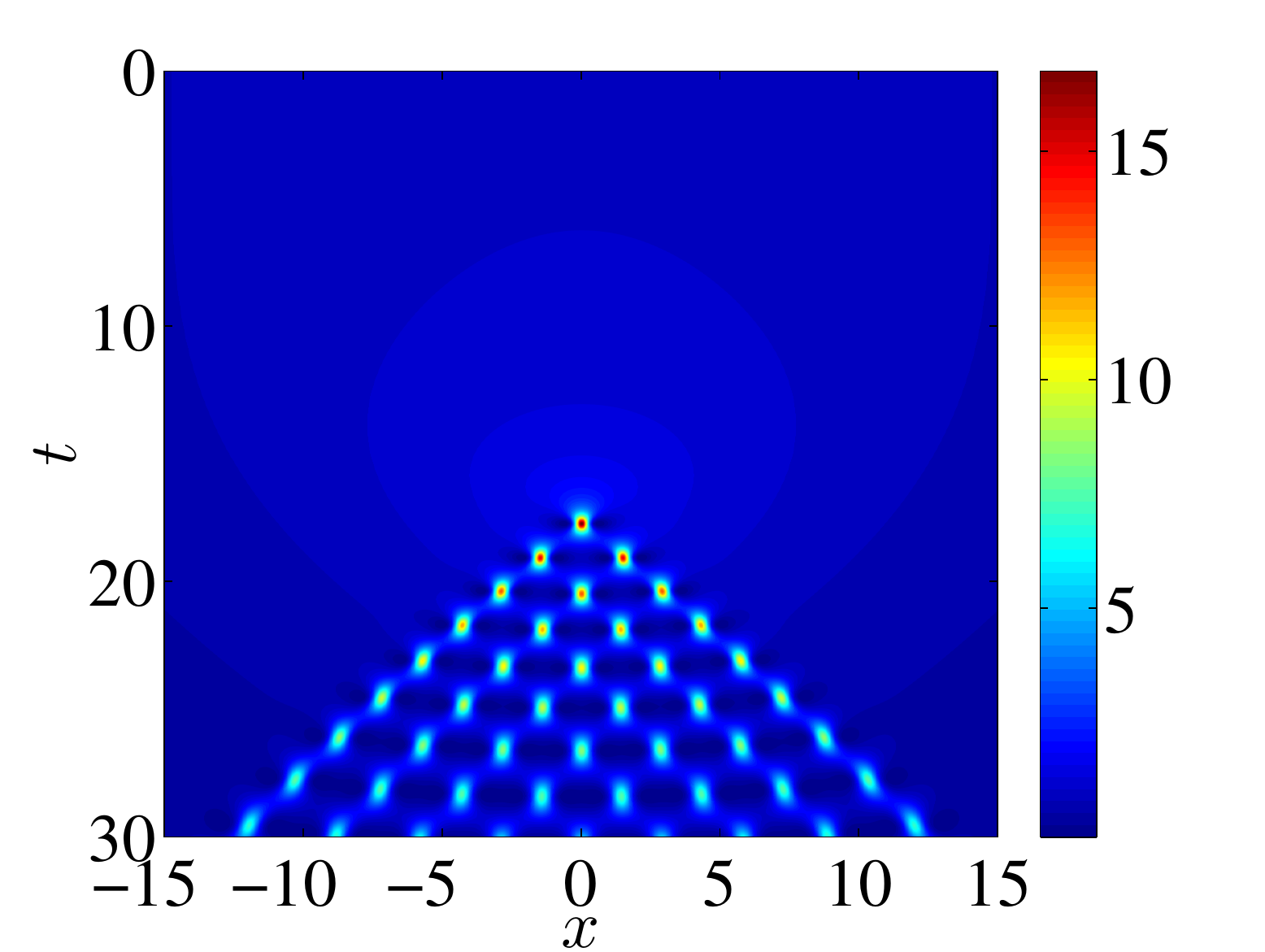}
\label{fig1a}
}
\subfigure[][]{\hspace{-0.5cm}
\includegraphics[height=.17\textheight, angle =0]{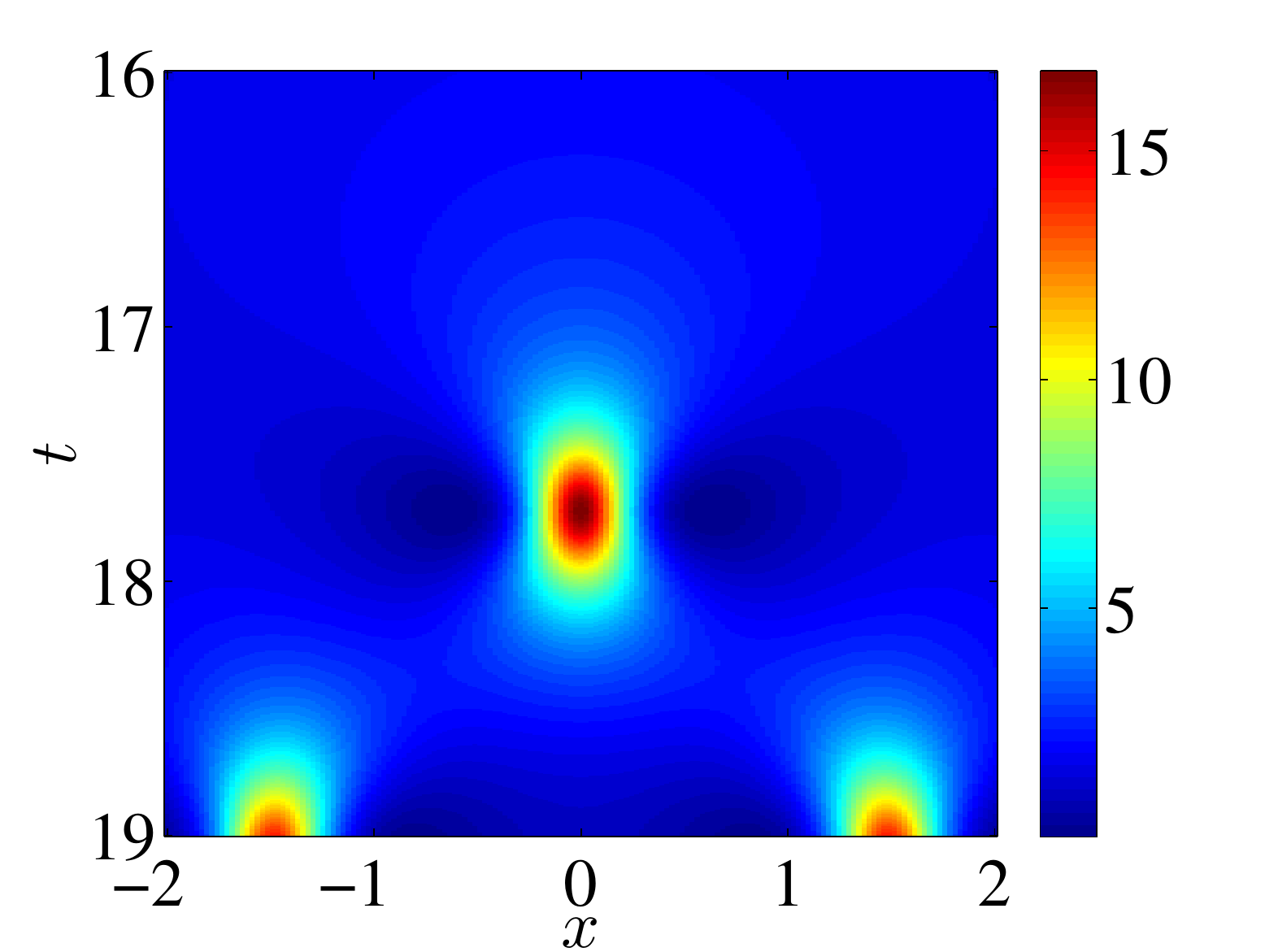}
\label{fig1b}
}
\subfigure[][]{\hspace{-0.5cm}
\includegraphics[height=.17\textheight, angle =0]{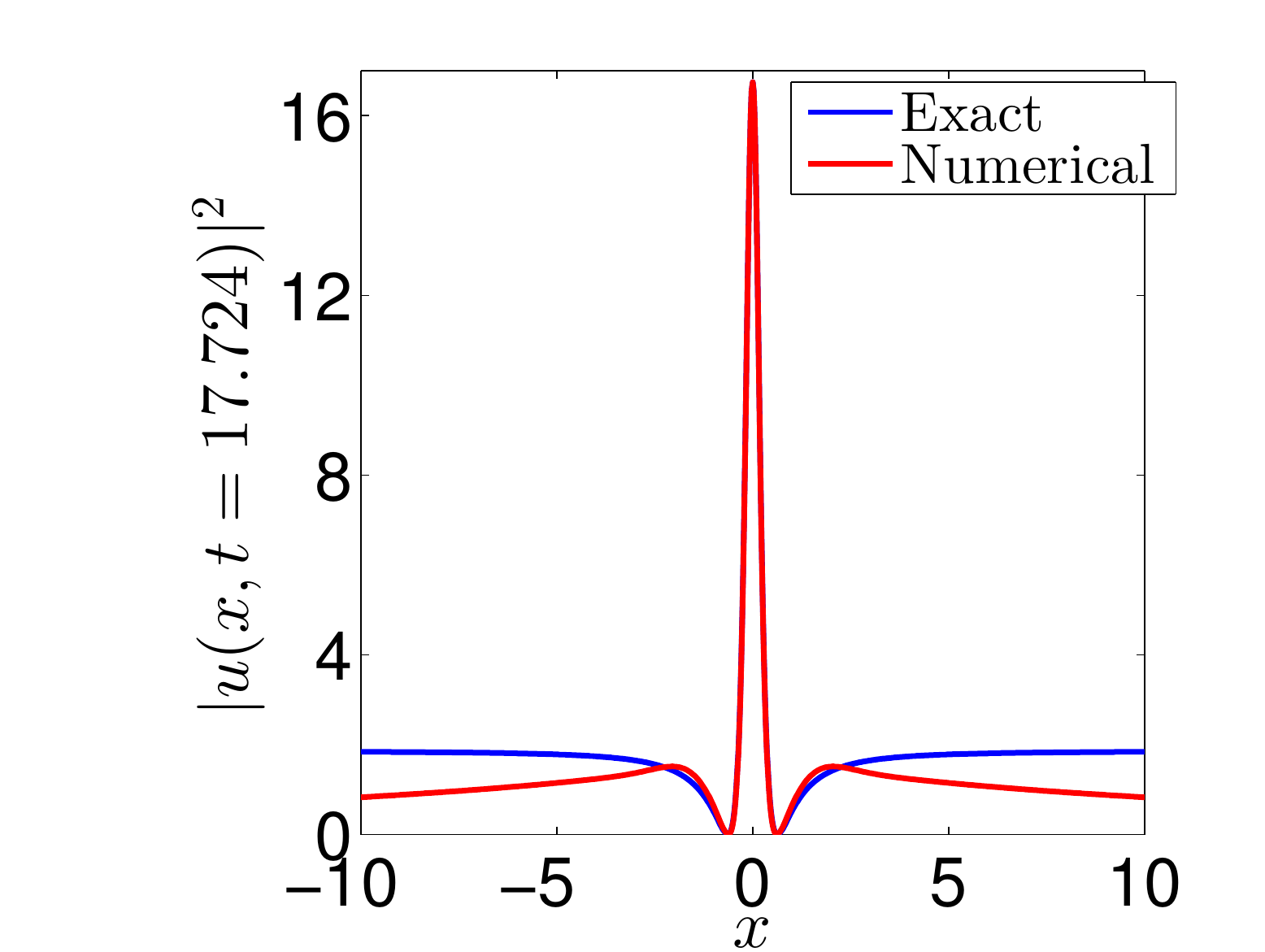}
\label{fig1c}
}
}
\vspace{0.0cm}
\mbox{\hspace{-0.5cm}
\subfigure[][]{\hspace{-1.0cm}
\includegraphics[height=.17\textheight, angle =0]{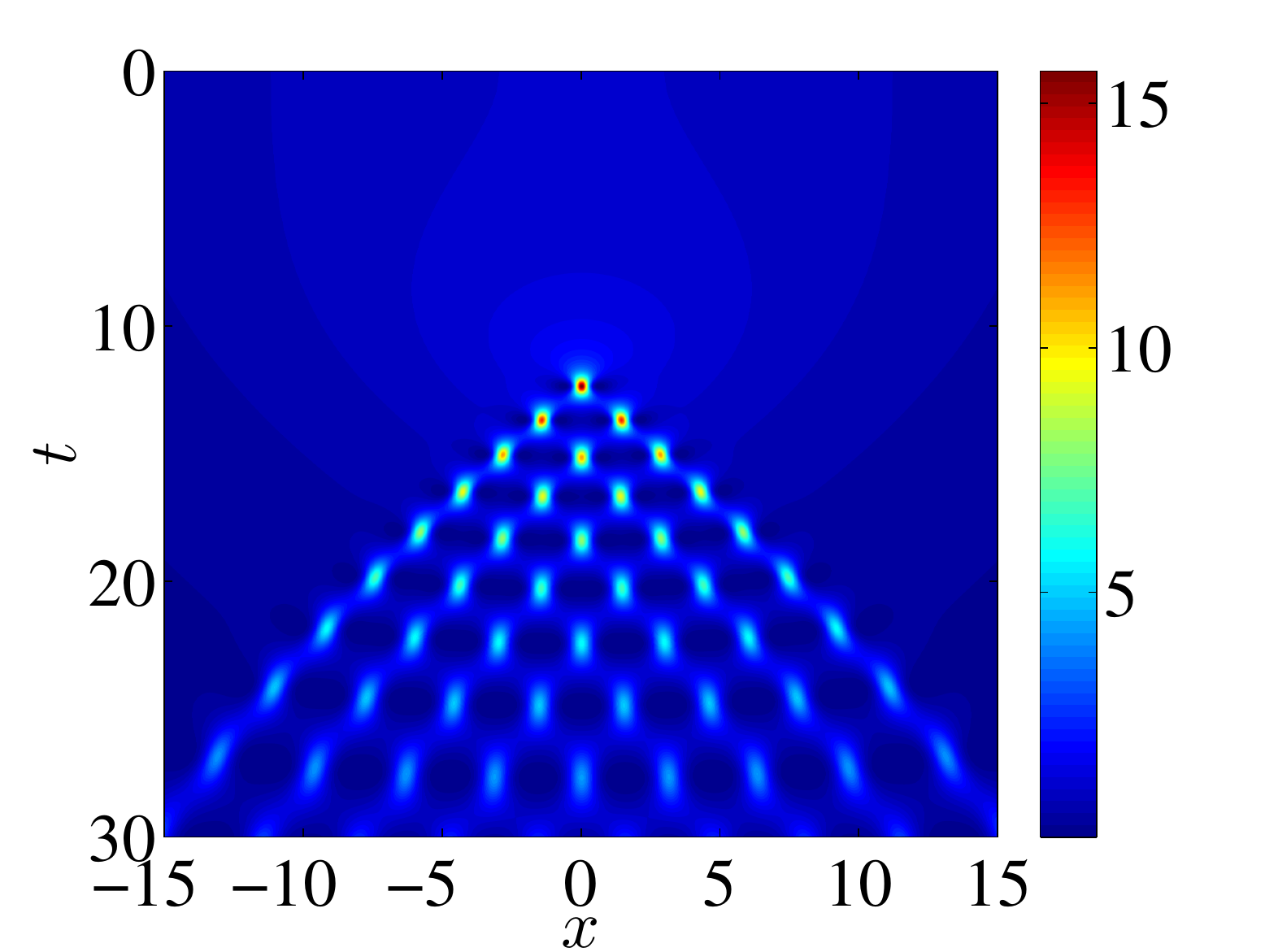}
\label{fig1d}
}
\subfigure[][]{\hspace{-0.5cm}
\includegraphics[height=.17\textheight, angle =0]{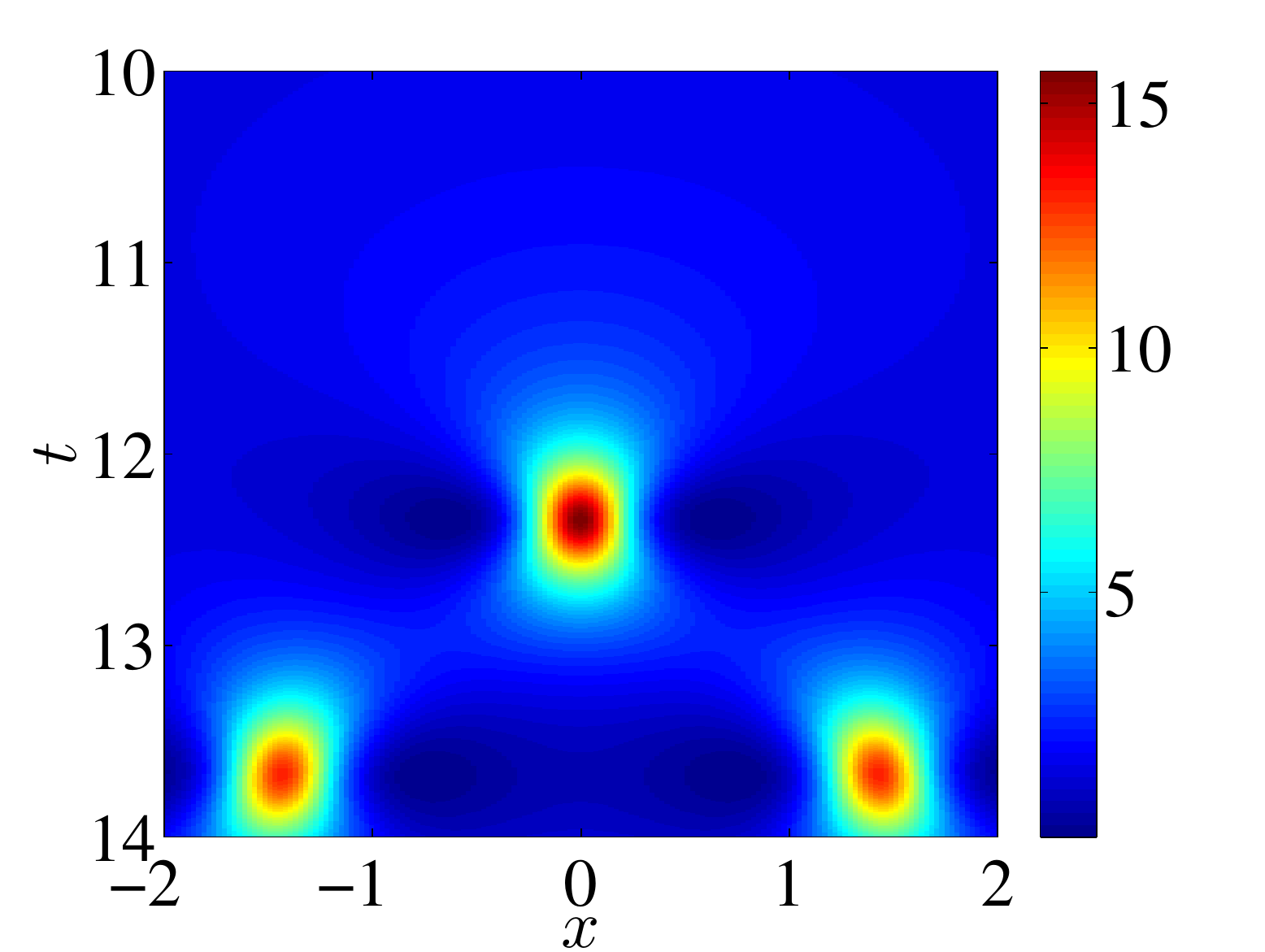}
\label{fig1e}
}
\subfigure[][]{\hspace{-0.5cm}
\includegraphics[height=.17\textheight, angle =0]{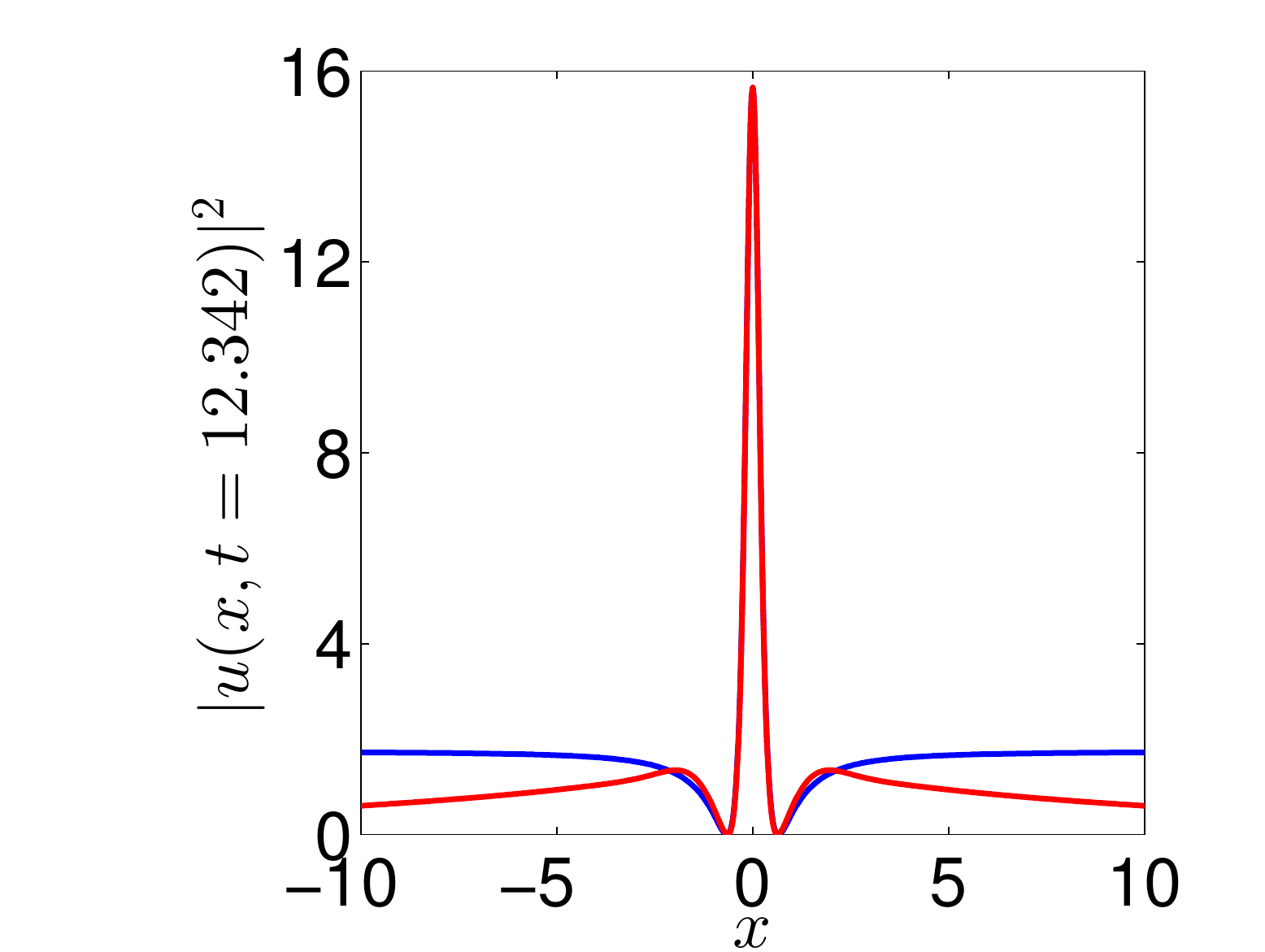}
\label{fig1f}
}
}
\end{center}
\caption{(Color online)
Summary of results corresponding to $\sigma=30$ (top row) and
$\sigma=20.1$ (bottom row). Spatiotemporal evolution of the
density $|u|^{2}$ and its zoom-ins are presented in the left
and middle panels, respectively. Spatial distribution of the
density at $t=17.724$ and $t=12.342$ (at which the first
peak is formed) is depicted by solid red
lines in the right panels. The densities originated from the
best-fit Peregrine for this extreme event
are plotted too by solid blue lines,
for comparison.
\label{fig1}
}
\end{figure}

We now present our results,
by considering first the rogue wave regime, that is, the interval of $\sigma$ where the
first high-amplitude waveform obtained numerically best fits
to a Peregrine soliton. 
The top and bottom panels of Fig.~\ref{fig1} highlight example
cases belonging to this class, with the monitored $|u(x,t)|^{2}$
corresponding to $\sigma=30$ and $\sigma=20.1$, respectively.
It can be discerned from the left panels of the figure, as well as their
respective zoom-ins in the middle panels, that the Gaussian pulse tends to focus and
is progressively transformed into a high-peaked wave (in its density),
surrounded by two local minima.
As the right panels illustrate, the resulting structure
can be deemed as faithfully representing
(the core of) a Peregrine soliton.
Specifically, the
densities at times $t=17.724$ (top) and $t=12.342$ (bottom) of the
respective cases are shown with solid red lines, 
while the exact ones, coming from the one-parameter family of rogue waves~(\ref{par_peregrine}),
are presented too with solid blue lines, for comparison. These panels
suggest a fairly good agreement between the two.

At this point, some clarification is necessary: the Peregrine structure ``lives''
on top of a finite background while our Gaussian has a decaying tail.
Hence, the fitting process is only considered in an interval around
the core, as is also evident by the disparity of the tails of the two
structures. For this reason, the Peregrine characterization should
be taken with a grain of salt. Clearly, the resulting feature
is an extreme event. However, it is not a ``genuine'' Peregrine,
but admittedly something that very strongly resembles a Peregrine
structure near its core. In fact, in this same core, it far more
resembles (both in its decay and in its non-monotonicity and
the structure of its side-blobs) to a Peregrine than to a best fit
soliton.

Furthermore, it should
be noted that such representations tend to appear at earlier times as
$\sigma$ decreases (see, the accompanying movie~\cite{nls_1_movie}).
However, an expanding structure (which we will refer to as a
``Christmas tree'' (CT)) appears to emerge
past the formation of the original (Peregrine-resembling) peak.
As the structure expands, progressively at the peak emergence
times more localized peaks arise (i.e., we go from one to two,
then to three, and so on.


%

\begin{figure}[!pt]
\begin{center}
\mbox{\hspace{-0.1cm}
\subfigure[][]{\hspace{-1.0cm}
\includegraphics[height=.17\textheight, angle =0]{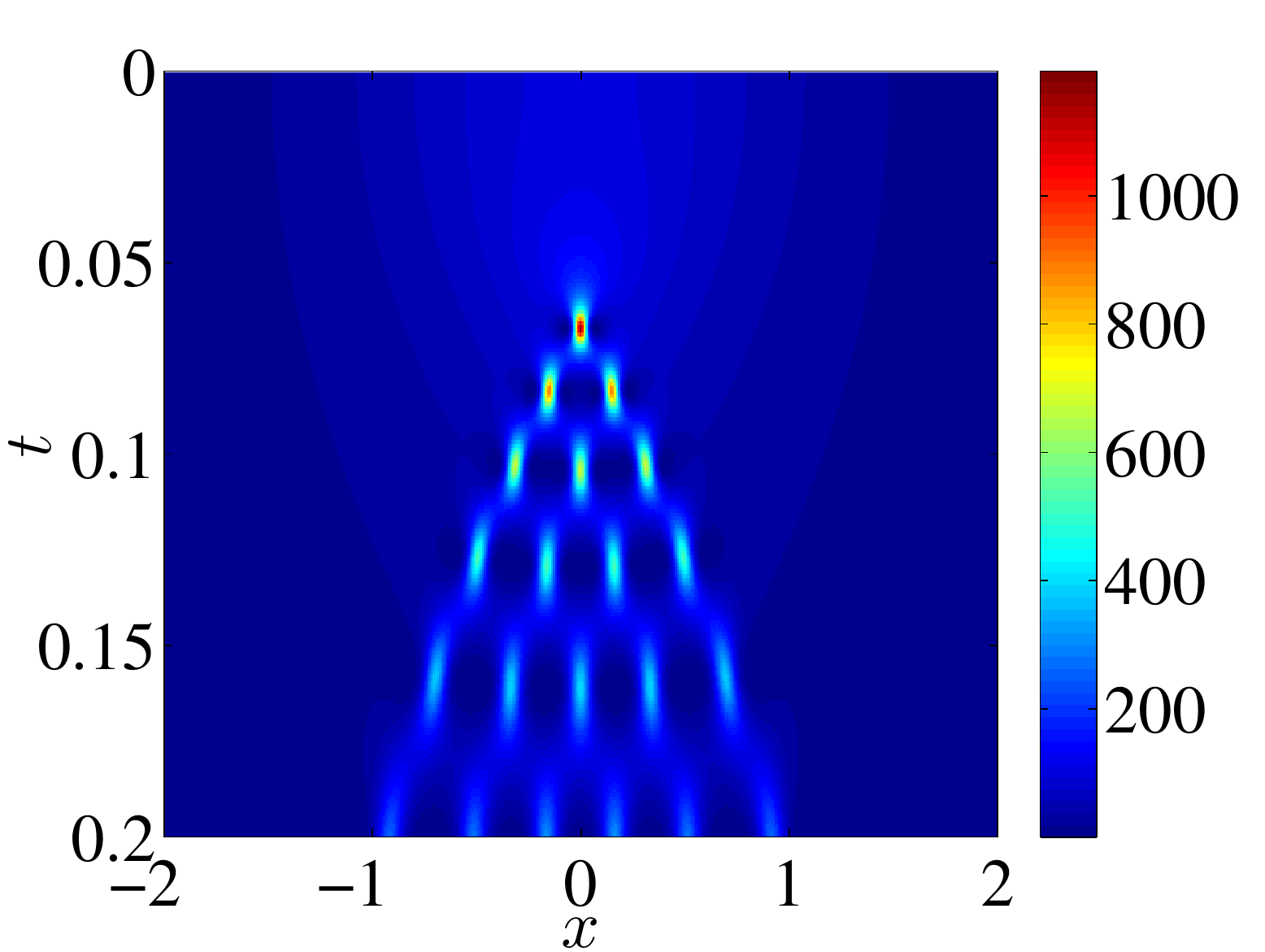}
\label{fig5a}
}
\subfigure[][]{\hspace{-0.2cm}
\includegraphics[height=.17\textheight, angle =0]{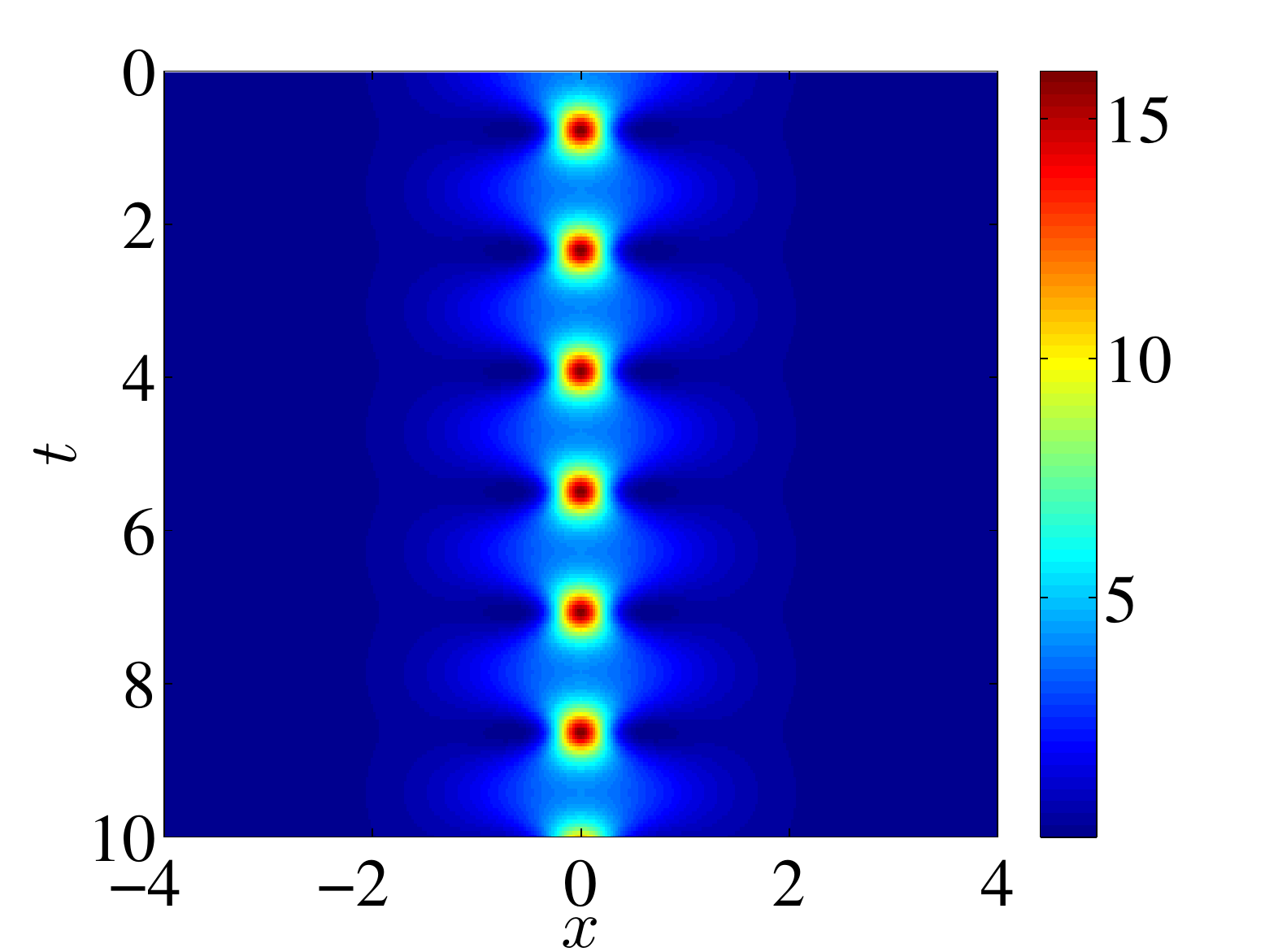}
\label{fig5b}
}
}
\end{center}
\caption{(Color online)
Spatiotemporal evolution of the density $|u|^{2}$ based on
initial data of the form of $u(x,t=0)=N\sech(x)$ \cite{Sats_Yaji}.
Left and right panels correspond to $N=10$ and $N=2$, respectively.
\label{fig5}
}
\end{figure}

A related remarkable (in our view) observation, however,
is that in addition to the strong resemblance of the
peak-structure in Fig.~\ref{fig1} with a Peregrine rogue wave
(at the time of its peak, as well as in its appearance and disappearance),
it is also strongly reminiscent of a multi-soliton pattern.
In particular, we compare these profiles to
$N$-soliton solutions~\cite{Sats_Yaji} initialized at $t=0$
as $u(x,t=0)=N\sech(x)$.
To establish comparisons with the $N$-soliton solution, we
report, for reference, the case with $N=10$ in Fig.~\ref{fig5a},
and also the case of $N=2$ in Fig.~\ref{fig5b}.
Notably, the CT structure is clearly present in such a multi-soliton
solution for $N$ sufficiently large. Nevertheless, it is remarkable
that the structure on the one hand has features strongly reminiscent
of $N$-soliton initial conditions, yet at the same time it has
this ``centaurian'' quality that its constituents are closely
approximated by the algebraically decaying Peregrine wave structures.

%

\begin{figure}[!pt]
\begin{center}
\mbox{\hspace{-0.5cm}
\subfigure[][]{\hspace{-1.0cm}
\includegraphics[height=.17\textheight, angle =0]{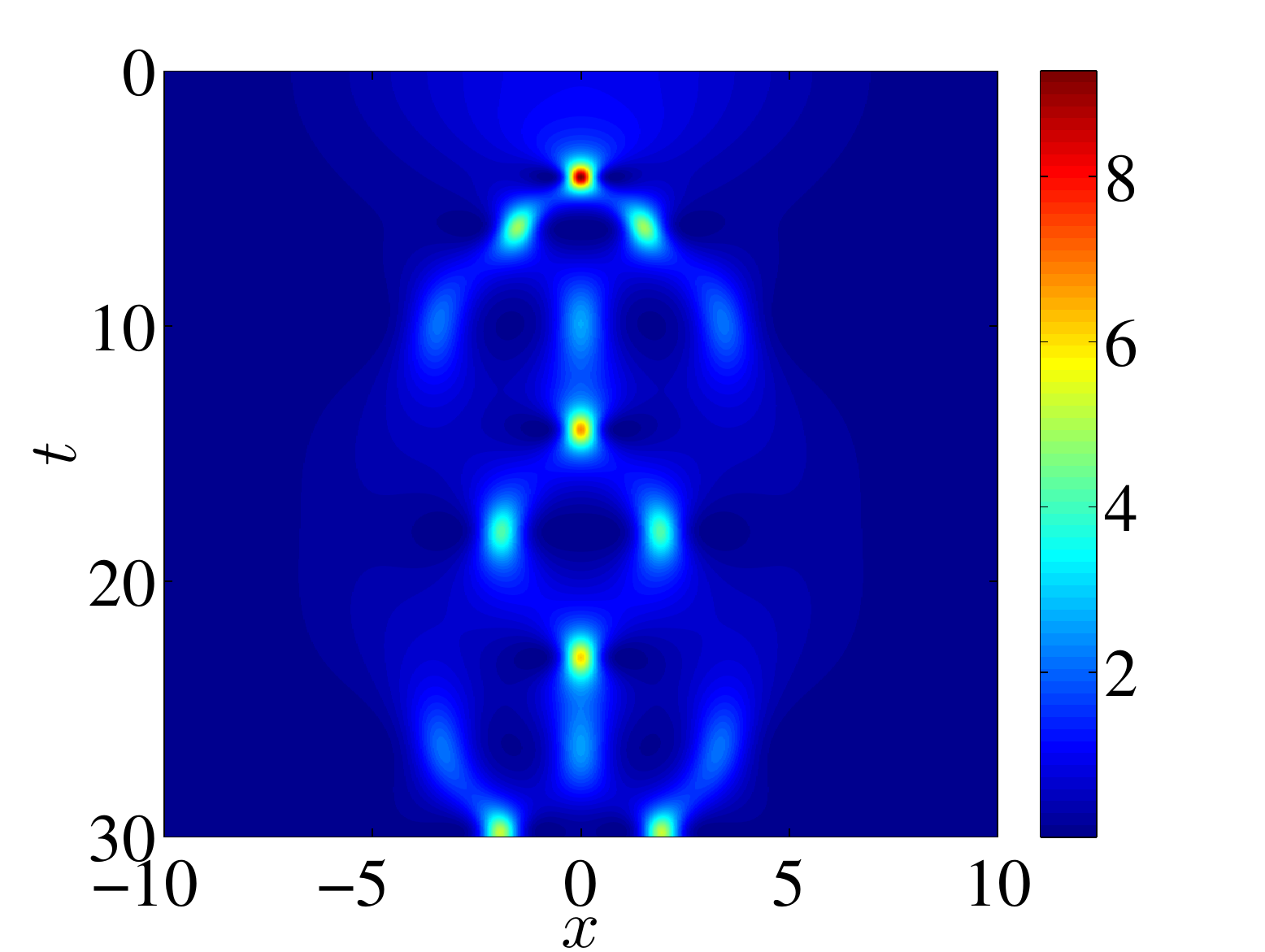}
\label{fig2a}
}
\subfigure[][]{\hspace{-0.5cm}
\includegraphics[height=.17\textheight, angle =0]{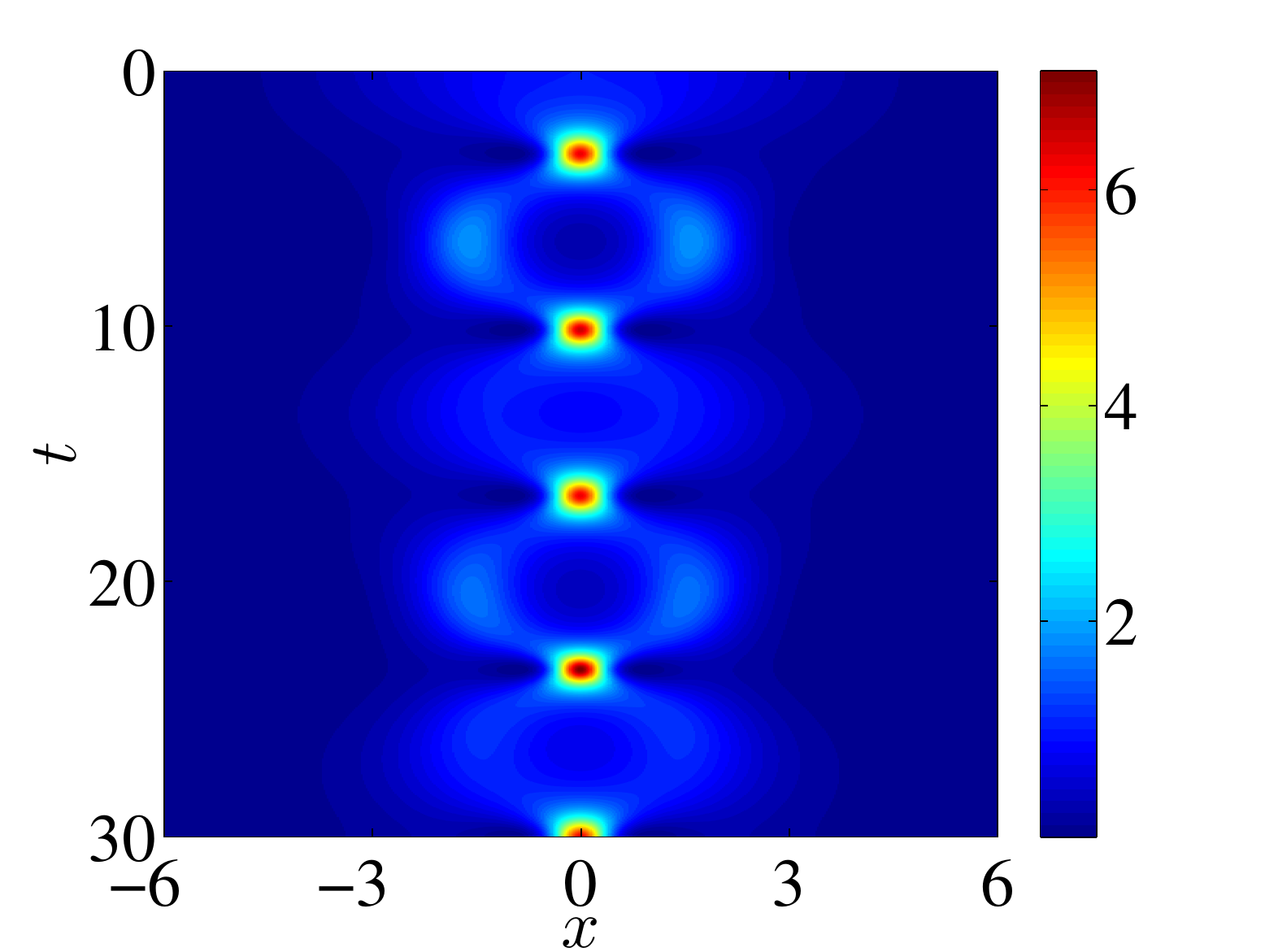}
\label{fig2b}
}
\subfigure[][]{\hspace{-0.5cm}
\includegraphics[height=.17\textheight, angle =0]{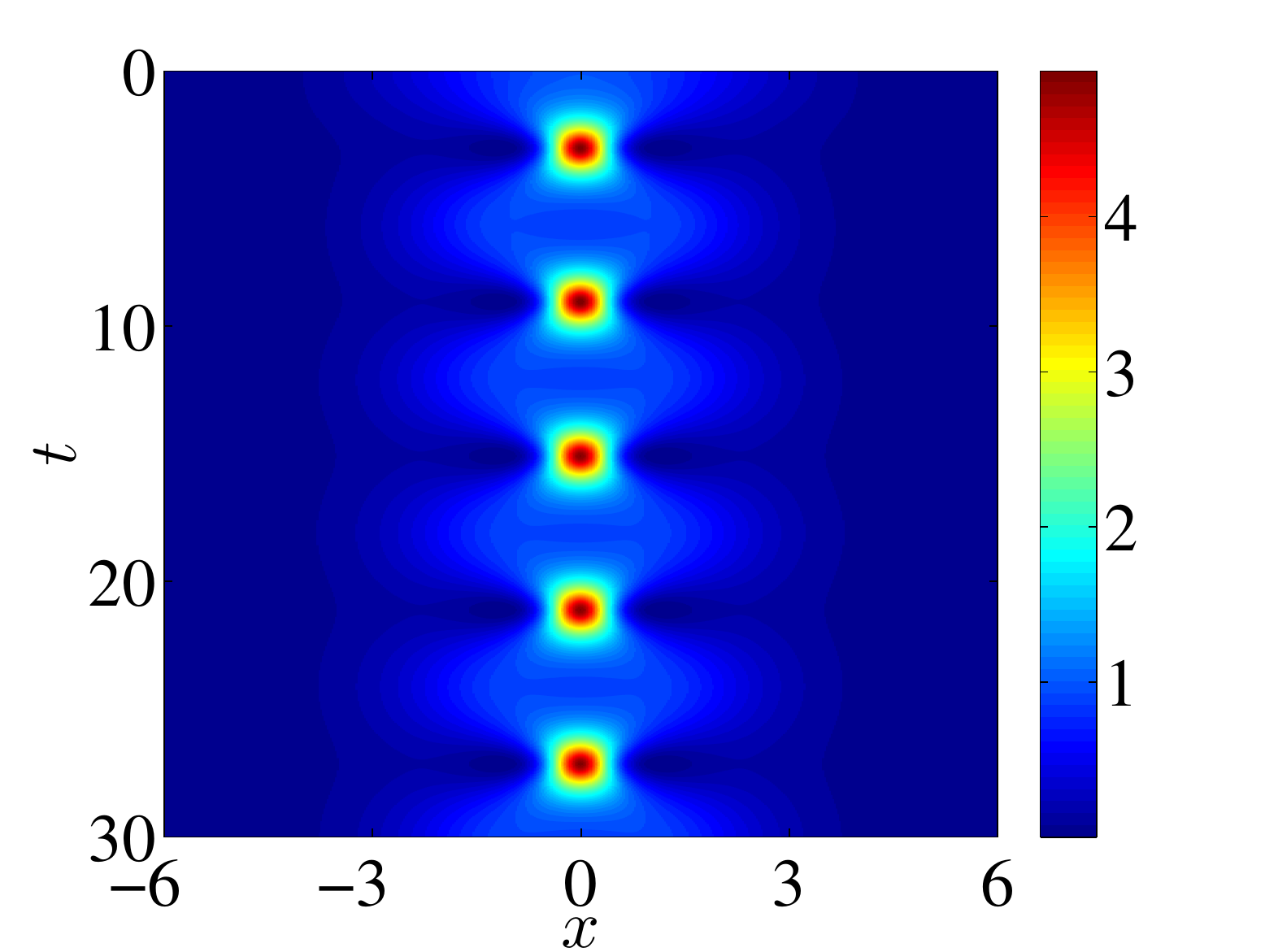}
\label{fig2c}
}
}
\vspace{0.0cm}
\mbox{\hspace{-0.5cm}
\subfigure[][]{\hspace{-1.0cm}
\includegraphics[height=.17\textheight, angle =0]{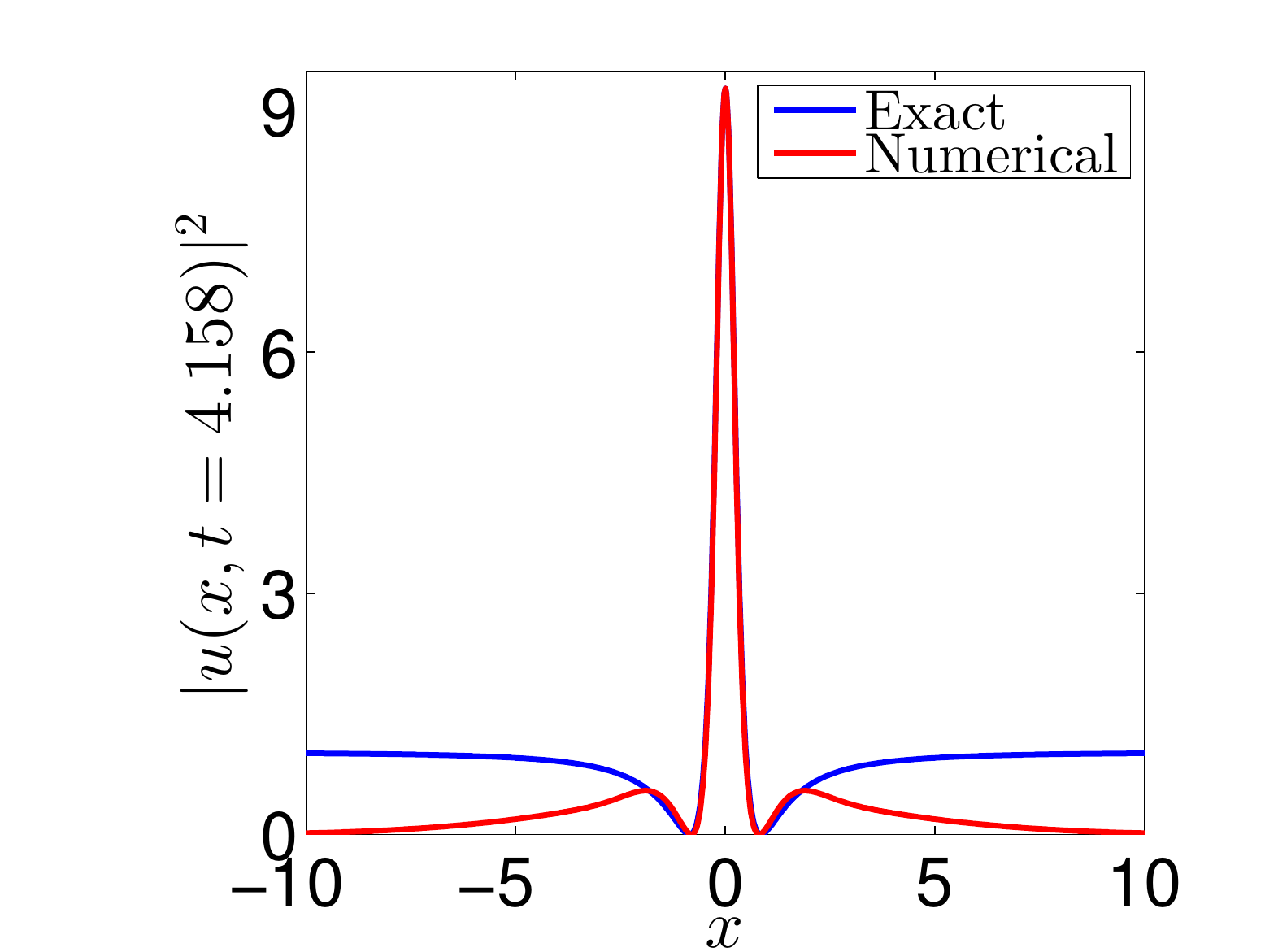}
\label{fig2d}
}
\subfigure[][]{\hspace{-0.5cm}
\includegraphics[height=.17\textheight, angle =0]{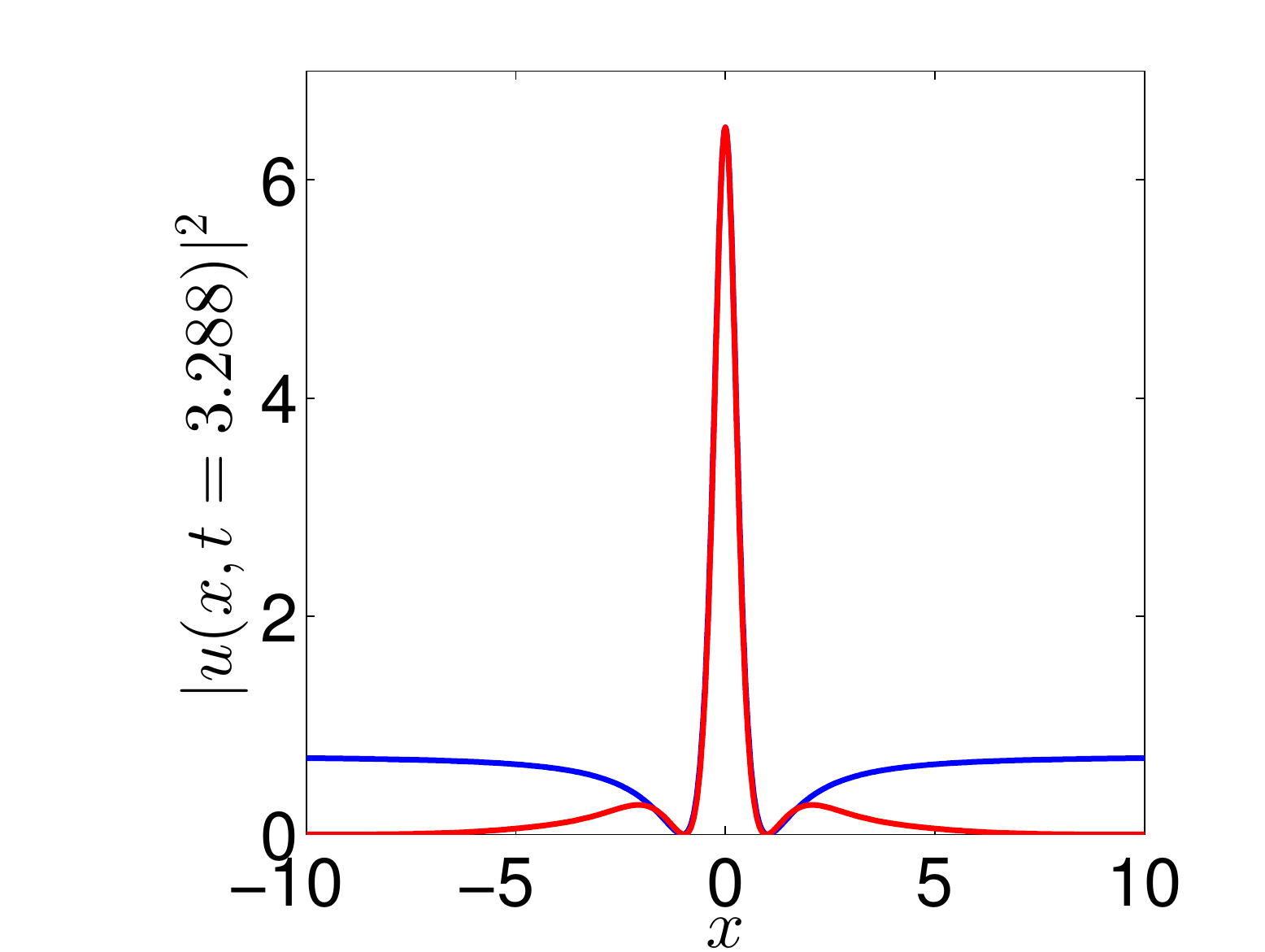}
\label{fig2e}
}
\subfigure[][]{\hspace{-0.5cm}
\includegraphics[height=.17\textheight, angle =0]{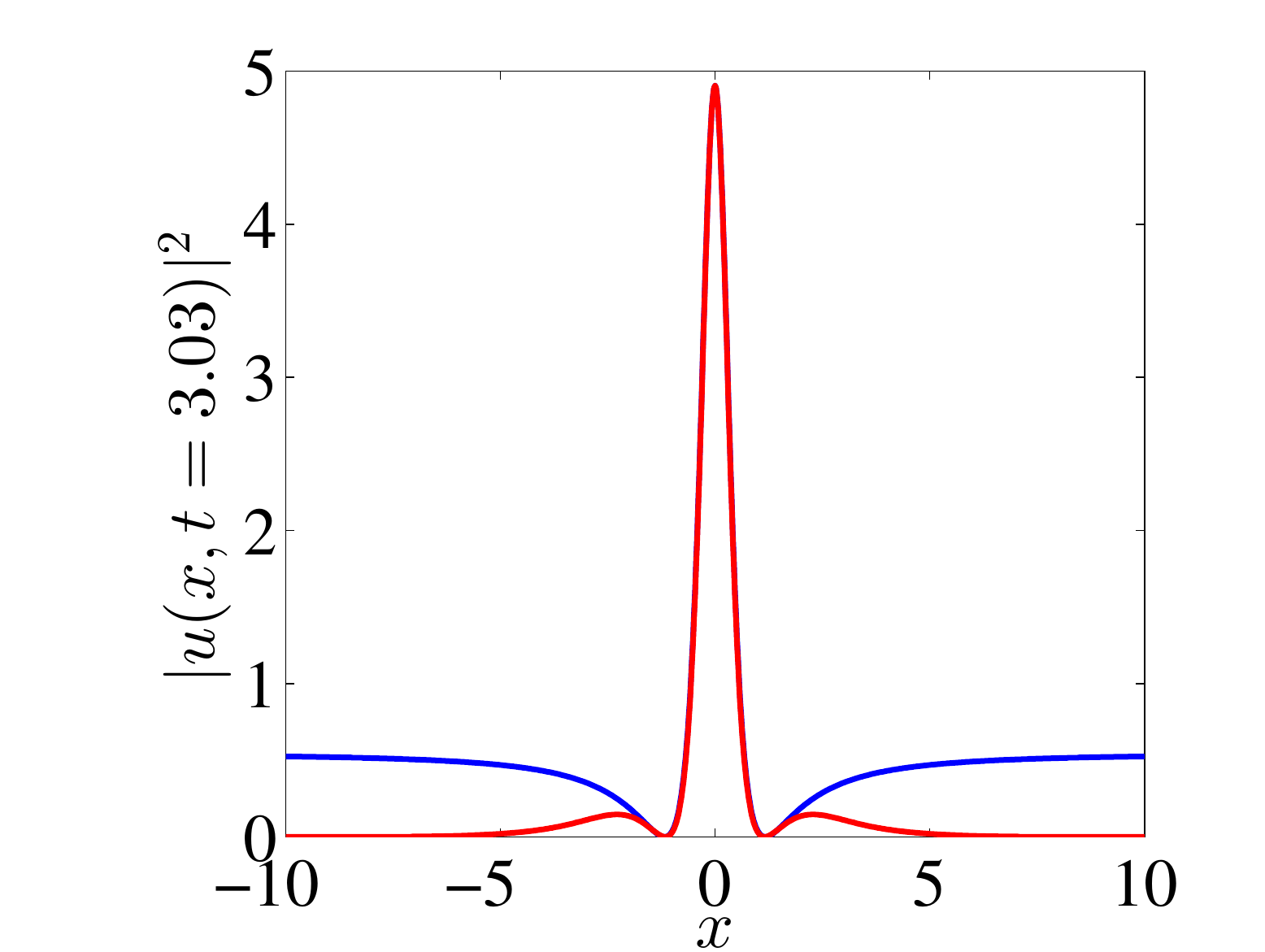}
\label{fig2f}
}
}
\vspace{0.0cm}
\mbox{\hspace{-0.5cm}
\subfigure[][]{\hspace{-1.0cm}
\includegraphics[height=.17\textheight, angle =0]{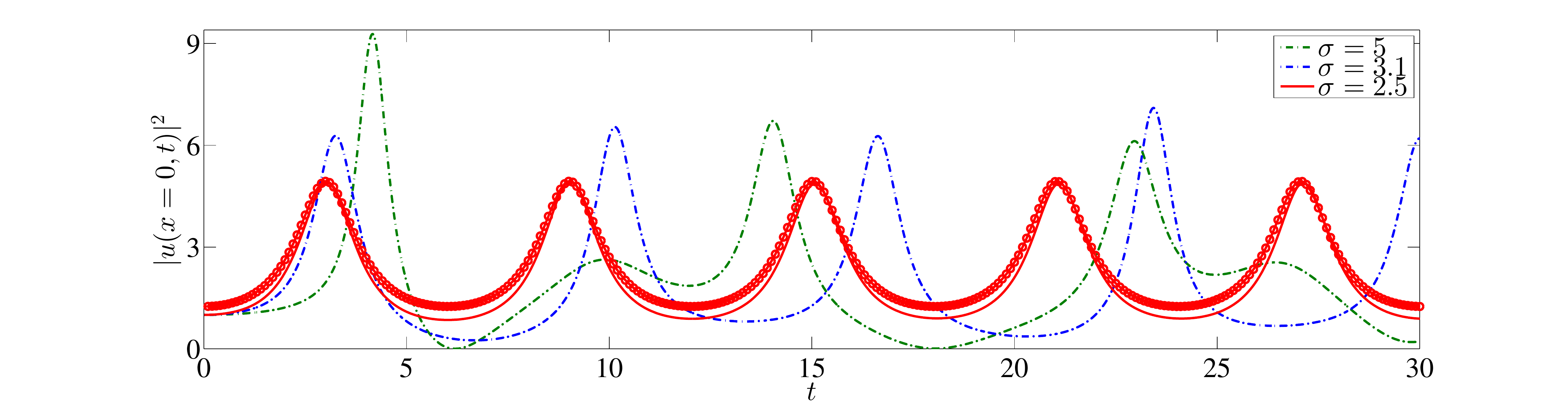}
\label{fig2g}
}
}
\end{center}
\caption{(Color online)
\textit{Top and middle rows}: Summary of results for
$\sigma=5$, $\sigma=3.1$, and $\sigma=2.5$ presented in left,
middle and right panels, respectively. Spatiotemporal evolution
of densities $|u|^{2}$ (top panels) as well as their spatial
distribution (middle panels) evaluated at $t=4.158$, $t=3.288$,
and $t=3.03$. Note that numerical and exact solutions are plotted
against each other with solid red and blue lines, respectively,
for comparison. \textit{Bottom row:} Temporal evolution of the
densities at $x=0$ for various values of $\sigma$. Red circles
correspond to best fit of the exact solution reported in~\cite{kamal_per}.
\label{fig2}
}
\end{figure}

We now turn to the case of smaller values of $\sigma$.
In particular, numerical results corresponding to
the cases with $\sigma=5$, $\sigma=3.1$, and $\sigma=2.5$ are
presented in the left, middle and right panels of Fig.~\ref{fig2},
respectively. It can be discerned from these panels that as the
value of $\sigma$ decreases, the $N$-soliton train (formed past
the first high-peaked wave) starts decreasing in $N$,
presumably going from a case associated with $N=3$ (left panels)
to ones with $N=2$ in the middle and right panels.
The latter two clearly feature the progression towards
a time-periodic solution (see especially the $\sigma=2.5$
case in the right panel). Yet, at the same time, in the spirit
of centaurian qualities, this structure too can be very
adequately approximated by the well-known Kuznetsov-Ma (KM)
breather~\cite{kuz,ma}.
A relevant fit of the center evolution to that of the center of a
KM breather~\cite{kamal_per} (red circles based on the
analytical expression) can be found in the bottom panel of the
figure in the case of $\sigma=2.5$.
At the same time, as indicated above, the $N=2$ solution features
a similar-looking periodic dependence illustrated for comparison
in Fig.~\ref{fig2c}. Thus, while we can clearly observe
the progressively smaller number of solitons as $\sigma$ decreases, both
a 2-soliton structure and the KM breather bear characteristics
closely resembling those of the solution for sufficiently small $\sigma$.

\begin{figure}[!pt]
\begin{center}
\mbox{\hspace{-0.5cm}
\subfigure[][]{\hspace{-1.0cm}
\includegraphics[height=.17\textheight, angle =0]{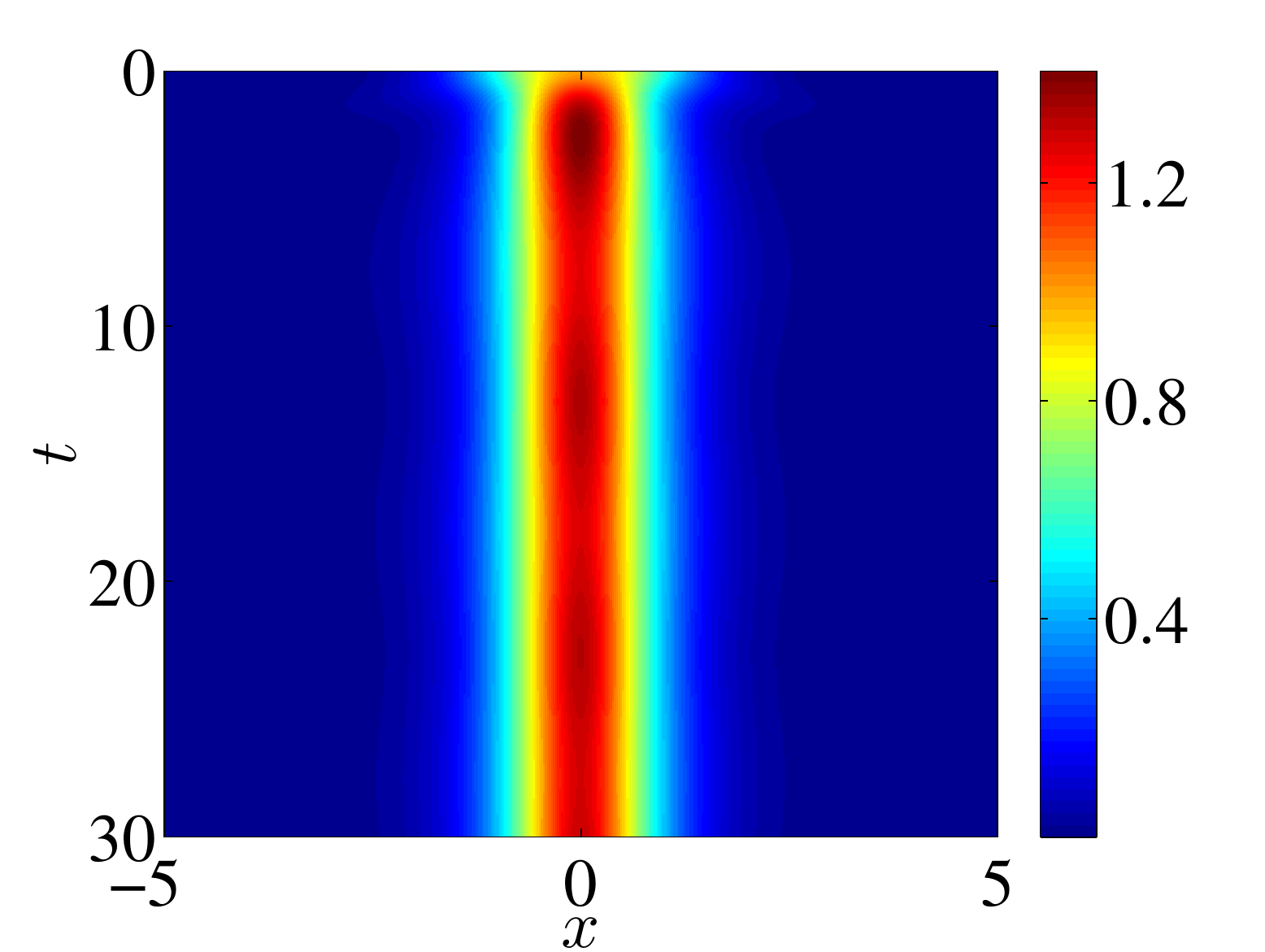}
\label{fig3a}
}
\subfigure[][]{\hspace{-0.5cm}
\includegraphics[height=.17\textheight, angle =0]{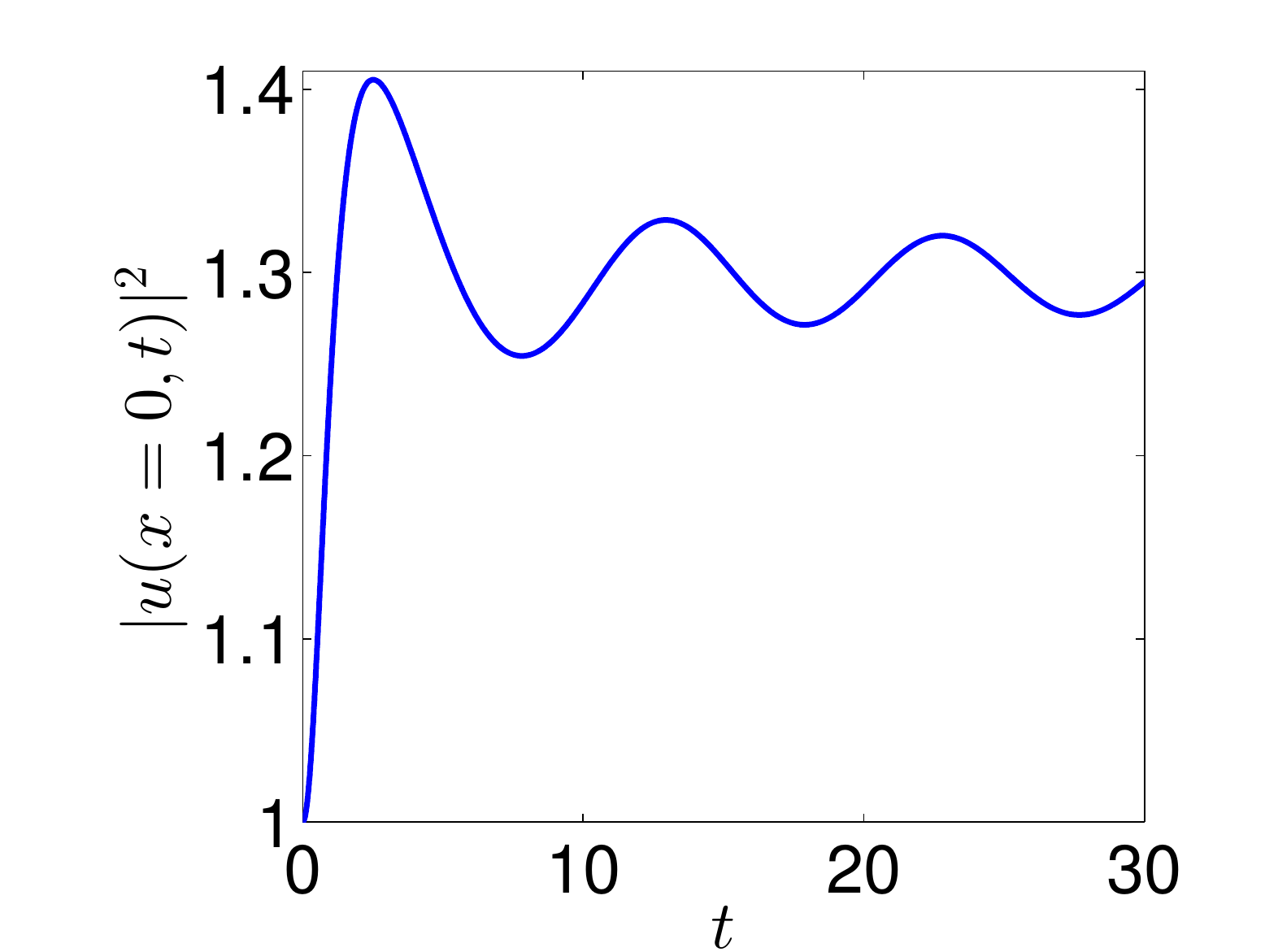}
\label{fig3b}
}
\subfigure[][]{\hspace{-0.5cm}
\includegraphics[height=.17\textheight, angle =0]{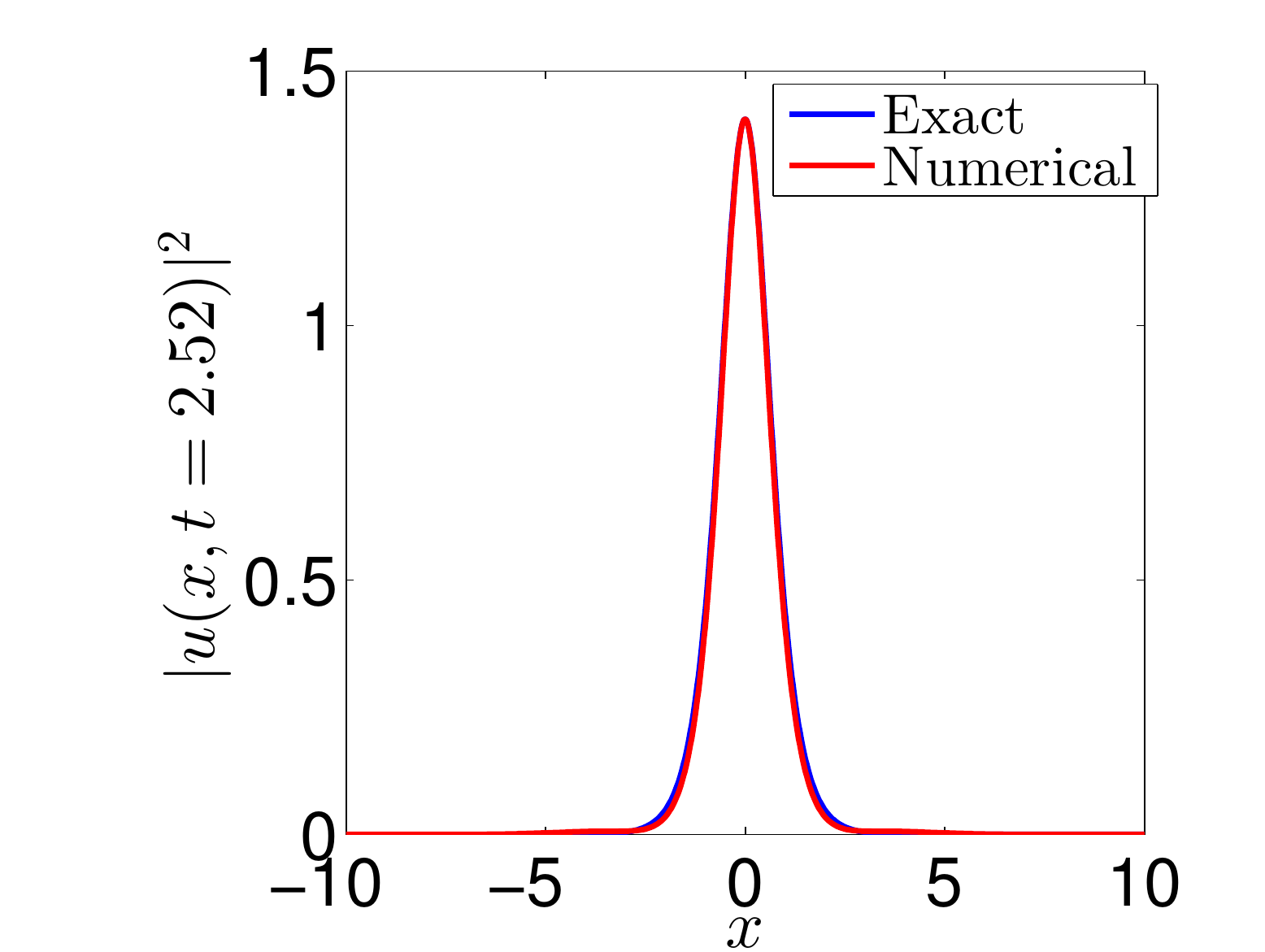}
\label{fig3c}
}
}
\end{center}
\caption{(Color online)
Summary of results corresponding to $\sigma=1.3$: The spatiotemporal
evolution of the density $|u|^{2}$ is presented in the left panel.
Case examples of the
temporal and spatial distributions of the density are depicted in
the middle and right panels, respectively. In the latter, a
 fit to a solitonic waveform is also presented (under ``Exact'').
\label{fig3}
}
\end{figure}

We now proceed to further decrease the value of $\sigma$,
to values of about $\approx 1$. In this case, the
high-peaked waveform emerging fits best into a (single)
bright soliton described by Eq.~%
(\ref{soliton_fit}) [in all previous examples depicted, the best fit
of the core as indicated by the relevant snapshots was to a Peregrine].
As an illustrative example, numerical results for the case of
$\sigma=1.3$ are shown in Fig.~\ref{fig3}. Specifically, the right panel
of the figure suggests a very good agreement between the numerically obtained
solution and the exact one. As a side note, temporal oscillations of the
density are observed which are clearly demonstrated in the middle panel of
the figure, i.e., the pattern oscillates around a single bright soliton,
given its Hamiltonian character, without relaxing fully to it.

\begin{figure}[!pt]
\begin{center}
\mbox{\hspace{-0.1cm}
\subfigure[][]{\hspace{-1.0cm}
\includegraphics[height=.15\textheight, angle =0]{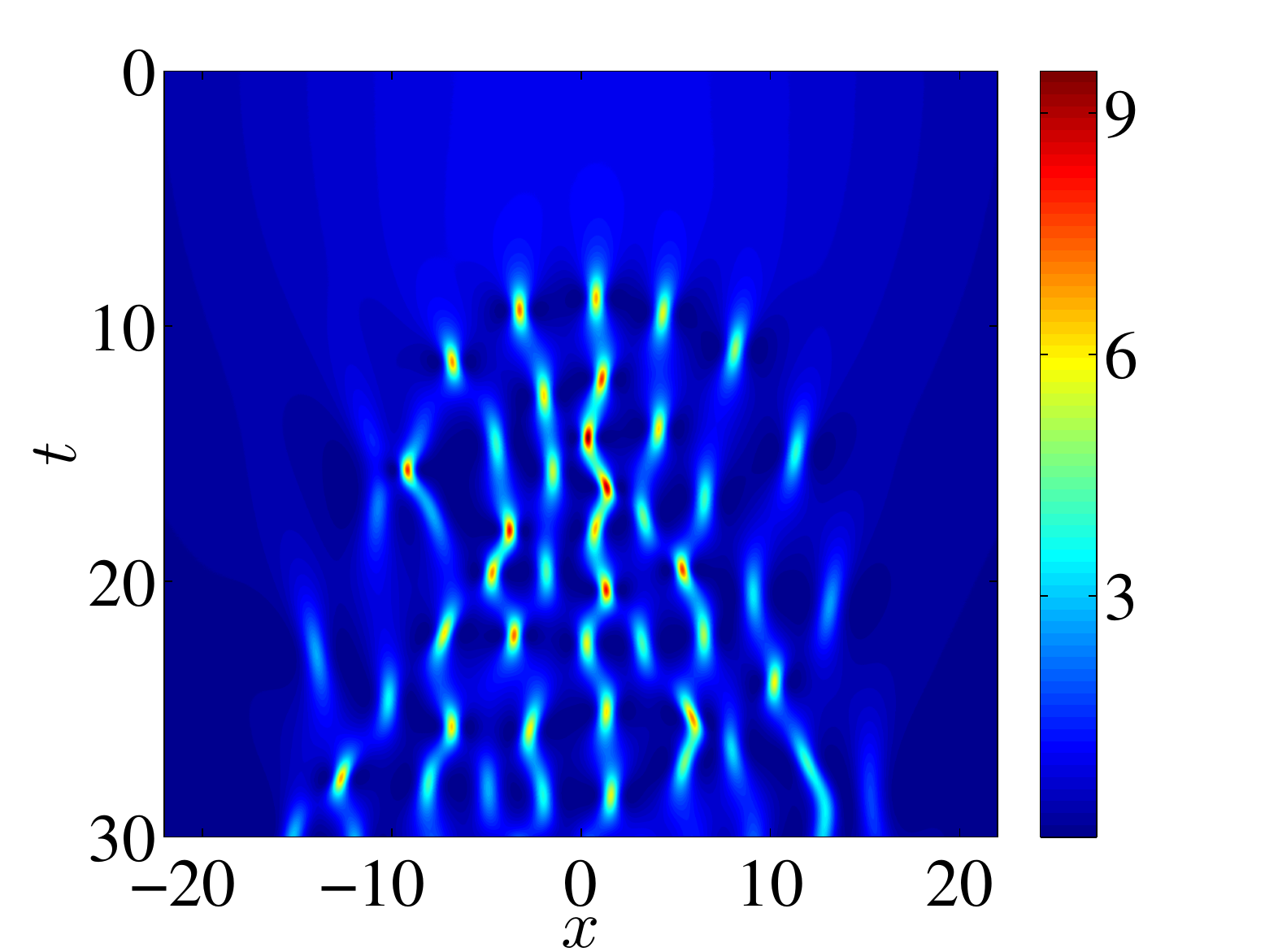}
\label{fig4a}
}
\subfigure[][]{\hspace{-0.5cm}
\includegraphics[height=.15\textheight, angle =0]{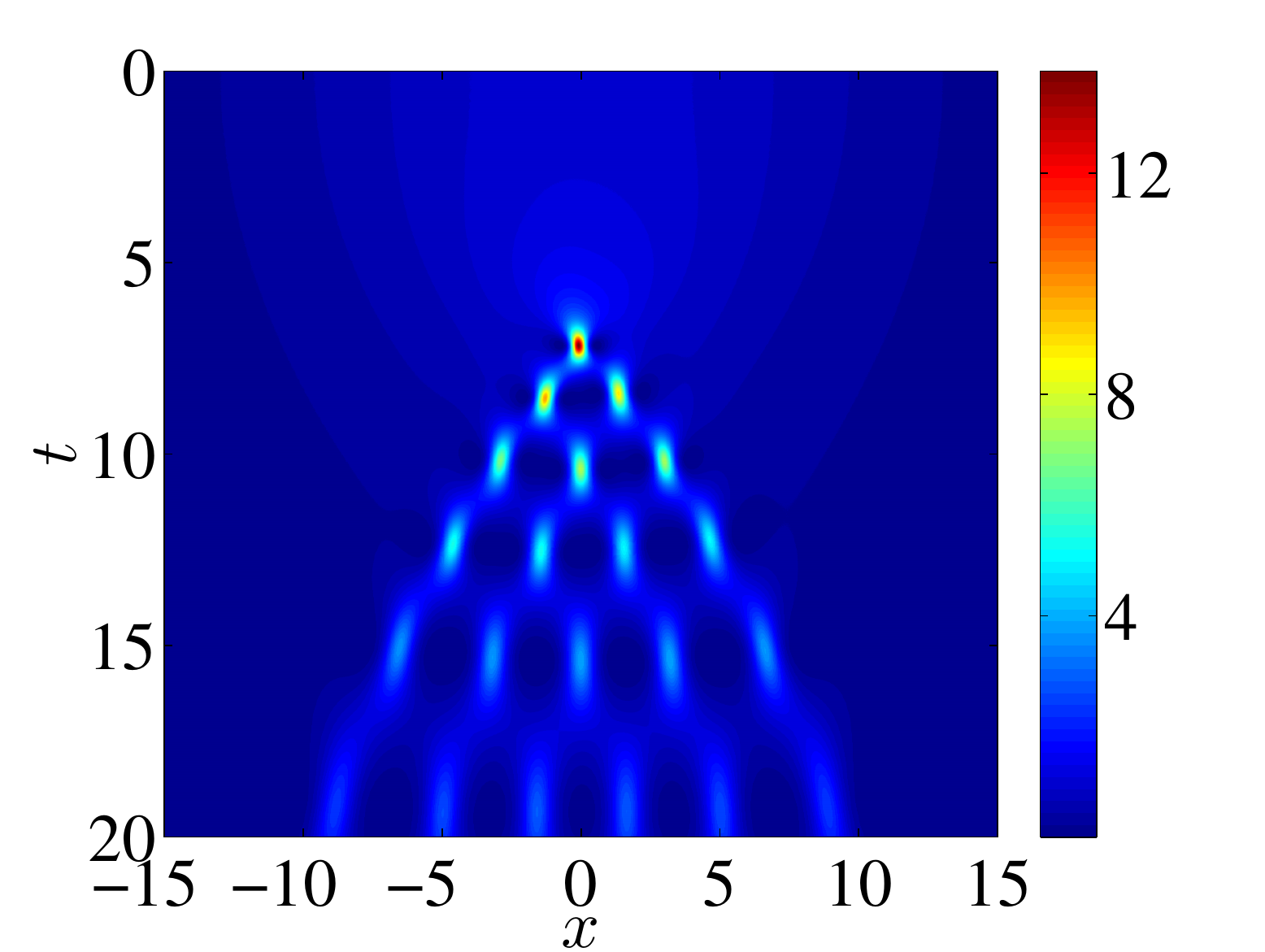}
\label{fig4b}
}
\subfigure[][]{\hspace{-0.5cm}
\includegraphics[height=.15\textheight, angle =0]{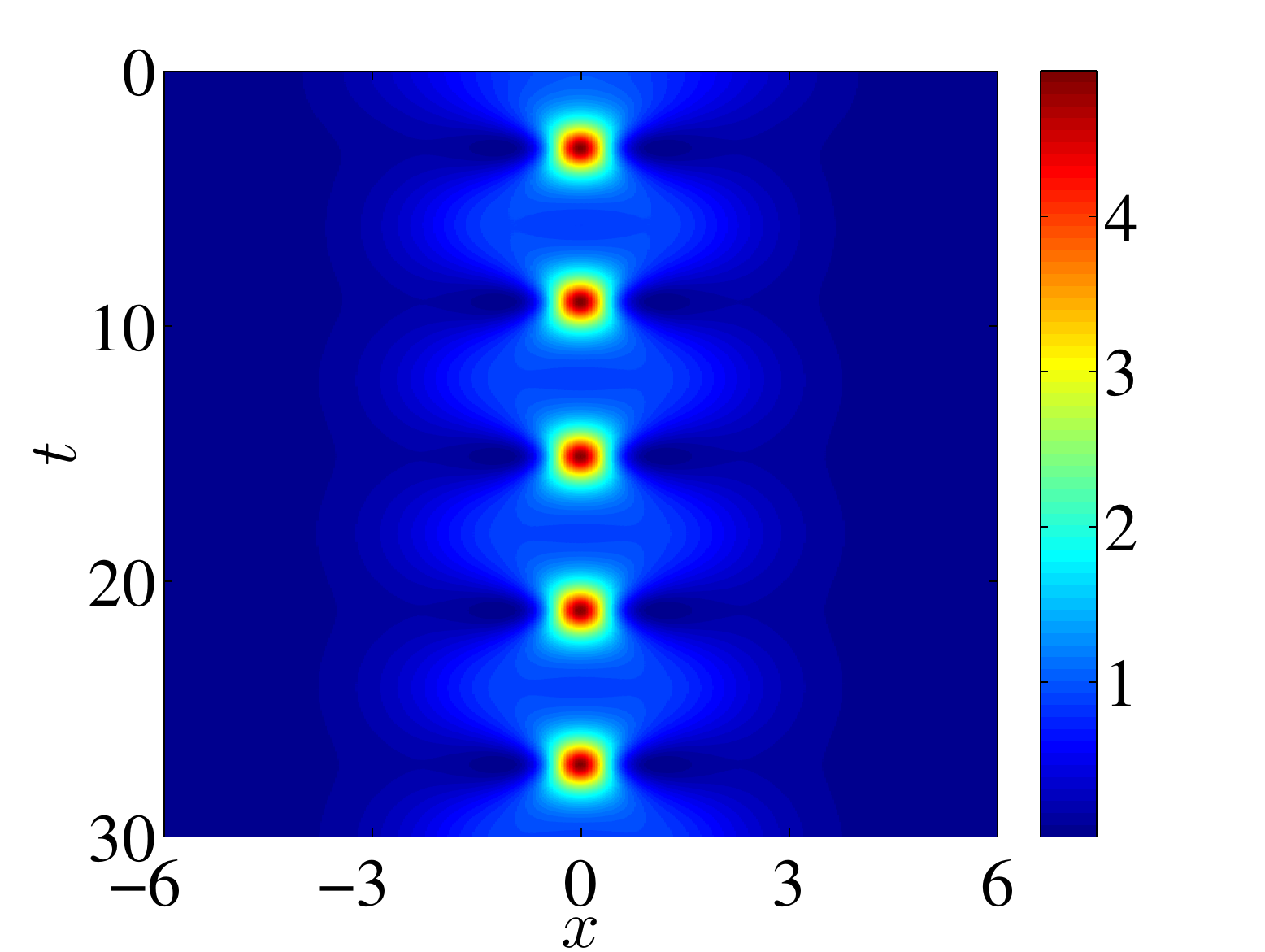}
\label{fig4c}
}
\subfigure[][]{\hspace{-0.5cm}
\includegraphics[height=.15\textheight, angle =0]{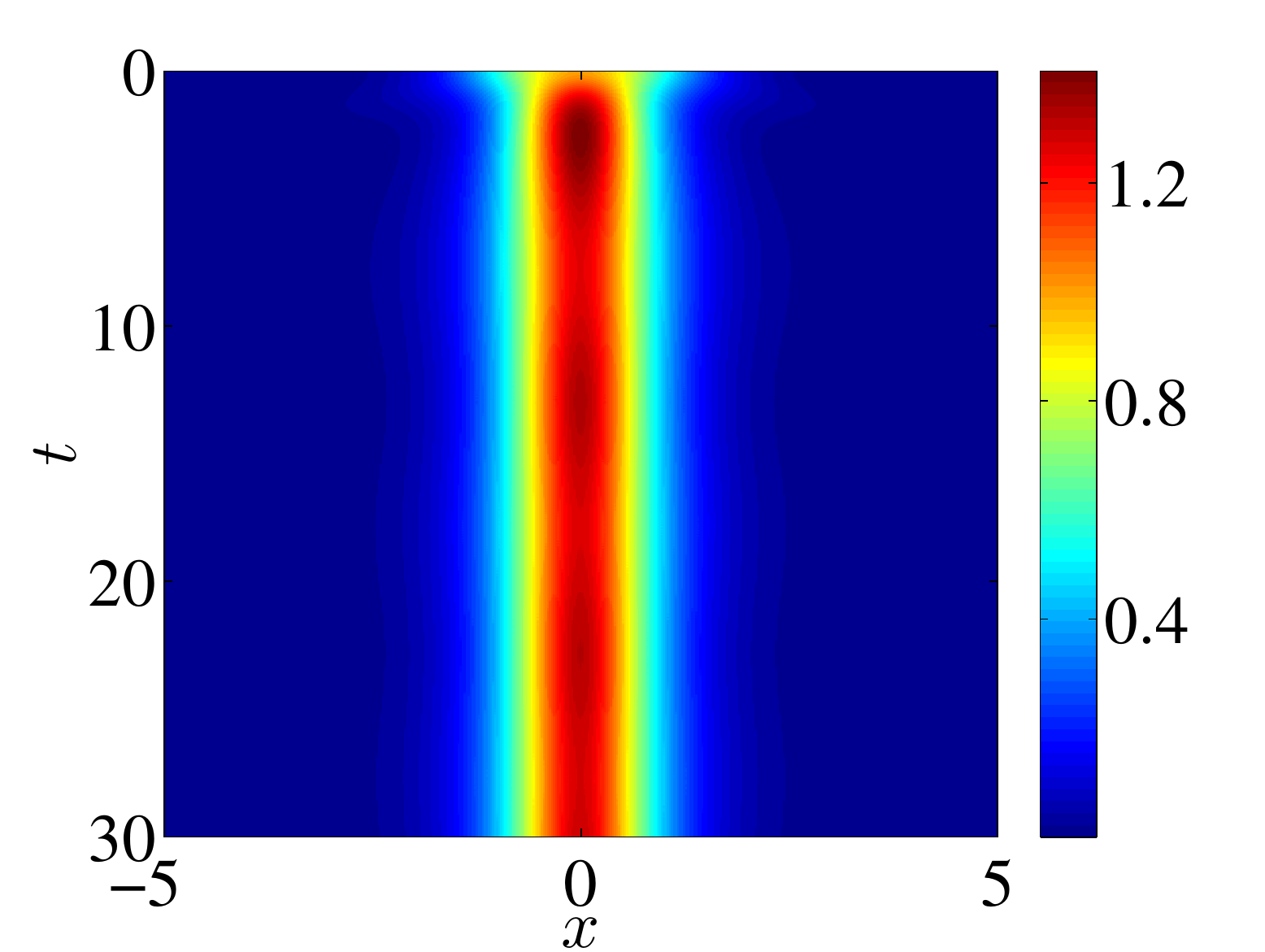}
\label{fig4d}
}
}
\end{center}
\caption{(Color online)
Spatiotemporal evolution of the density $|u|^{2}$ corresponding
to perturbed cases with (a) $\sigma=20.1$, (b) $\sigma=10.5$,
(c) $\sigma=2.5$, and (d) $\sigma=1.3$. The first one appears
to destroy the CT structure, while the second one to preserve it.
In (c) and (d), the patterns closely resemble the $N=2$
and $N=1$ soliton states.
\label{fig4}
}
\end{figure}

Finally, we study the robustness of the reported numerical
evolution results to perturbations in the initial data
(induced, e.g., by imperfections in the initial state preparation).
To do so, we perturb the dynamics of the NLS Eq.~(\ref{nlse_1d})
by adding a $50\si{\deci\bel}$ (signal-to-noise ratio per sample) white
noise to the localized region of the Gaussian pulse~(\ref{gauss_init})
specified by the full width at half maximum (FWHM); see, also,
Ref.~\cite{nls_2_movie} for a complete movie of the dynamics in this
case. Highlights of our findings are shown in Fig.~\ref{fig4}. The left
panel of the figure showcases the evolution dynamics for
$\sigma=20.1$, to be compared with the unperturbed case of Fig.~\ref{fig1d}.
Clearly, the results resembling the orderly CT structure
(or the $N$-soliton solution that it may be representing) seem
to lack persistence qualities in the present setting, as the
resulting pattern clearly seems highly disordered, bearing
little resemblance to its ordered origin. {\it Nevertheless},
its structural ingredients in the form of emergent peaks are
still present and we have confirmed that they can still be well
approximated (in fact, optimally approximated in comparison to
single soliton) 
by Peregrine waveforms near the core.
Hence, in some sense, the Peregrine-like features of the pattern
appear to be robust. The above features were found to be manifested
for a wide parametric window of $\sigma$ values.

Nevertheless, as we gradually approach
$\sigma\approx 10.5$, the dynamics are no longer affected
dramatically by the perturbation. We report the case with $\sigma=10.5$ in Fig.~\ref{fig4b},
where the formation of a rogue wave is followed by the previously mentioned
expanding CT structure potentially attributable to an
$N$-soliton solution. By decreasing the
value of $\sigma$ further, the noise does not affect the dynamics
essentially at all.
This is clearly illustrated in the cases featured in
Fig.~\ref{fig4c} and Fig.~\ref{fig4d}, where the apparent
$N=2$ and $N=1$ solitons seem to clearly persist.

\subsection{The nonintegrable case: $\delta>1$}

Having studied the integrable case, we now turn to the nonintegrable one
corresponding to $\delta>1$. We are interested in identifying parametric
case examples of $\sigma$ and $\delta$ for which collapsing events may take place
and, perhaps more importantly, understanding the underlying mechanisms which
are responsible for creating such eventual evolutionary phenomena.
That is, as $\delta$ is increased beyond the quintic term of $\delta=2$,
the focusing becomes amenable to wave collapse~\cite{sulem} through
the bifurcation of self-similar waveforms~\cite{siettos}.
It is then intriguing to explore to what degree the possibility of
extreme events, like Peregrine solitons,
may interplay with the self-similar structures in producing such collapse events.

Here, it should be pointed out that the solutions given by Eqs.~(\ref{soliton_fit})
and~(\ref{par_peregrine}) cannot be used in the subsequent analysis,
due to the fact that neither constitutes a solution to the NLS equation for $\delta>1$.
On the other hand, the soliton solution
can be generalized in this setting, 
taking the form:
\begin{eqnarray}
u(x,t)=\left(\frac{A}{2}\right)^{\frac{1}{2 \delta}} \left[{\rm sech}\left(\delta
\sqrt{\frac{A}{\delta+1}} (x-x_0) \right) \right]^{\frac{1}{\delta}} \exp(i \beta t),
\label{deltasol}
\end{eqnarray}
with $\beta=A/(2 (1+\delta))$. This fact naturally raises the interesting
question of whether Peregrine waveforms may also generalize in this setting.
This is a question particularly interesting in its own right, which, to the
best of our knowledge, has not been addressed as of yet.

\begin{figure}[!pt]
\begin{center}
\mbox{\hspace{-0.1cm}
\subfigure[][]{\hspace{-1.0cm}
\includegraphics[height=.15\textheight, angle =0]{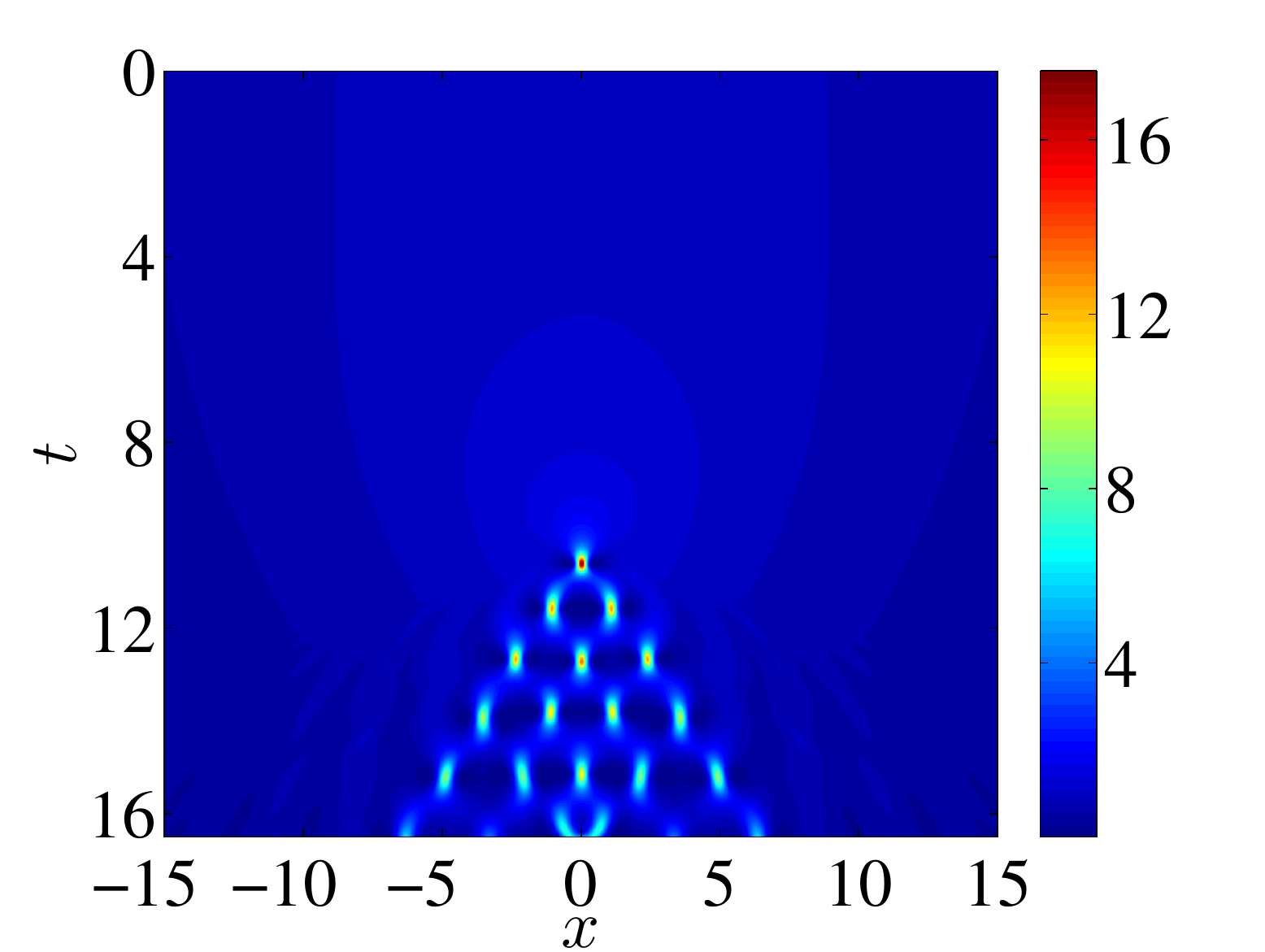}
\label{fig6a}
}
\subfigure[][]{\hspace{-0.5cm}
\includegraphics[height=.15\textheight, angle =0]{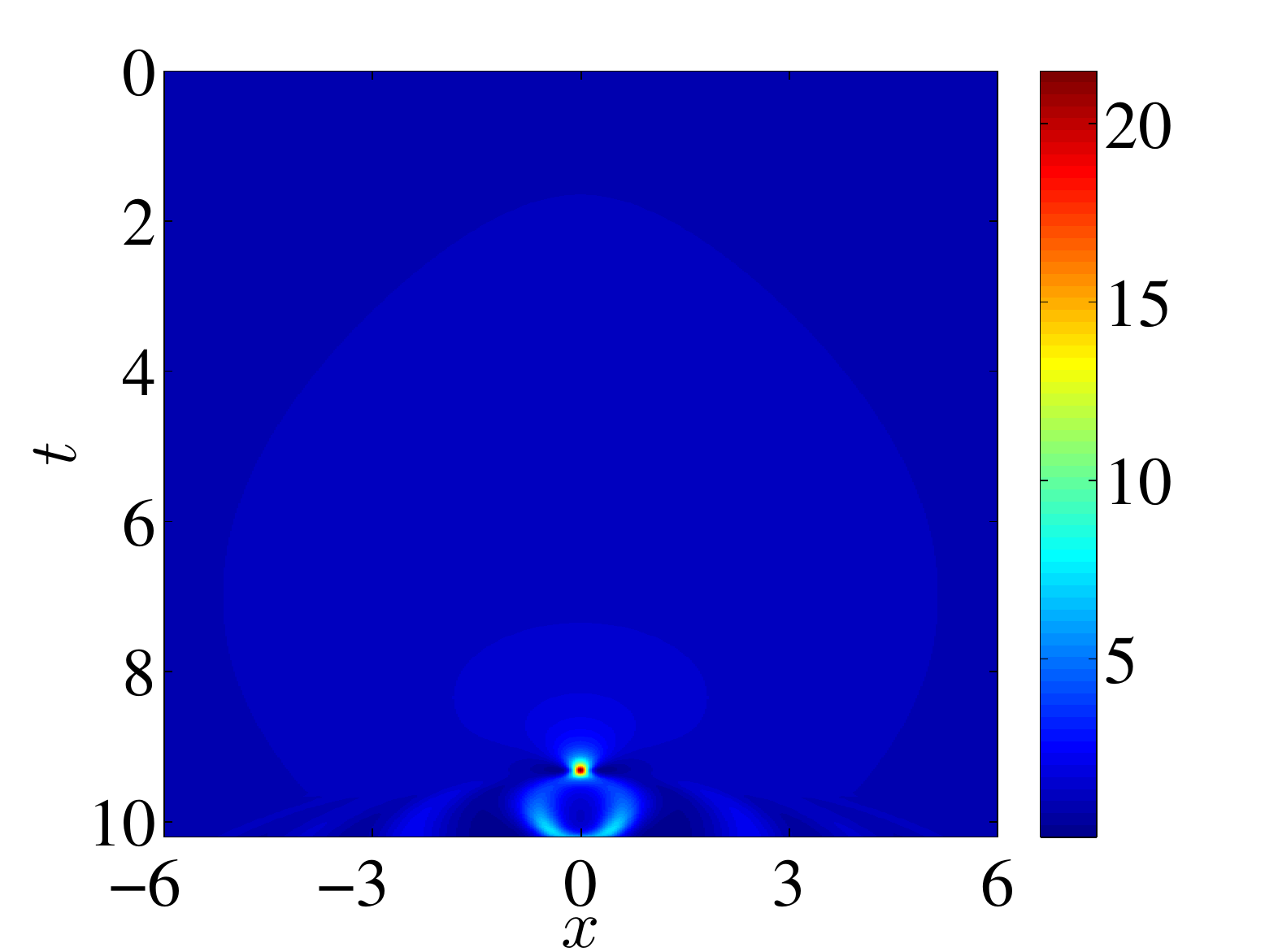}
\label{fig6b}
}
\subfigure[][]{\hspace{-0.5cm}
\includegraphics[height=.15\textheight, angle =0]{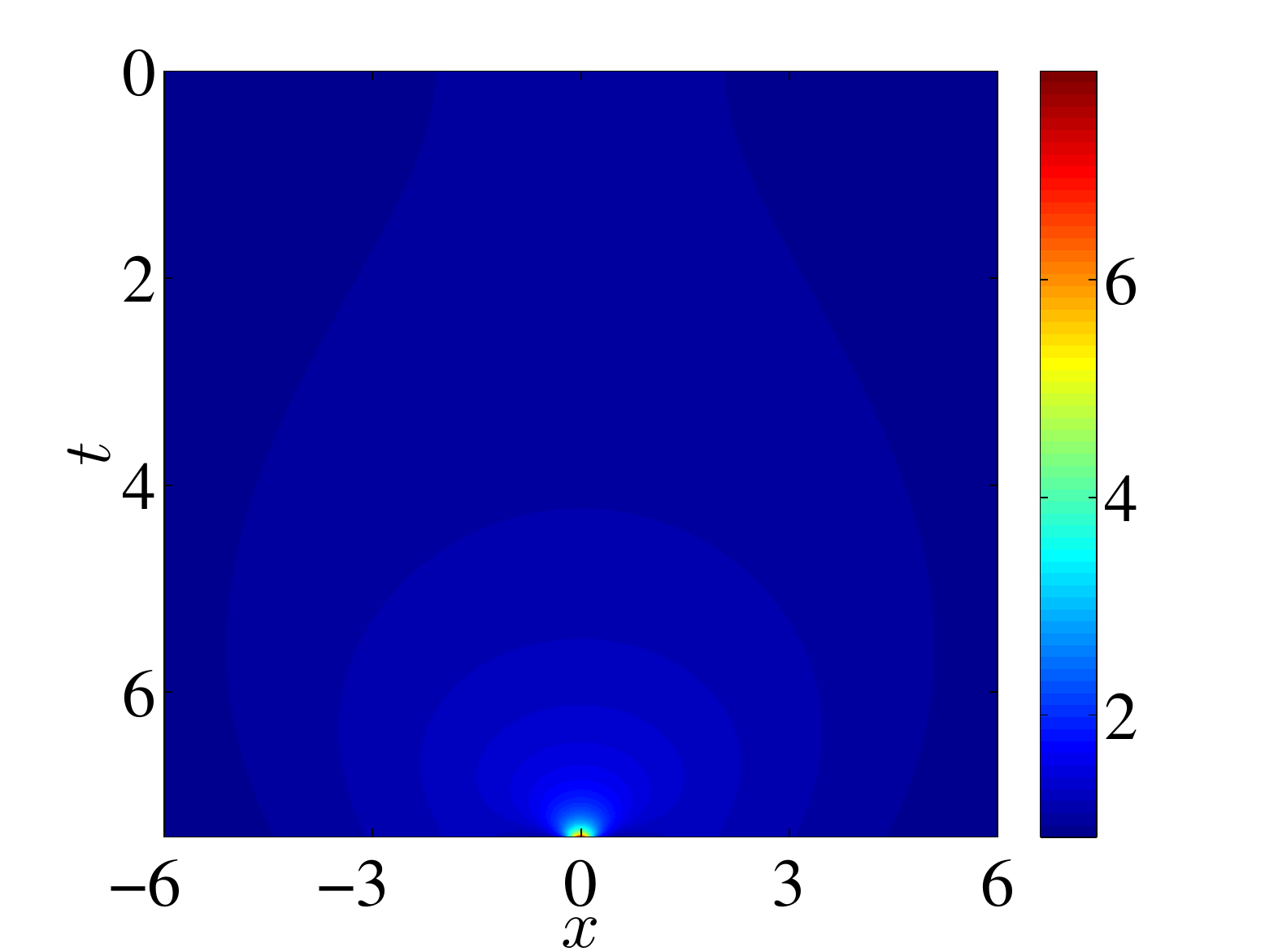}
\label{fig6c}
}
\subfigure[][]{\hspace{-0.5cm}
\includegraphics[height=.15\textheight, angle =0]{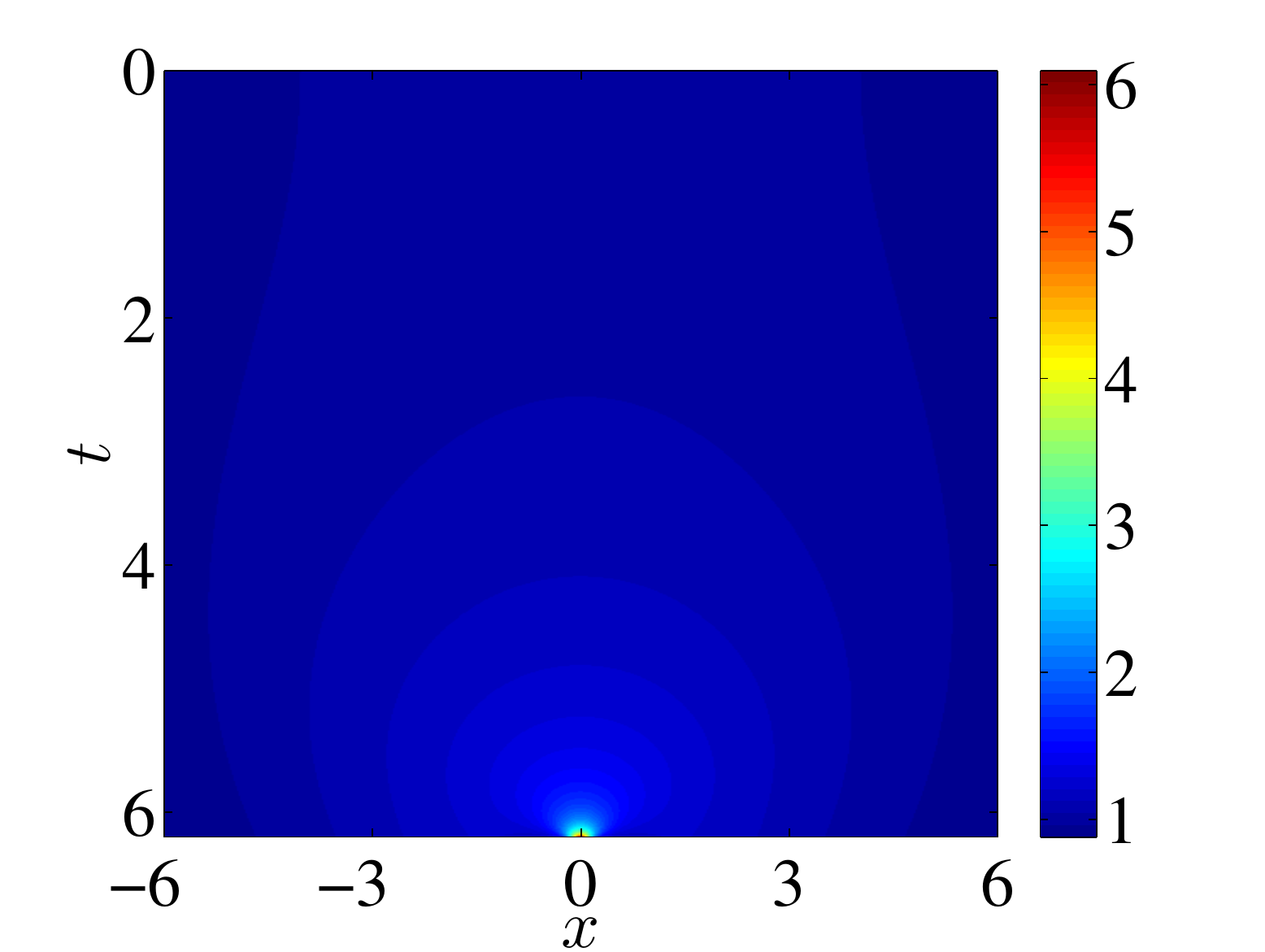}
\label{fig6d}
}
}
\mbox{\hspace{-0.1cm}
\subfigure[][]{\hspace{-1.0cm}
\includegraphics[height=.15\textheight, angle =0]{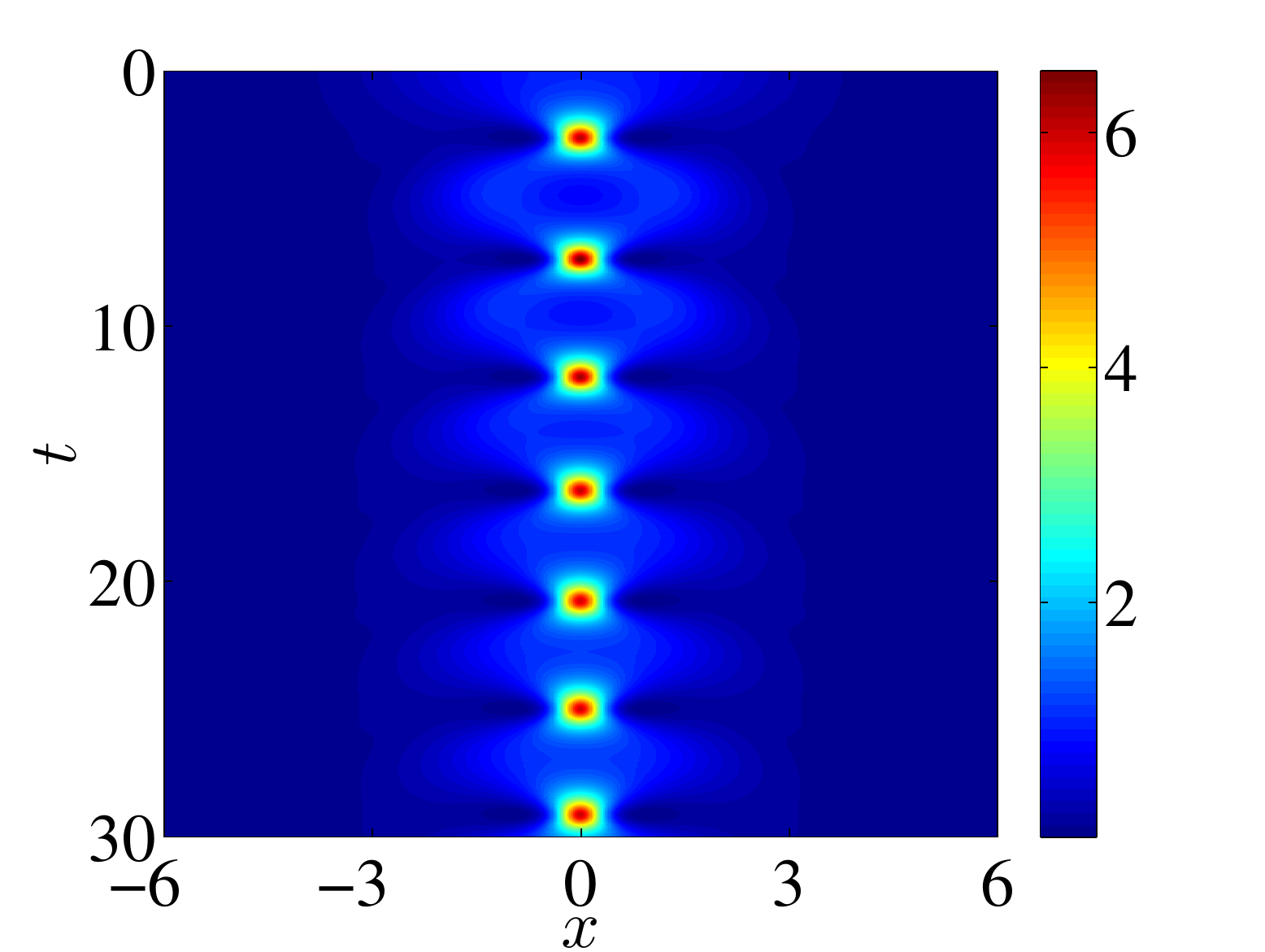}
\label{fig6e}
}
\subfigure[][]{\hspace{-0.5cm}
\includegraphics[height=.15\textheight, angle =0]{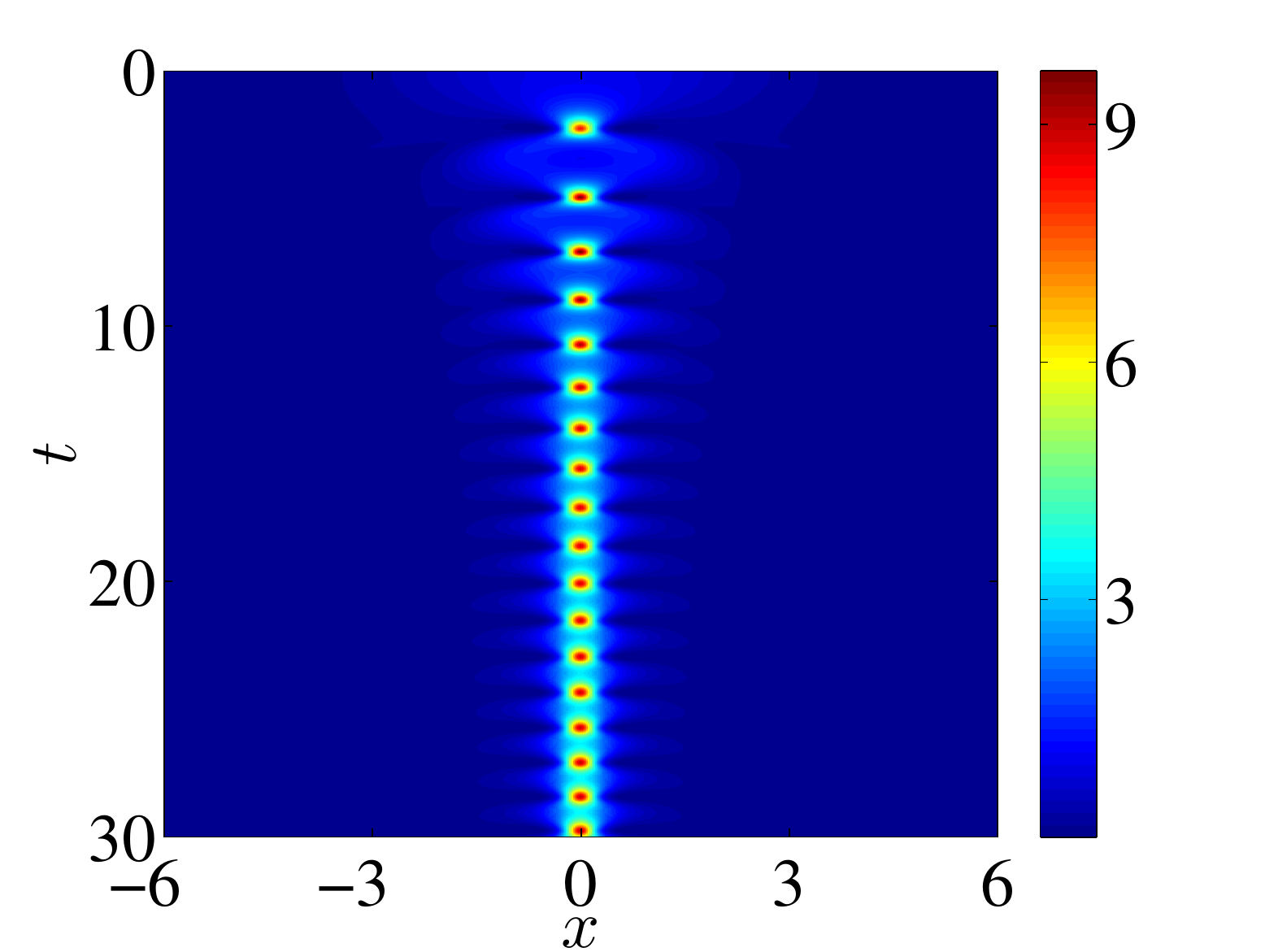}
\label{fig6f}
}
\subfigure[][]{\hspace{-0.5cm}
\includegraphics[height=.15\textheight, angle =0]{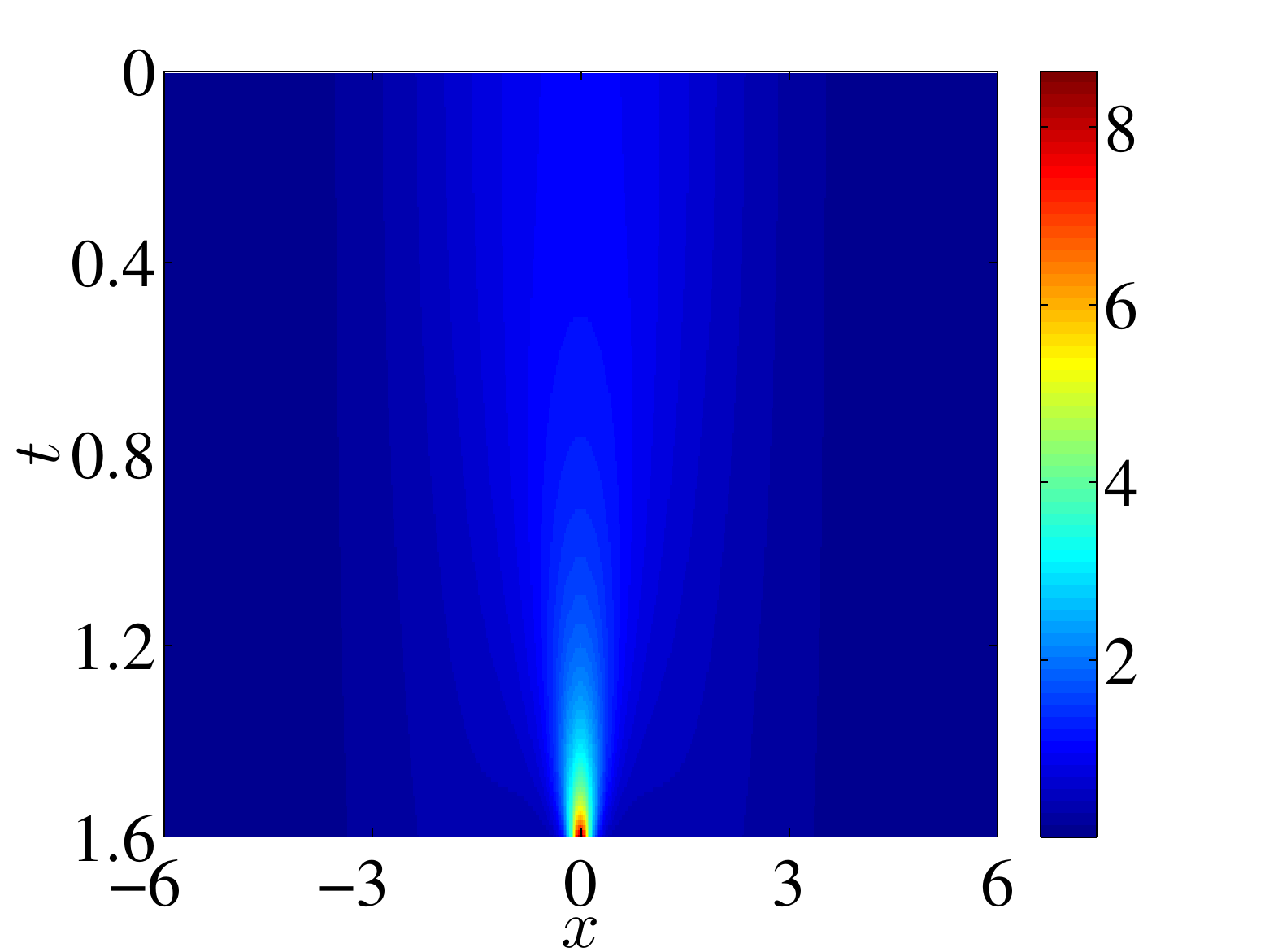}
\label{fig6g}
}
\subfigure[][]{\hspace{-0.5cm}
\includegraphics[height=.15\textheight, angle =0]{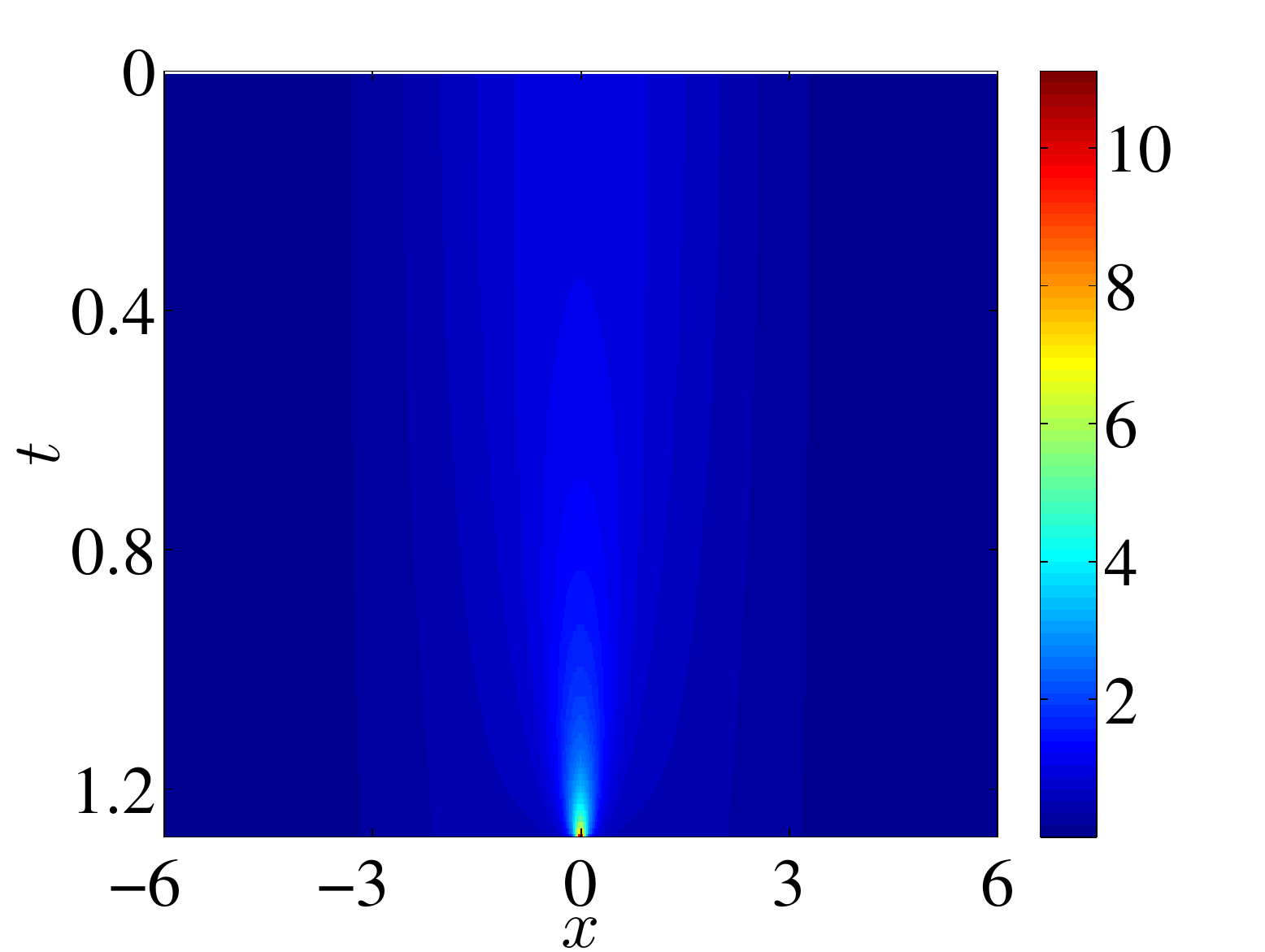}
\label{fig6h}
}
}
\mbox{\hspace{-0.1cm}
\subfigure[][]{\hspace{-1.0cm}
\includegraphics[height=.15\textheight, angle =0]{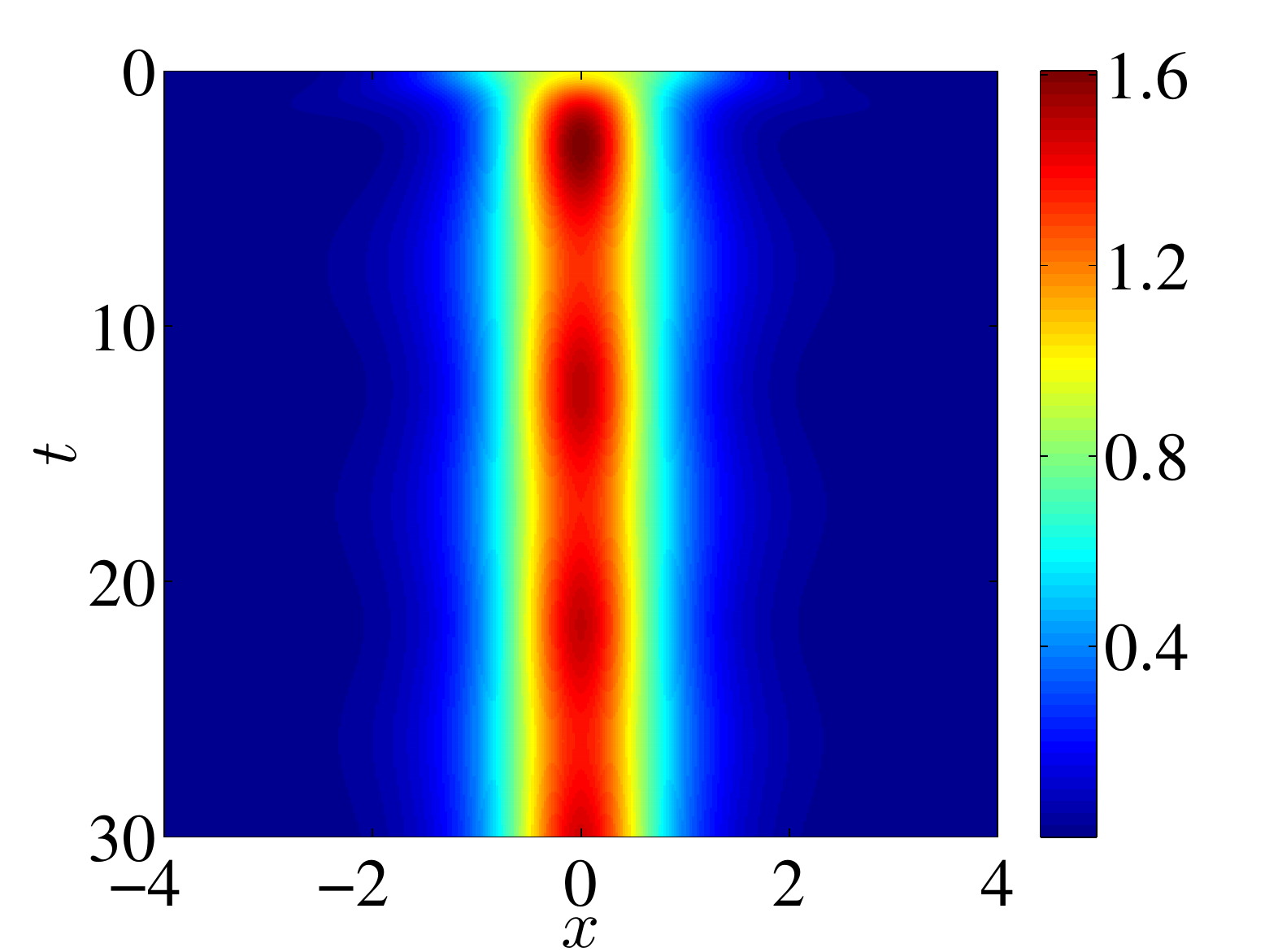}
\label{fig6i}
}
\subfigure[][]{\hspace{-0.5cm}
\includegraphics[height=.15\textheight, angle =0]{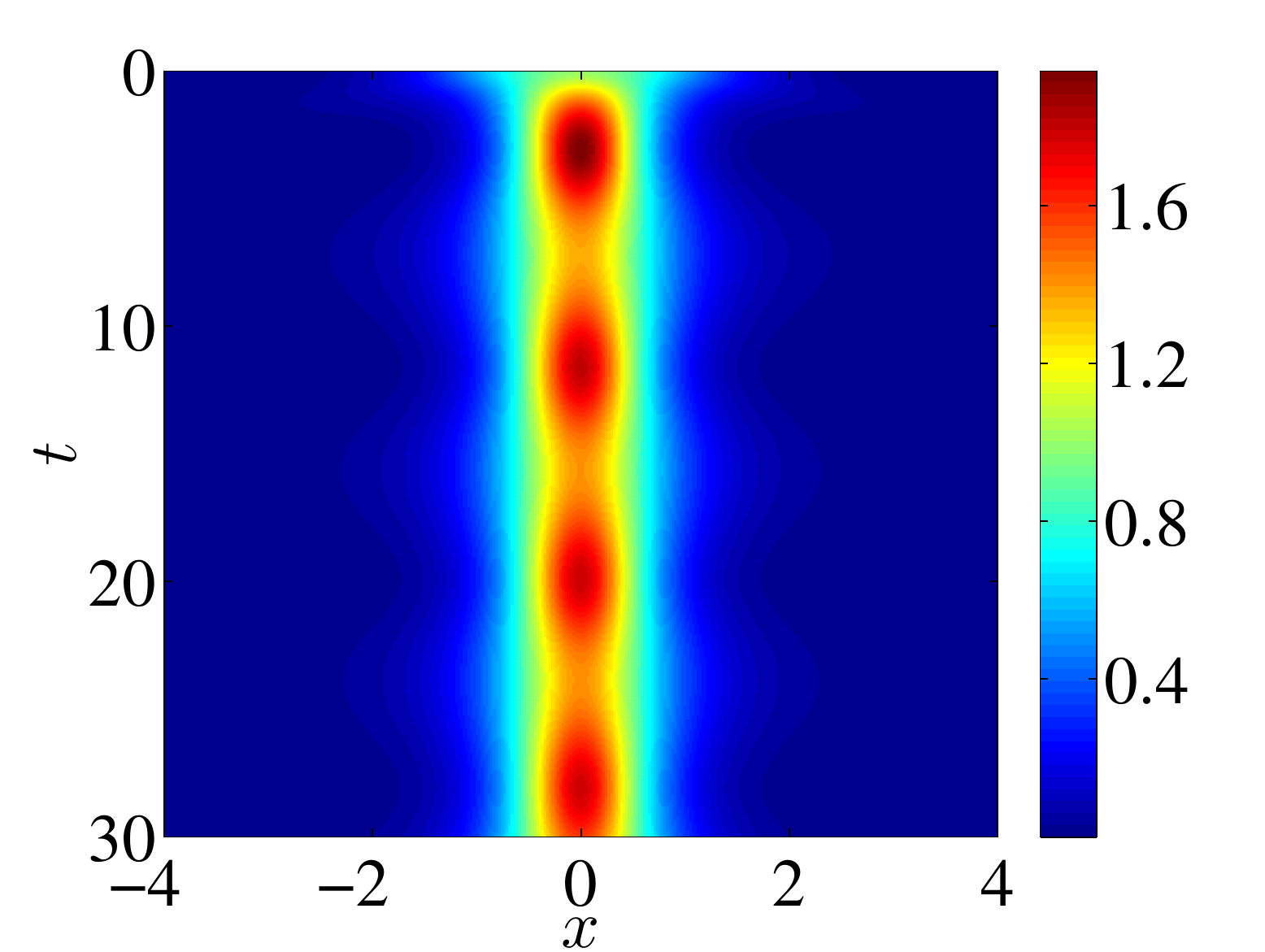}
\label{fig6j}
}
\subfigure[][]{\hspace{-0.5cm}
\includegraphics[height=.15\textheight, angle =0]{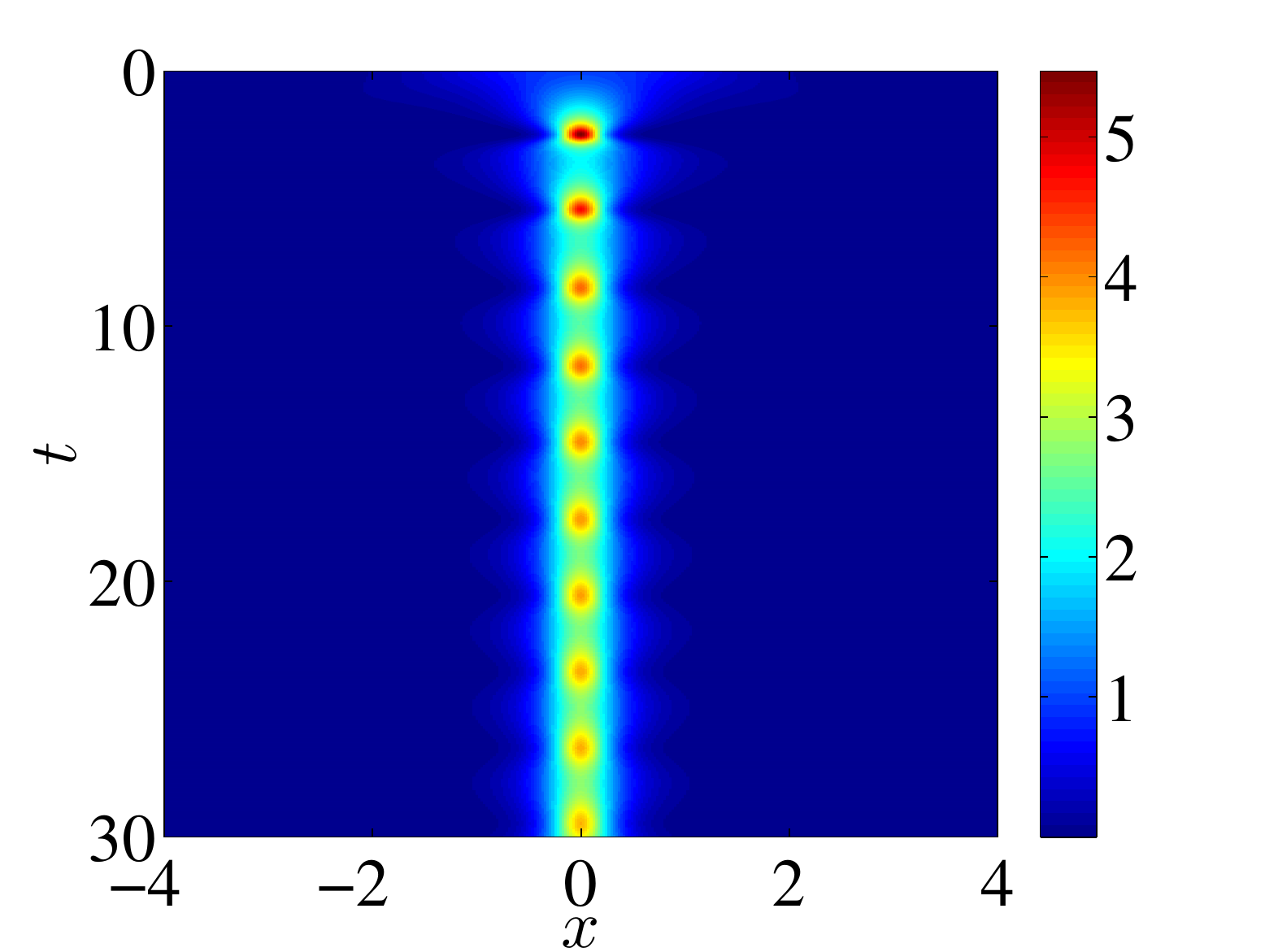}
\label{fig6k}
}
\subfigure[][]{\hspace{-0.5cm}
\includegraphics[height=.15\textheight, angle =0]{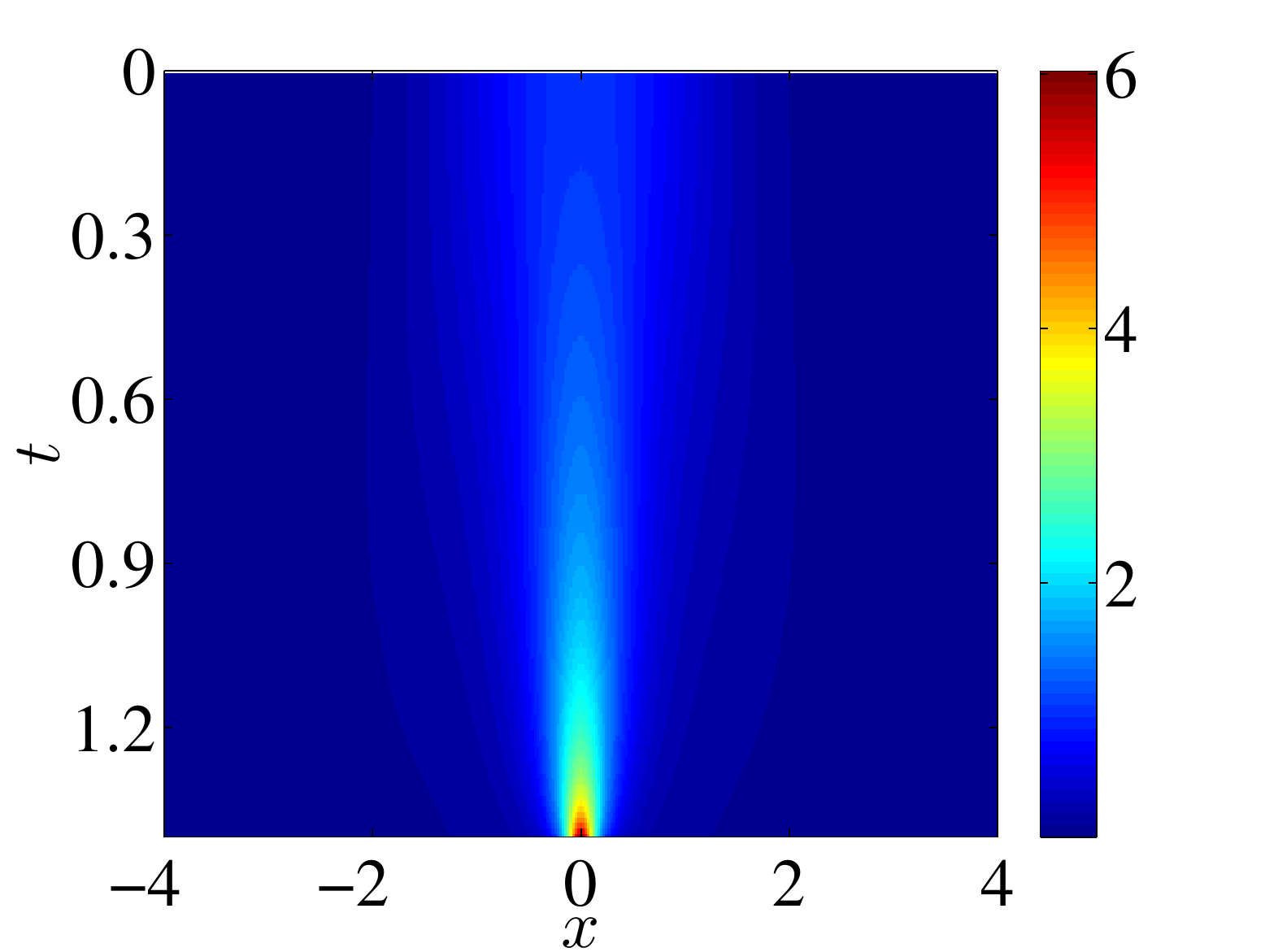}
\label{fig6l}
}
}
\end{center}
\caption{(Color online)
Spatiotemporal evolution of the density $|u|^{2}$ corresponding
to the nonintegrable NLS with (a)-(d) $\sigma=20.1$, (e)-(h) $\sigma=2.5$
and (i)-(l) $\sigma=1.3$. First, second, third, and fourth columns correspond
to values of nonlinearity power of $\delta=1.2$, $\delta=1.4$, $\delta=1.8$
and $\delta=2.2$, respectively. A detailed interpretation of these
observations is given in the text.
\label{fig6}
}
\end{figure}

To present a broad perspective of the corresponding interplay,
we showcase numerical results for different values of the nonlinearity power $\delta$
by employing the same Gaussian initial data~(\ref{gauss_init}), also varying
the width $\sigma$ (for fixed amplitude, $\alpha=1$); this is a map,
presented in Fig.~\ref{fig6}, of the two parameter space of the system and
its corresponding dynamical response. Specifically, we present results with $\sigma=20.1$
(that is, within the interval of $\sigma$ where the CT structure
previously emerged), $\sigma=2.5$ (where a breathing pattern reminiscent of
both the KM breather and the $N=2$ soliton arose) and $\sigma=1.3$ (the fundamental soliton regime) in Fig.~\ref{fig6} for different values of $\delta$. 
See, also, Refs.~\cite{nls_3_movies} for complete movies of the dynamics
corresponding to example cases with $\delta=1.2$, $\delta=1.4$ as well as
$\delta=2.2$, together with their perturbed versions in Refs.~\cite{nls_4_movies}.
It should be stressed that the final times showed in Fig.~\ref{fig6} have been
selected in a way such that our numerical scheme is still expected to
properly capture the dynamics. That is to say, when the solution of the IVP
presented focusing events
such that the width of the solution became comparable to the (selected to be
very small) grid spacing, the simulation was stopped before the occurrence
of such an event. While, admittedly, techniques based on adaptive mesh
refinement and dynamic rescaling exist~\cite{renwang} and may allow to continue these
events further, for the purposes of this study, we consider these events
to be faithful precursors of very strong focusing (conducive to collapse
or in any event regimes where the NLS model would no longer be applicable
in physical settings and higher-order terms would come into play).

We can see in the top row of Fig.~\ref{fig6}, corresponding to $\sigma=20.1$ (see
also, for comparison, the bottom row of Fig.~\ref{fig1}), 
that as we progressively
increase the value of $\delta$, 
the formation of the CT structure is
persistent in panel (a) for $\delta=1.2$, while panel (b)
for $\delta=1.4$ also supports a peak reminiscent
of the extreme events discussed previously, that can be mapped
adequately by a local core fit to a Peregrine structure.
However, in panels (c) and (d) (for $\delta=1.8$ and
$\delta=2.2$, respectively) the evolution of the density leads to
strong focusing events as discussed previously although, importantly, in the
former case we are not in the regime (of $\delta \geq 2$) where regular
solitons become unstable towards collapse; notice nevertheless
the structural similarity of these two events.
This suggests that, in some way, the formation of these extreme
events may promote strong focusing even in cases where the self-similar
collapsing solutions (and the instability of regular solitons) are not
supported, i.e., 
these events may be {\it triggered}
by the Peregrine-like entities observed herein.
To further enhance this perspective, the middle row of the figure showcases results
for $\sigma=2.5$ corresponding to the
breathing solution (of considerably smaller 
number of atoms/squared $L^2$ norm).
The density, through its deformation (see panel (e)) for $\delta=1.2$, starts
becoming more localized in the middle of the spatial grid, i.e., at $x=0$ (see, panel
(f) corresponding to $\delta=1.4$ therein) with a rapidly
decreasing vibration period, until the dynamics
leads again to strong focusing (see panels (g) and (h) of the figure).
Yet, once again in (g), we are below the threshold of $\delta=2$.
Finally, and as per the
soliton regime (see Fig.~\ref{fig3} corresponding to $\sigma=1.3$, for comparison),
the bottom panels reveal that the fundamental soliton progressively becomes localized
at $x=0$ as $\delta$ increases until it collapses for $\sigma=2.2$.
In this case, collapse-resembling features are not apparent, except for
the supercritical case of $\delta=2.2$.

\begin{figure}[!pt]
\begin{center}
\mbox{\hspace{-0.1cm}
\subfigure[][]{\hspace{-1.0cm}
\includegraphics[height=.17\textheight, angle =0]{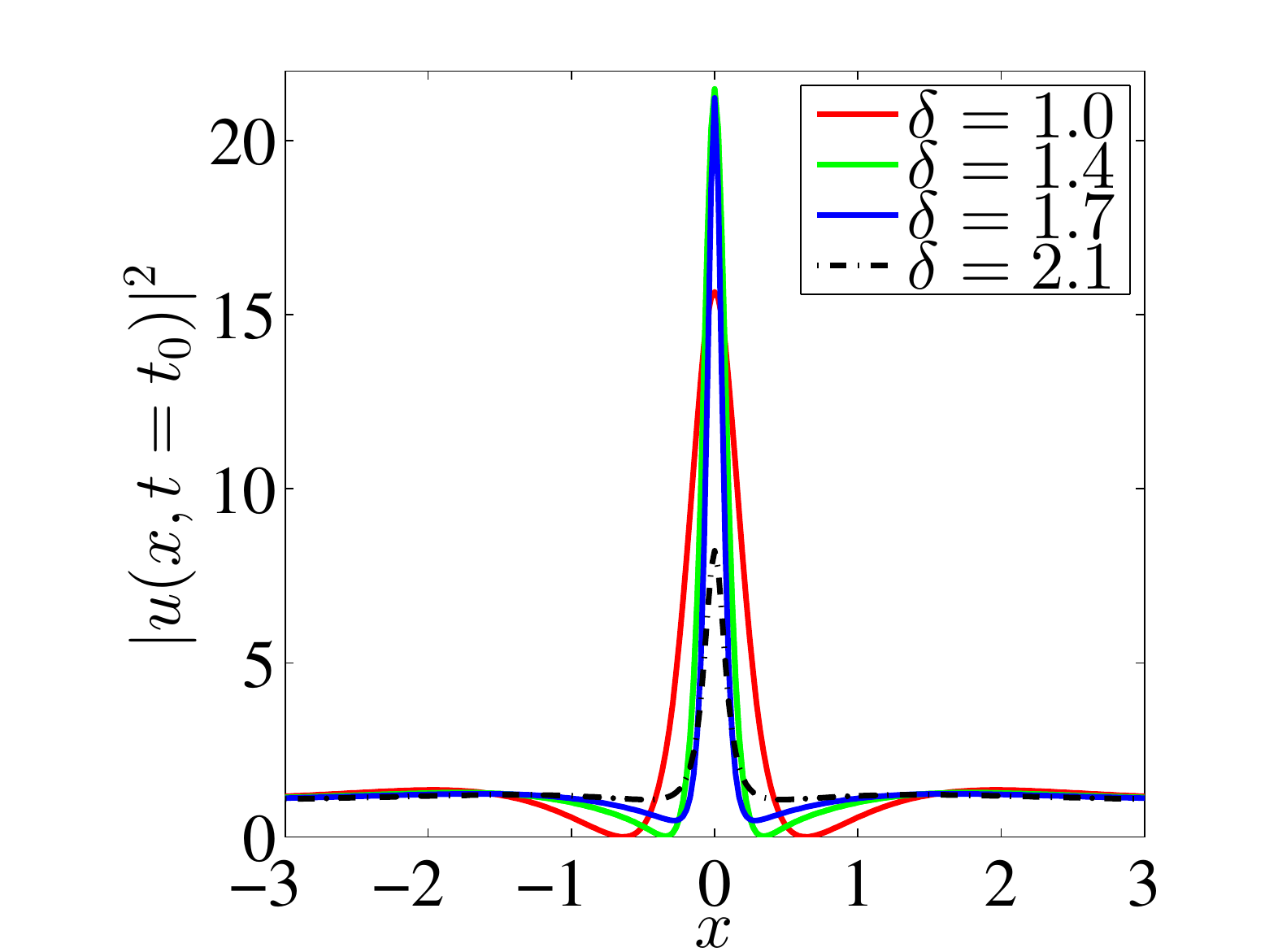}
\label{fig8a}
}
\subfigure[][]{\hspace{-0.5cm}
\includegraphics[height=.17\textheight, angle =0]{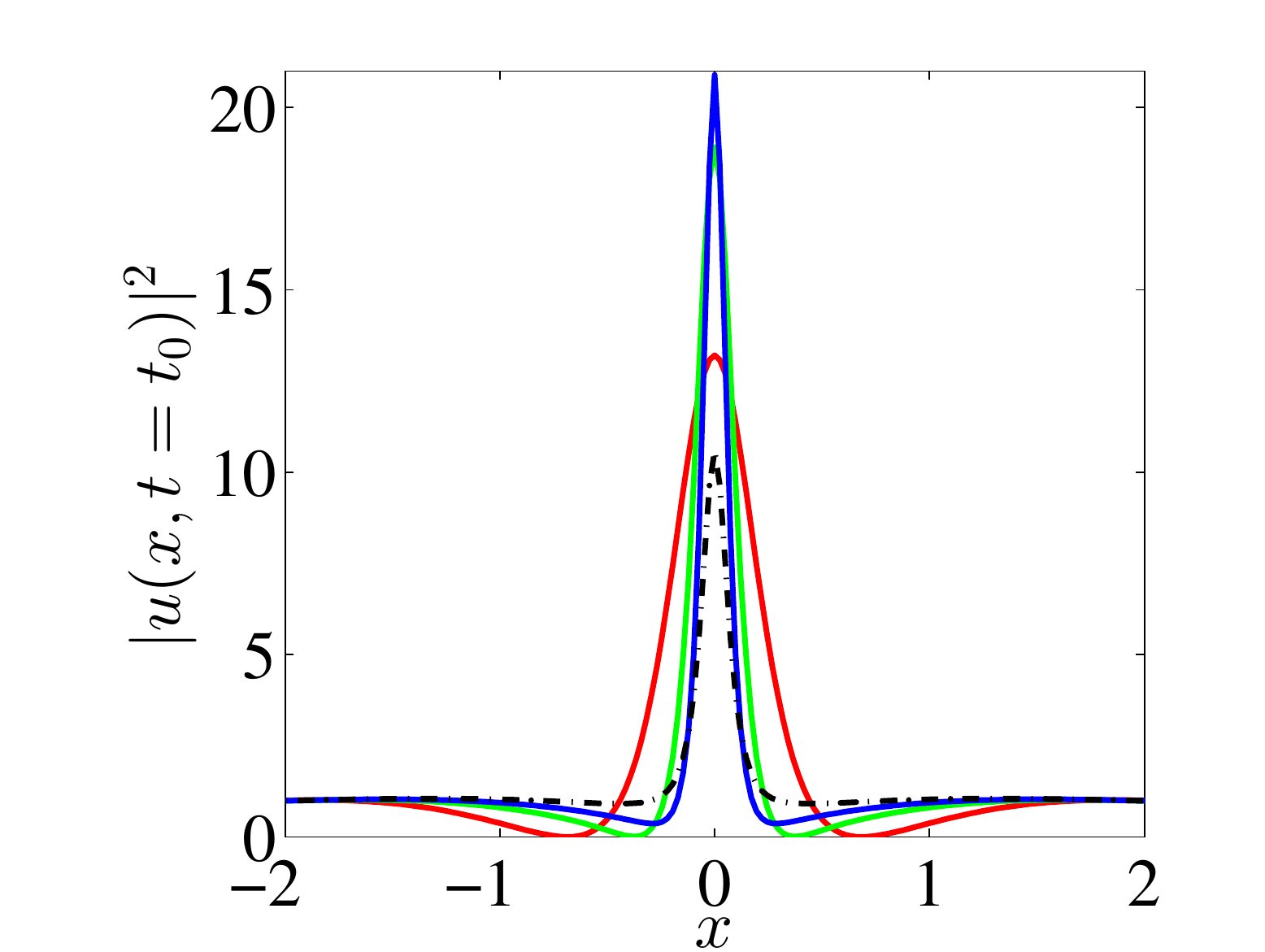}
\label{fig8b}
}
\subfigure[][]{\hspace{-0.5cm}
\includegraphics[height=.17\textheight, angle =0]{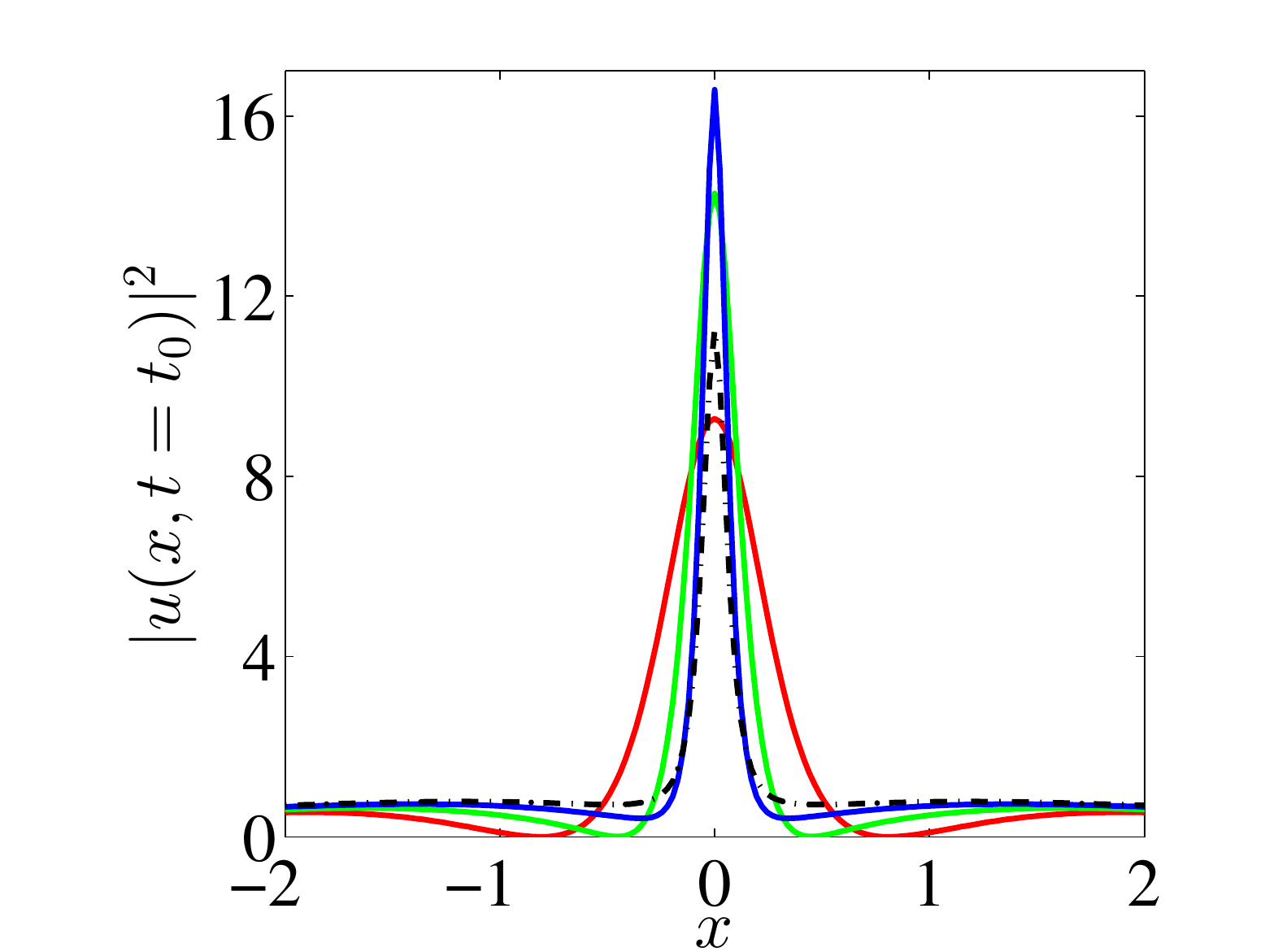}
\label{fig8c}
}
}
\mbox{\hspace{-0.1cm}
\subfigure[][]{\hspace{-1.0cm}
\includegraphics[height=.17\textheight, angle =0]{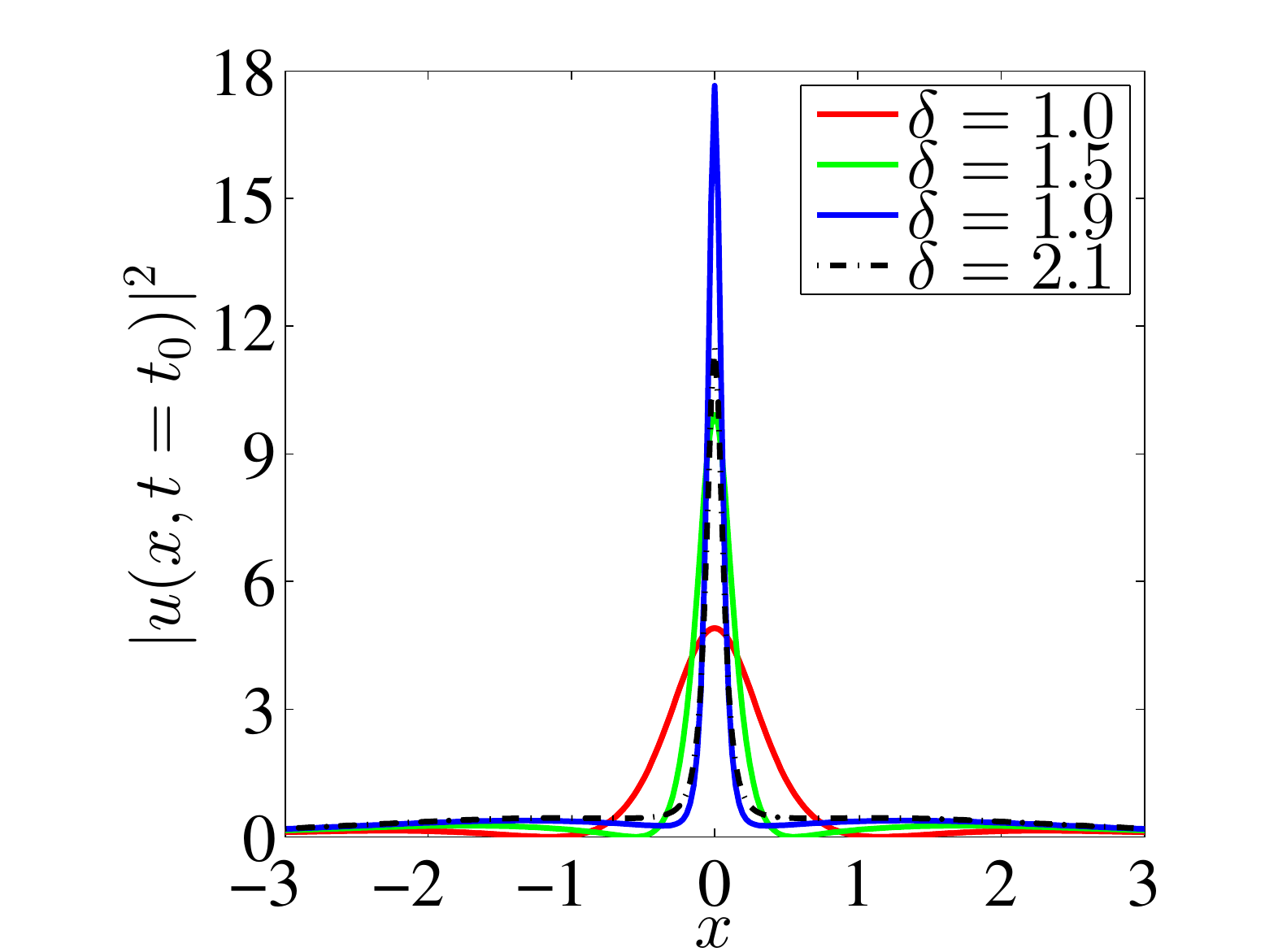}
\label{fig8d}
}
\subfigure[][]{\hspace{-0.5cm}
\includegraphics[height=.17\textheight, angle =0]{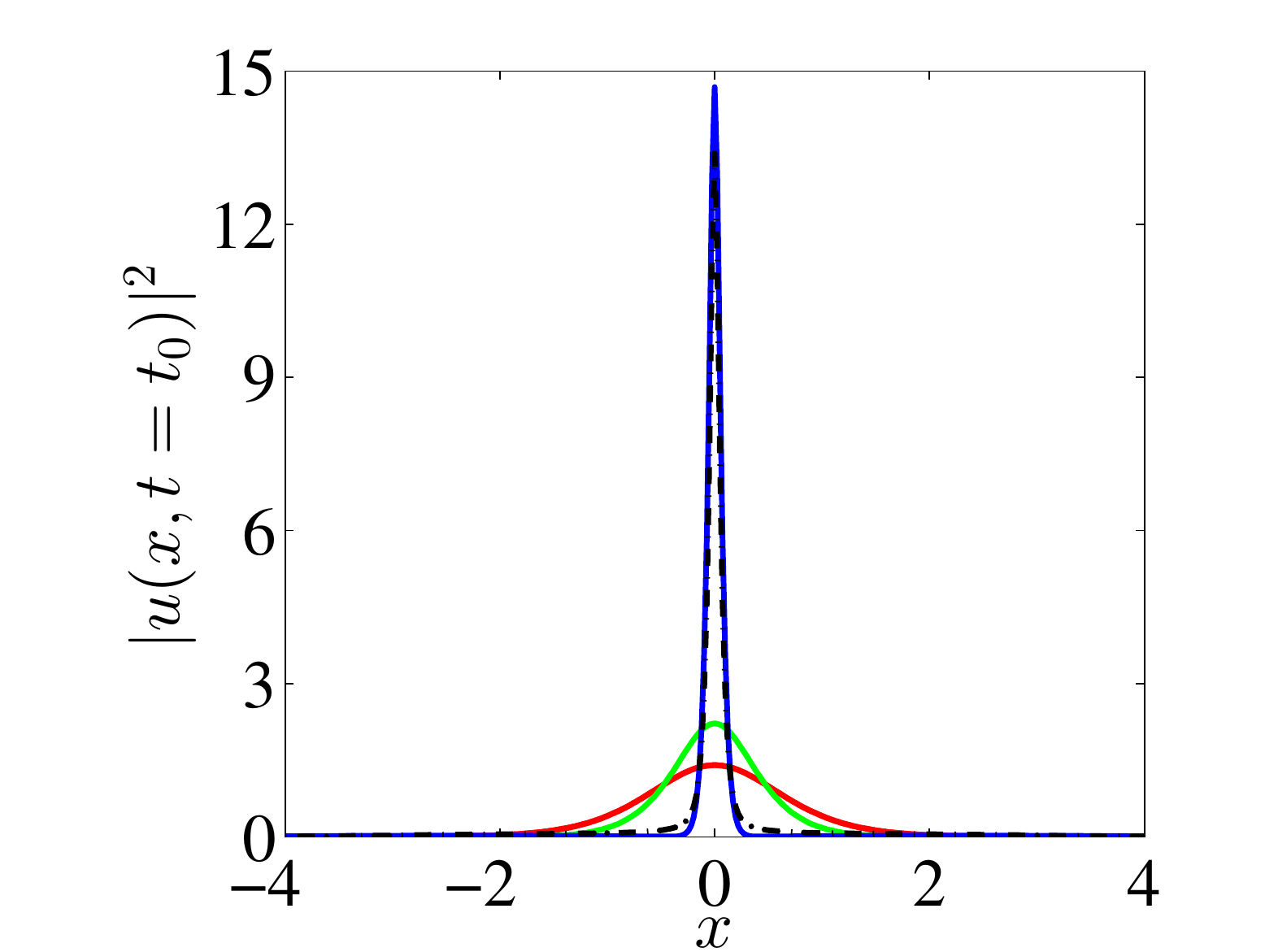}
\label{fig8e}
}
}
\end{center}
\caption{(Color online)
Snapshots of the densities $|u|^{2}$ at $t=t_{0}$ (see, text for details)
and for various values of $\delta$ (see legends) and $\sigma$. Panels (a),
(b), and (c) correspond to values of $\sigma$ of $\sigma=20.1$, $10.5$ and
$5$, whereas panels (d) and (e) to $\sigma=2.5$ and $1.3$, respectively.
\label{fig8}
}
\end{figure}

We have used different types of diagnostics in order to capture the
trends of the variation over $\sigma$ and $\delta$. We have found,
for instance, that the times $t_{0}$ associated with the appearance
of the first peak structure typically decrease (with a notable
exception within the so-called soliton regime of very small
$\sigma$) as $\delta$ is increased, i.e., the effect of $\delta$
increasing clearly promotes focusing. The strength of the focusing
(the intensity of the peak event) is more substantial too when
$\delta$ is increased or when $\sigma$ (and the overall power) is decreased
(data not shown). However, as an additional diagnostic here,
snapshots of densities for various values of $\sigma$ are shown
in Fig.~\ref{fig8} with the aim to shed some light on the nature
of the structures that trigger the high intensity events.
It can be discerned from these plots that the formation of a
high-amplitude wave surrounded by two minima corresponding to zeros of the density
is evident for $\delta=1$ (denoted by solid red lines in the figure) and for all the
cases with $\sigma>1.3$. Such localized structures are reminiscent of the
Peregrine soliton, although for larger $\delta~(>1)$ the locations of the minima are
approaching to each other and no longer correspond to zeros of the density (see
how the relevant spatial distribution has been lifted up).
Thus, our results suggest that the Peregrine-like pattern emerging
(and rapidly disappearing) in the dynamics is crucially responsible for
the strong focusing featured by this generalized NLS dynamics even
for $\delta < 2$. Progressively as $\delta$ approaches (and especially
surpasses) 2, and even more
so as $\sigma$ decreases (where the pattern resembles more a regular soliton),
we encounter the familiar bell-shaped collapse. However, we believe that
this numerical evidence makes a strong case for extreme events
not sharing the permanence of single solitons yet being significant promoters
of large amplitude dynamics resembling collapse even below the critical
point of the (generalized) NLS model.

%

%

\section{Similarities and differences in the NPSE and the 3D GPE}

Our aim in this Section is to briefly illustrate some of the similarities,
as well as differences between the more standard NLS and generalized
NLS models of the previous Section and the results of the more accurate,
in the context of atomic condensates, 1D NPSE and 3D GPE. The two equations
were generally found to provide similar results qualitatively, hence,
we only present selected case examples from each one.
In order to analyze the outcome from the initial condition (\ref{gauss_init}),
we have taken several values of $\sigma$. Collapse takes place as long
as the amplitude $\alpha$ is 
larger than a critical value that differs between the cases of
$\Omega=0$ and $\Omega\neq0$. We have considered the dynamics
for $\alpha$ slightly below the critical one to again showcase the phenomenology
as large amplitude extreme events are approached.

\begin{figure}[!ht]
\begin{center}
\begin{tabular}{ccc}
\includegraphics[width=.3\textwidth]{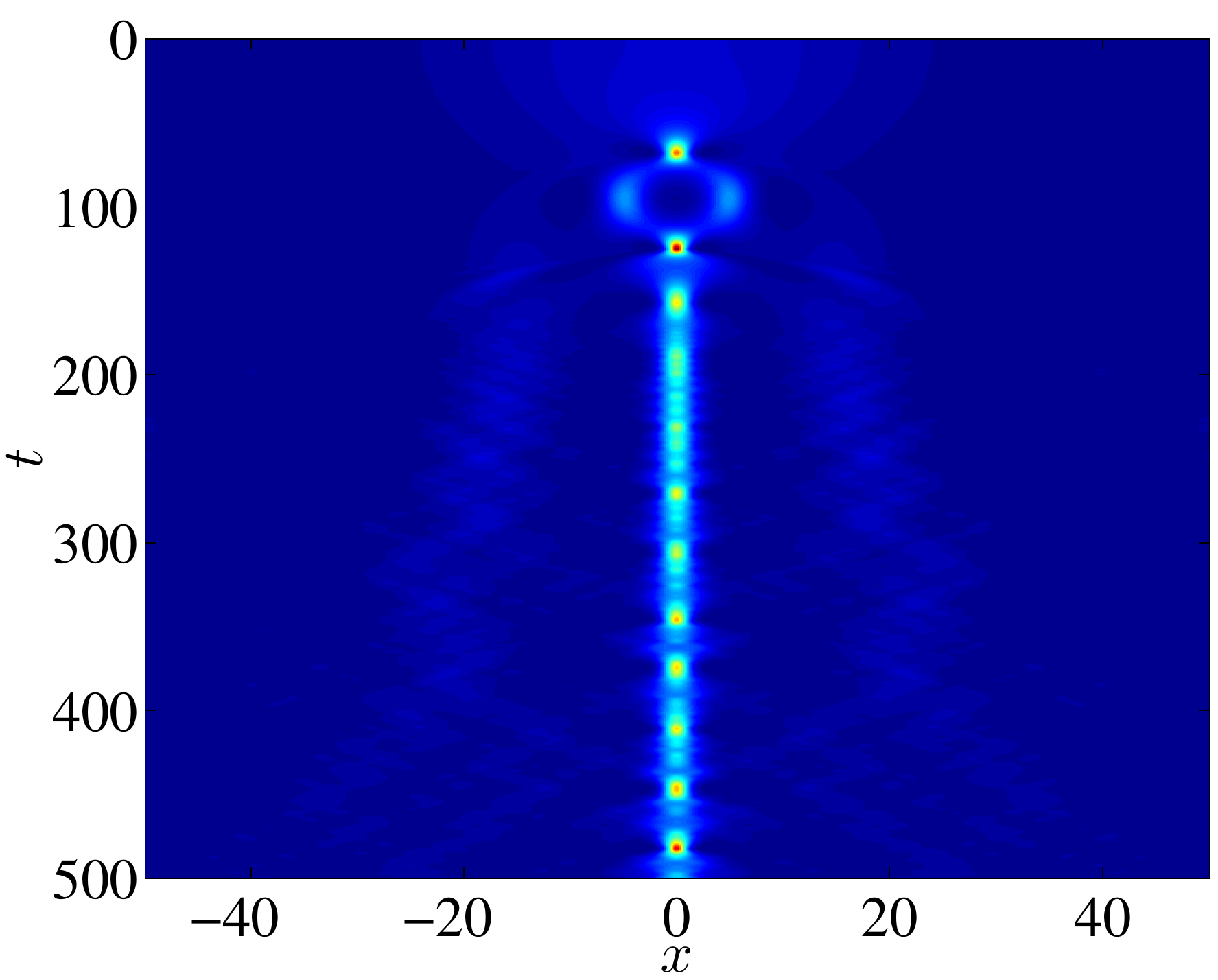} &
\includegraphics[width=.3\textwidth]{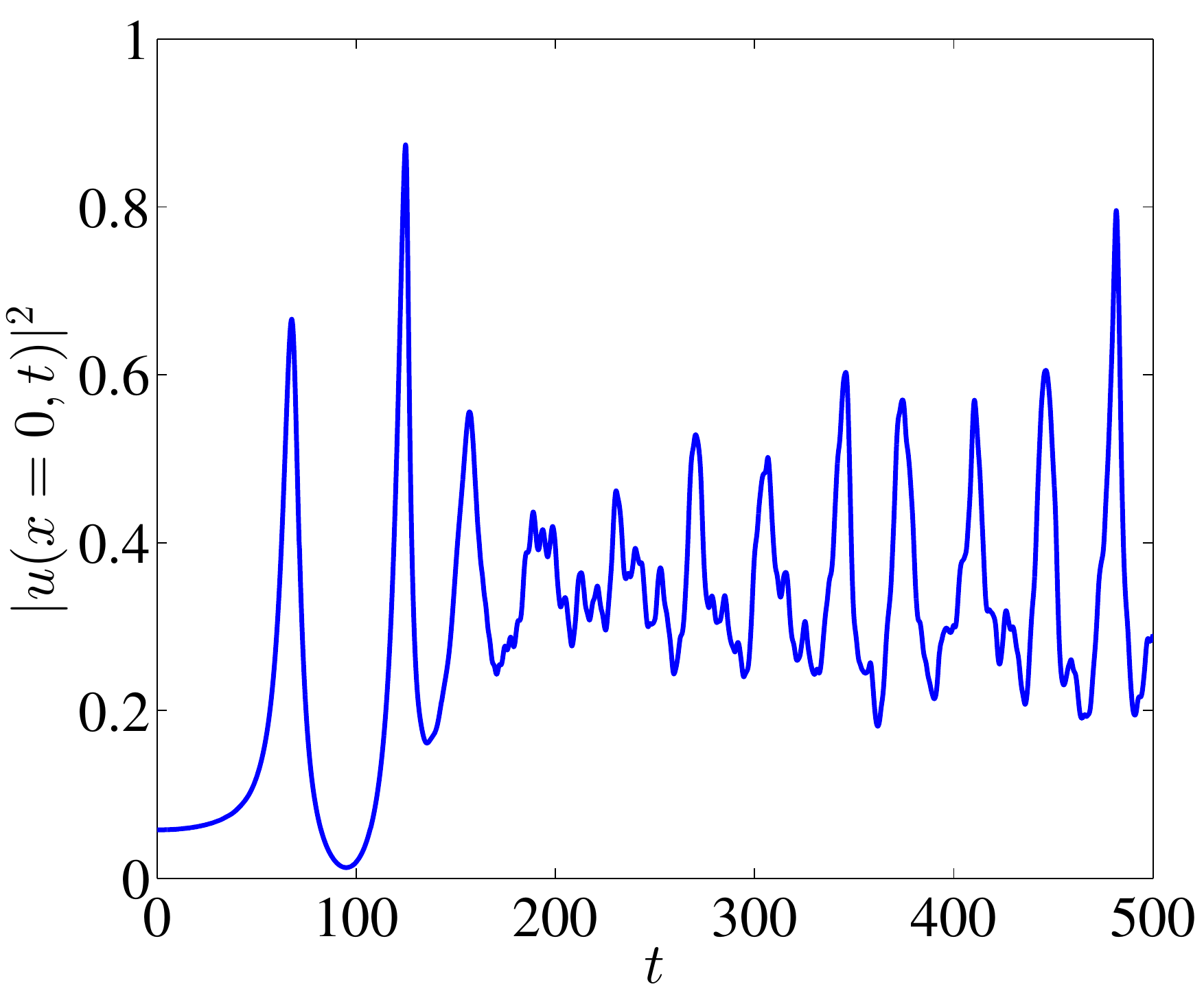} &
\includegraphics[width=.3\textwidth]{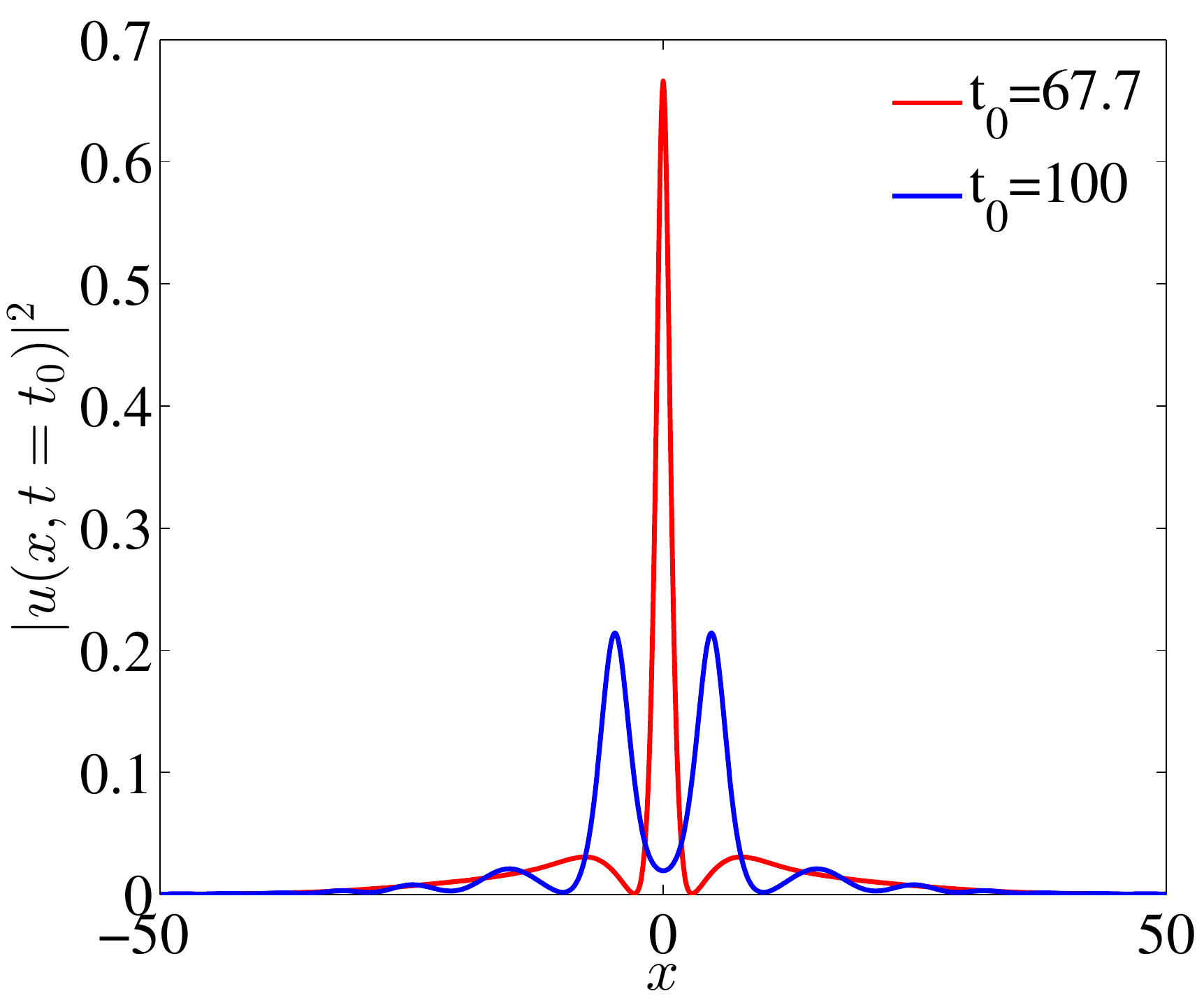} \\
(a) & (b) & (c) \\
\includegraphics[width=.3\textwidth]{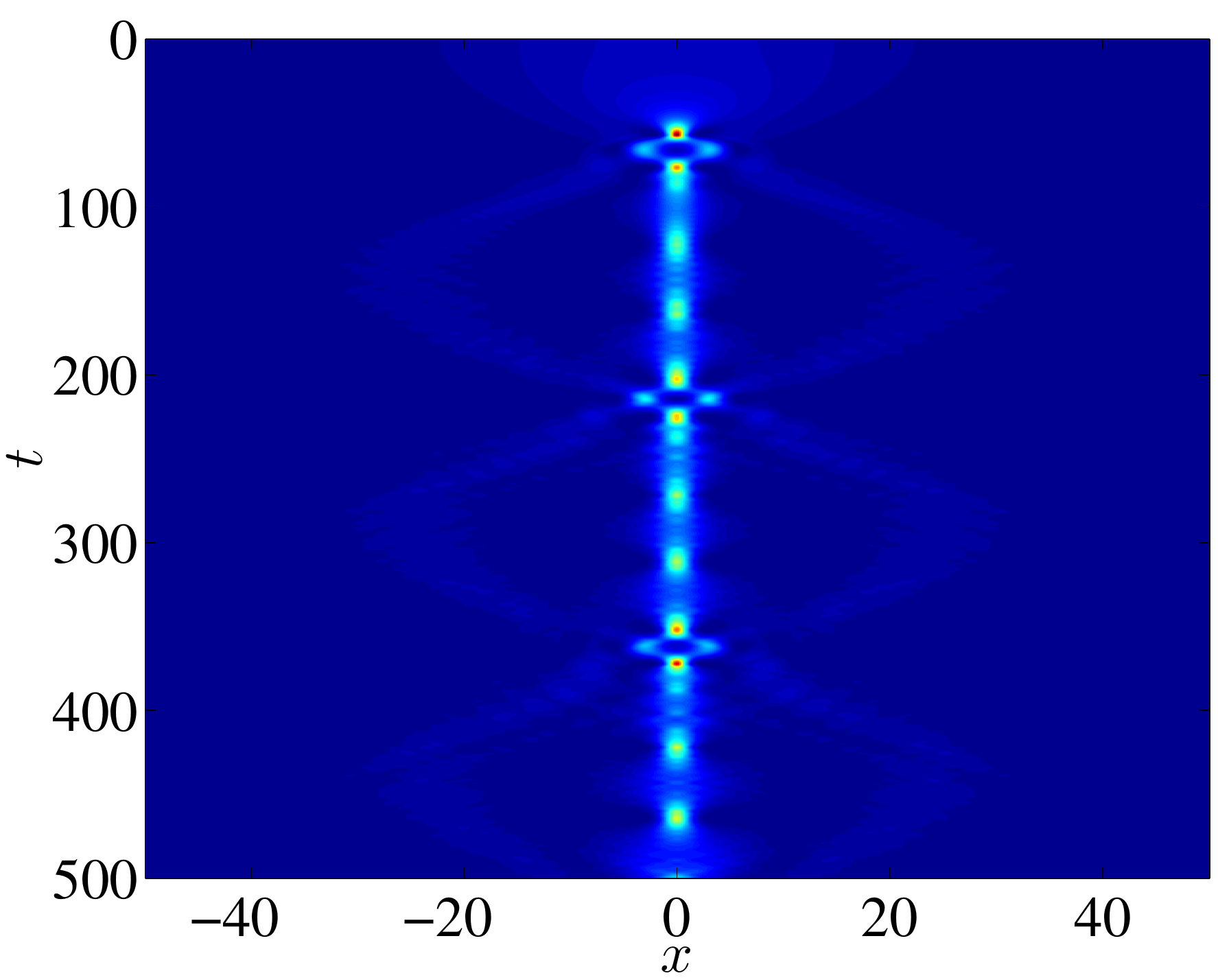} &
\includegraphics[width=.3\textwidth]{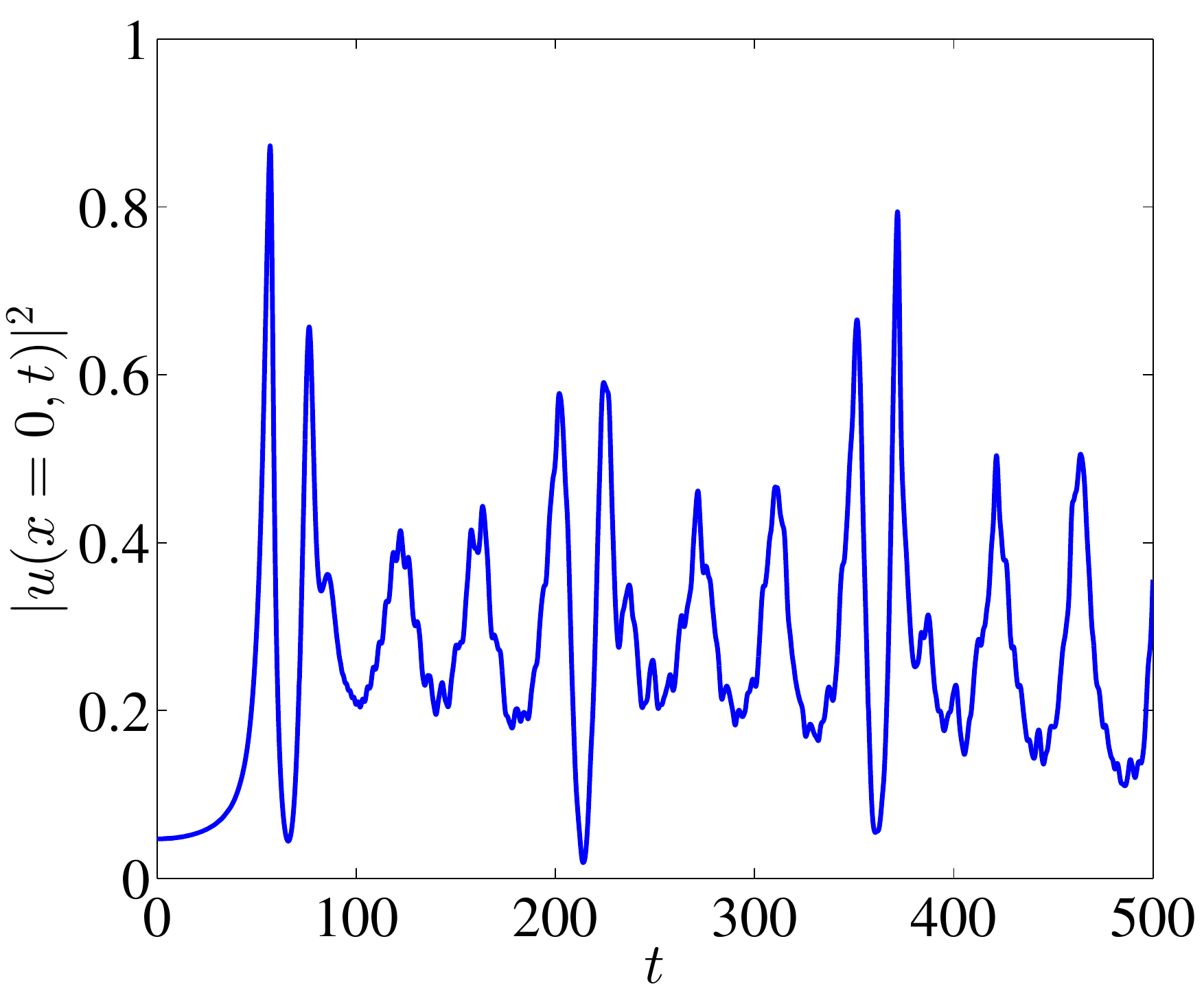} &
\includegraphics[width=.3\textwidth]{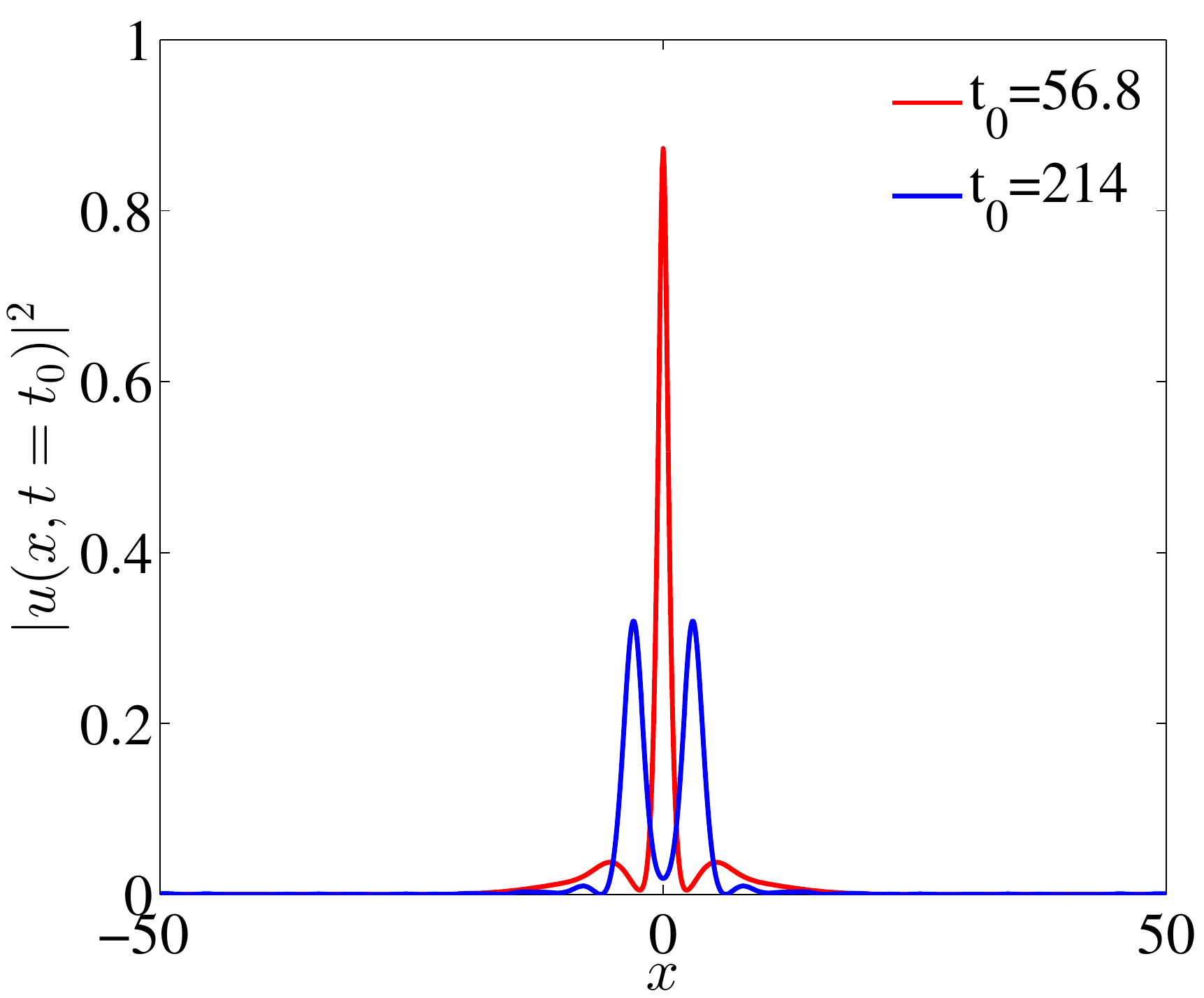} \\
(d) & (e) & (f) \\
\end{tabular}
\end{center}
\caption{
Summary of results corresponding to the NPSE for $\sigma=20$ with ($\alpha=0.2405$, $\Omega=0$) [top] and ($\alpha=0.2172$, $\Omega=\Omega_0$) [bottom]. (a,d) Spatiotemporal evolution of the density $|u(x,t)|^{2}$. (b,e) The global maximum of the density $|u(x,t)|^{2}$ evaluated at $x=0$ as a function of time $t$. (c,f) Density profile at the first maximum in panels (b,e) and a later time where the maximum separation between the humps is observed.
\label{npse3}
}
\end{figure}

A prototypical example of our results for the NPSE model is provided
in Fig.~\ref{npse3}. Here, the value of $\sigma$ chosen was $20$, although
we also performed similar runs for $\sigma=10$ and $\sigma=5$ without dramatically
different results. What can be seen for different amplitudes in the
figure is that the solution has a clearly breathing character (this is
especially evident in the bottom panels), with large focusing events
similar to the events we classified in the previous Section.
That being said, much of the recurrence phenomenology in this case is
lost -- except if the trap induces it, as seems to be partially the
case in the bottom panel. Even more importantly prototypical structures
that much of our analysis of the previous Section was based on -- such as
the CT waveform -- are completely absent, suggesting that the strong non-integrability
of the NPSE model is partially detrimental to such features.

\begin{figure}[!ht]
\begin{center}
\begin{tabular}{cc}
\includegraphics[width=.3\textwidth]{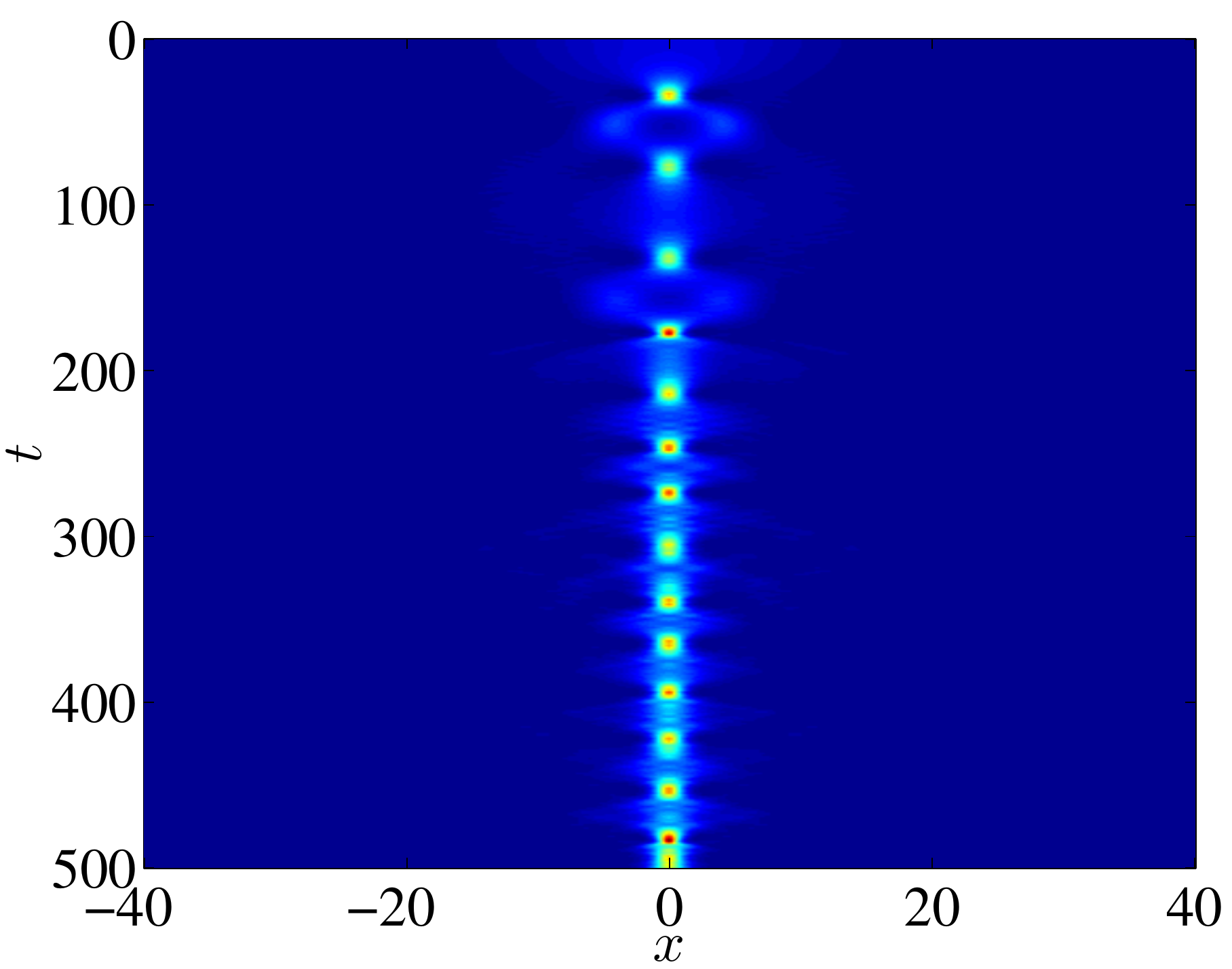} &
\includegraphics[width=.3\textwidth]{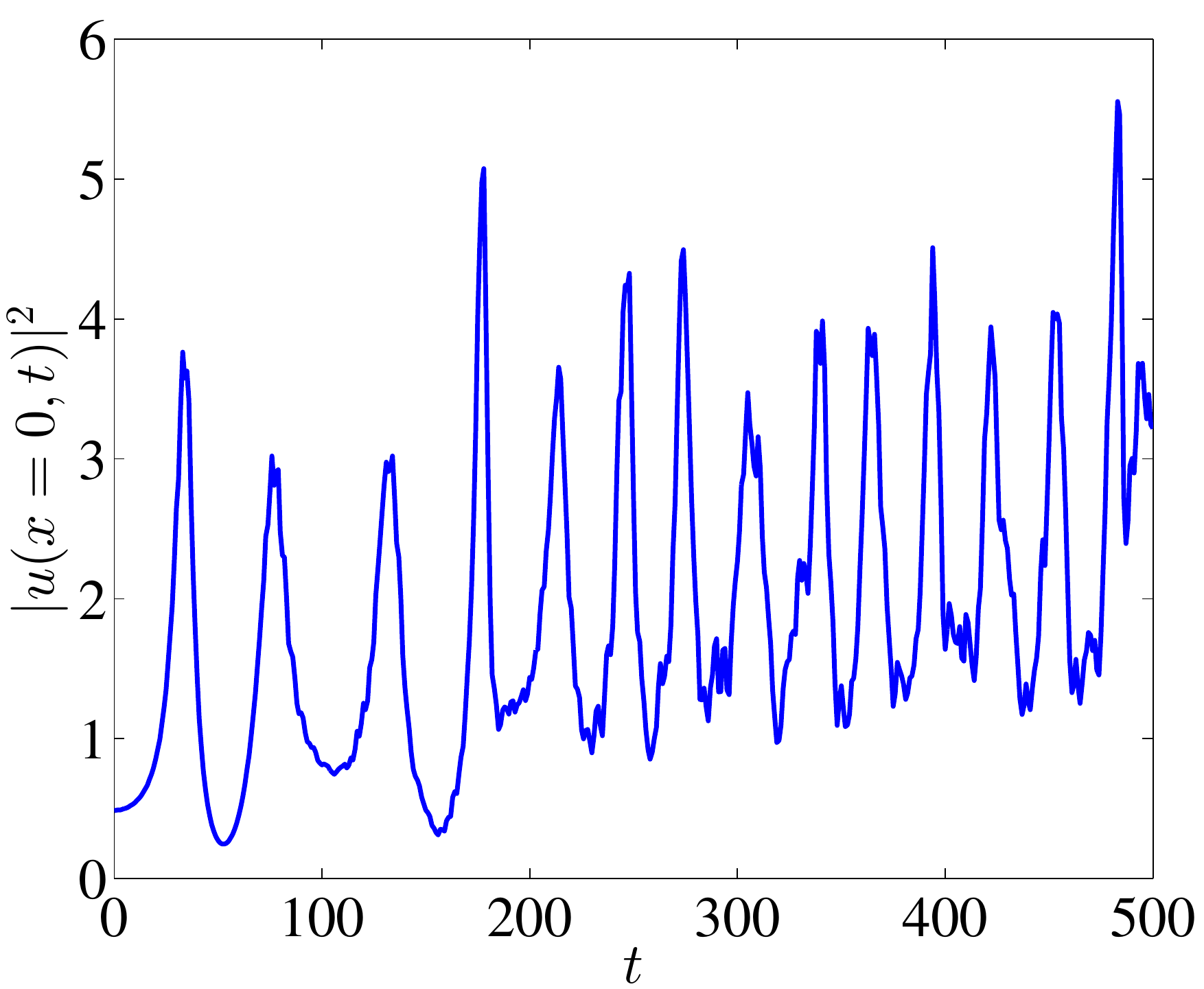} \\
(a) & (b) \\
\includegraphics[width=.3\textwidth]{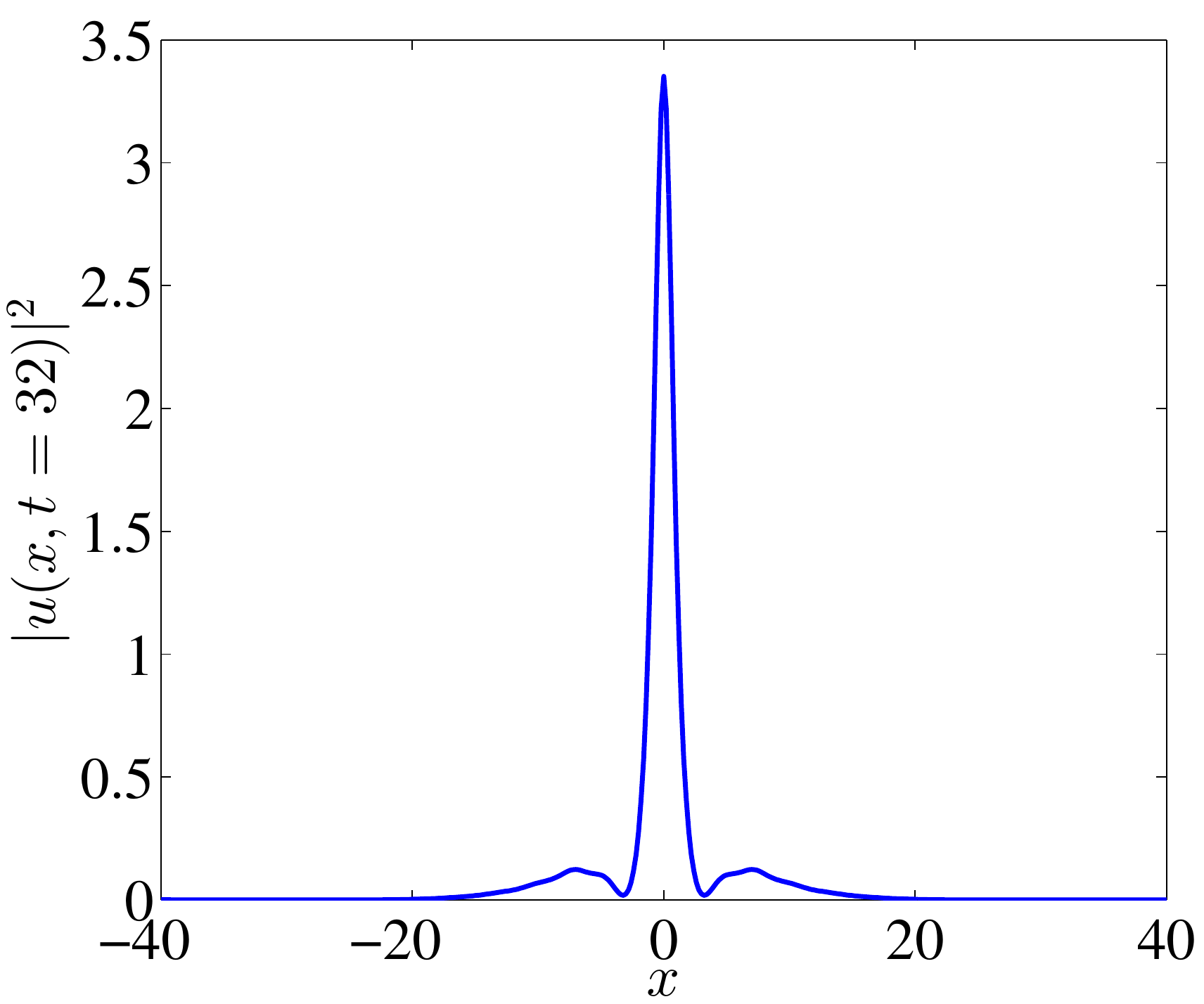} &
\includegraphics[width=.3\textwidth]{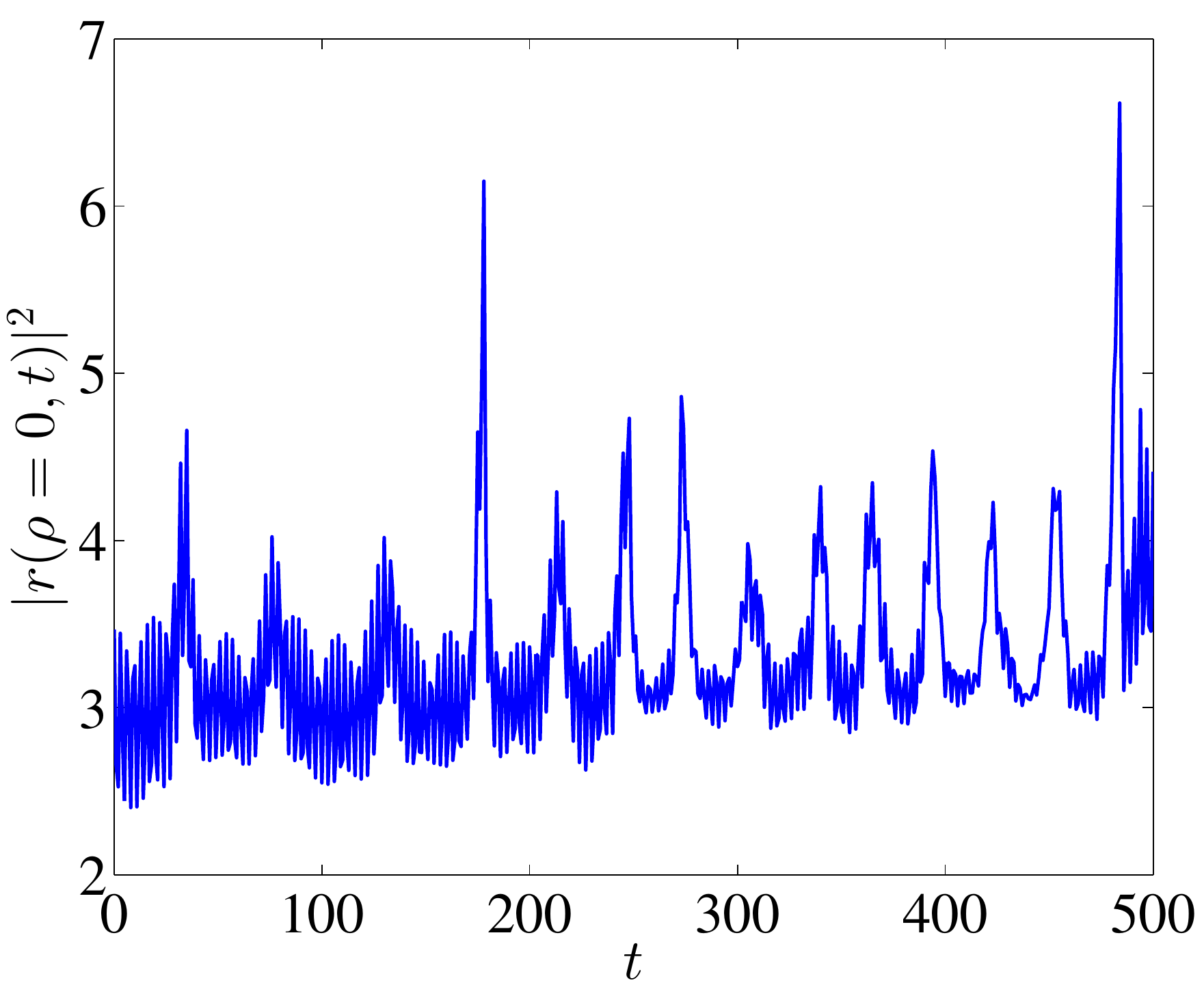} \\
(c) & (d) \\
\end{tabular}
\end{center}
\caption{Summary of results corresponding to the 3D GPE
for $\sigma=10$ and $\alpha=0.6697$ (just below the collapse threshold).
(a) Spatiotemporal evolution of the longitudinal excitations density $|u(x,t)|^{2}$, with $u(x,t)=2\pi\int_{0}^{\infty }\rho d\rho |\psi(\rho,x,t) |^{2}$.
(b) The global maximum of the longitudinal excitations density $|u(x,t)|^{2}$
evaluated at $x=0$ as a function of time $t$.
(c) Density profile at the first maximum in panel (b).
(d) The global maximum of the transverse excitations density $|r(\rho,t)|^2$, with $r(\rho,t)=\int_{-\infty}^{+\infty }dx\, |\psi(\rho,x,t) |^{2}$,
evaluated at $\rho=0$ as a function of time $t$ for the 3D equation.
\label{3d5}
}
\end{figure}

To corroborate that these types of features also appear in the
3D GPE, we have selected a prototypical example of the latter as well,
shown in Fig.~\ref{3d5}. Here, we observe that again despite the
relatively large value of $\sigma$, only a beating oscillation is
observed, i.e., there is no evidence of the CT structure. Nevertheless,
the formation of the peak again seems to have structural characteristics
[through its brief appearance and subsequent disappearance (top left
and top right panels), the formation of the (vanishing density in the
bottom left panel) local minima etc]
of Peregrine-like patterns
rather than of permanent solitonic ones. At the same time, though,
in the precursors to large density focusing in this case, features
of significant transverse excitation are amply evident, e.g., in
the bottom right panel, hence here the role of higher dimensionality
is also important towards appreciating the full
dynamical evolution.

\section{Conclusions and Future Challenges}

The intention of this work was, to a large extent, to raise intriguing
questions and perhaps to a smaller degree, to provide partial answers.
One question was/is: are there initial data whose evolution will resemble
an extreme event (and perhaps more concretely a Peregrine soliton)
in NLS systems ? The answer seems to generally
be yes, in fact, even stemming from rather generic initial data, such
as a Gaussian. The NLS itself and ``mild'' variations thereof
(such as the generalized power model, for exponents close to the cubic
case) seem to in fact bear more complex features, such as the ``Christmas
tree'' structure that we observed. Another question was/is: do these
rogue-wave-like patterns have a role in collapse-type phenomena ?
Again, we believe that numerical simulations seem to be strongly
suggestive in that direction, in fact promoting such phenomena
for models that do not feature self-similarly collapsing solutions
(such as the generalized NLS model with nonlinearities of exponent
below the critical quintic case). Do these features appear to
persist in applications including 
three-dimensional systems inspired by the physics of atomic condensates ?
Our numerical experiments
here suggest that this is only partially the case -- i.e., Peregrine-like waveforms
may seem persistent but other features such as the ``Christmas tree''
patterns are definitely absent in the latter setting.

On the other hand, there are many more questions that are either
not answered or that are perhaps created by these results.
For example, we cannot distinguish definitively if the patterns
that emerge resemble more arrays of Peregrine solitons or whether they
are more $N$-soliton like solutions for $N$ large; what is the
connection between the two ? Similarly,
questions involving understanding a potential
connection between structures like the 2-soliton and Kuznetsov-Ma
breather also arise. Characterizing the outcome of the inverse
scattering problem for the NLS (Zakharov-Shabat) for a prototypical
Gaussian initial state naturally emerges as an important problem to solve.
Another question for theoretical consideration is whether there is a
explicit analogue of the Peregrine soliton for the case of the generalized NLS
model with the exponent $\delta$. We argued that the Peregrine-like
patterns seem responsible for the enhanced focusing, but could one
``distinguish'' the contributing role of a Peregrine and that of a soliton
in inducing collapse phenomena~? Also, do collapse events mathematically truly
arise for $\delta < 2$~? Of course, the over-arching (and perhaps
over-shadowing) query is: would it be possible to observe
experimentally rogue waves in such ultracold bosonic seas ?
These questions suggest an intriguing array of investigations
lying ahead.

\begin{acknowledgments}
P.G.K. and D.J.F. gratefully acknowledge the support of QNRF Grant
No. NPRP8-764-1-160. E.G.C. thanks Theodoros Horikis (University
of Ioannina) for providing help in connection with the spectral
code for the NLS equation P.G.K. and D.J.F. also thank Theodoros Horikis
for multiple fruitful discussions at the beginning of this
project. P.G.K. also thanks G. Biondini, T. Sapsis, and A. Ludu
for discussions. P.G.K. acknowledges support from the
National Science Foundation under Grant DMS-1312856 and from
FP7-People under Grant No. IRSES-605096.
\end{acknowledgments}


\begin{thebibliography}{99}
\bibitem{foot1} 
Contrary to the NLSE given by Eq.~(\ref{nlse_1d}), the model
~(\ref{NPSE}) results in collapse dynamics when $|u|^2\rightarrow1$,
that is, when the density of the field $u$ becomes sufficiently
close to unity. Thus, this limit imposes an upper bound on the
amplitude of the initial condition employed [see Eq.~(\ref{gauss_init})].
However, it should be noted in passing that this limitation
will be bypassed when we consider the full
3D model from which the NPSE originates from (see, the
corresponding discussion below).
\end{thebibliography}
\end{document}